\documentclass[11pt]{article}

\usepackage[disable=true,
textsize=scriptsize,textwidth=\marginparwidth]{todonotes} \newcommand{\todoi}[1]{\todo[inline]{#1}}

\usepackage{chngcntr}
\usepackage{apptools}
\AtAppendix{\counterwithin{theorem}{section}}

\usepackage{amsmath}
\usepackage{amssymb}
\usepackage{ifpdf}
\usepackage{cite}
\usepackage{color}
\usepackage{comment}
\usepackage{wrapfig}
\usepackage{authblk}

\usepackage{url}
\usepackage{amsfonts}

\usepackage{commands-tam}

\usepackage{graphicx}
\usepackage{rotating}
\graphicspath{ {images/} }

\usepackage[colorlinks=true,citecolor=blue,linkcolor=blue]{hyperref}
\usepackage[margin=1in,centering]{geometry}

\usepackage{amsthm}

\numberwithin{equation}{section}
\newtheorem{theorem}{Theorem}[section]

\newtheorem{corollary}[theorem]{Corollary}

\newtheorem{lemma}[theorem]{Lemma}
\newtheorem{thm}{Theorem}[section]
\newtheorem{cor}[theorem]{Corollary}
\newtheorem{lem}[theorem]{Lemma}

\newtheorem{obs}[theorem]{Observation}

\newcommand{\bdd}{\mathsf{bdd}}
\newcommand{\unbdd}{\mathsf{unbdd}}
\newcommand{\post}{\mathsf{post}}
\newcommand{\amax}{\mathrm{amax}}

\newcommand{\annotatestate}[1]{\overline{#1}}
\newcommand{\ans}[1]{\annotatestate{#1}}
\newcommand{\ins}{\ans{x}}
\newcommand{\outs}{\ans{y}}
\newcommand{\as}{\ans{a}}
\newcommand{\os}{\ans{s}}
\newcommand{\qs}{\ans{q}}

\newcommand{\ds}{\ans{d}}
\newcommand{\gs}{\ans{g}}

\renewcommand{\vec}[1]{\mathbf{#1}}

\newcommand{\polylog}{\mathrm{polylog}}

\usepackage{url}
\usepackage{amsfonts}

\newcommand{\ignore}[1]{}

\newcommand{\rest}{\upharpoonright}

\usepackage[normalem]{ulem}

\def\longrightharpoonup{\relbar\joinrel\rightharpoonup}
\def\longleftharpoondown{\leftharpoondown\joinrel\relbar}
\def\longrightleftharpoons{\mathop{\vcenter{\hbox{\ooalign{\raise1pt\hbox{$\longrightharpoonup\joinrel$}\crcr\lower1pt\hbox{$\longleftharpoondown\joinrel$}}}}}}
\newcommand{\calE}{\mathcal{E}}

\newcommand{\vx}{\vec{x}}
\newcommand{\ve}{\vec{e}}
\newcommand{\vm}{\vec{m}}

\newcommand{\vd}{\vec{d}}
\newcommand{\vi}{\vec{i}}
\newcommand{\vs}{\vec{s}}

\newcommand{\vo}{\vec{o}}
\newcommand{\vp}{\vec{p}}
\newcommand{\vq}{\vec{q}}

\renewcommand{\vb}{\vec{b}}
\newcommand{\vc}{\vec{c}}

\newcommand{\vw}{\vec{w}}
\newcommand{\vz}{\vec{z}}
\newcommand{\vu}{\vec{u}}
\newcommand{\vv}{\vec{v}}

\newcommand{\vt}{\vec{t}}

\newcommand{\vC}{\vec{C}}
\newcommand{\tvC}{\tilde{\vec{C}}}
\newcommand{\tvD}{\tilde{\vec{D}}}
\newcommand{\tvT}{\tilde{\vec{T}}}
\newcommand{\tvt}{\tilde{\vec{t}}}
\newcommand{\tvc}{\tilde{\vec{c}}}
\newcommand{\vT}{\vec{T}}
\newcommand{\vS}{\vec{S}}
\newcommand{\vG}{\vec{G}}

\renewcommand{\Pr}{\mathsf{Pr}}
\renewcommand{\time}[2]{\mathsf{T}\left[ #1 \reach #2 \right]}

\usepackage{centernot}
\def\reach{\mathop{\Longrightarrow}\nolimits}

\title{Hardness of computing and approximating predicates and functions with leaderless population protocols\footnote{The first and second authors were supported by NSF grant $1619343$, and the third author by NSF grant $1618895$.}}

\author[1]{Amanda Belleville}
\author[1]{David Doty}
\author[2]{David Soloveichik}
\affil[1]{Department of Computer Science, University of California, Davis}
\affil[2]{Department of Electrical and Computer Engineering, University of Texas, Austin}

\begin{document}
\date{}
\maketitle

\begin{abstract}
Population protocols are a distributed computing model appropriate for describing massive numbers of agents with very limited computational power (finite automata in this paper),
such as sensor networks or programmable chemical reaction networks in synthetic biology.
A population protocol is said to require a leader if every valid initial configuration contains a single agent in a special ``leader'' state that helps to coordinate the computation.
Although the class of predicates and functions computable with probability $1$ (stable computation) is the same whether a leader is required or not (semilinear functions and predicates),
it is not known whether a leader is necessary for fast computation.
Due to the large number of agents $n$ (synthetic molecular systems routinely have trillions of molecules), efficient population protocols are generally defined as those computing in polylogarithmic in $n$ (parallel) time.
We consider population protocols that start in leaderless initial configurations,
and the computation is regarded finished when the population protocol reaches a configuration from which a different output is no longer reachable.

In this setting we show that a wide class of functions and predicates computable by population protocols are not \emph{efficiently} computable (they require at least linear time to stabilize on a correct answer), nor are some linear functions even efficiently \emph{approximable}.
For example, our results for predicates immediately imply that the widely studied parity, majority, and equality predicates cannot be computed in sublinear time.
(Existing arguments specific to majority were already known).
Moreover, it requires at least linear time for a population protocol even to approximate division by a constant or subtraction (or any linear function with a coefficient outside of $\N$), in the sense that for sufficiently small $\gamma > 0$, the output of a sublinear time protocol can stabilize outside the interval $f(m) (1 \pm \gamma)$
on infinitely many inputs $m$.
We also show that it requires linear time to exactly compute a wide range of semilinear functions
(e.g., $f(m)=m$ if $m$ is even and $2m$ if $m$ is odd).

In a complementary positive result,
we show that
with a sufficiently large value of $\gamma$,
a population protocol \emph{can} approximate any linear $f$ with nonnegative rational coefficients,
within approximation factor $\gamma$,
in $O(\log n)$ time.
\end{abstract}

\section{Introduction}

Population protocols were introduced by Angluin, Aspnes, Diamadi, Fischer, and Peralta\cite{angluin2006passivelymobile} as a model of distributed computing in which the agents have very little computational power %(being finite state machines) 
and no control over their schedule of interaction with other agents.
They can be thought of as a special case of 
a model of concurrent processing introduced in the 1960s,
known alternately as 
vector addition systems\cite{karp1969parallel}, 
Petri nets\cite{petri1966communication}, 
or commutative semi-Thue systems (or, when all transitions are reversible, ``commutative semigroups'')\cite{cardoza1976exponential, mayr1982complexity}.
As well as being an appropriate model for electronic computing scenarios such as sensor networks, 
they are a useful abstraction of ``fast-mixing'' physical systems such as 
animal populations\cite{Volterra26}, 
gene regulatory networks\cite{bower2004computational},
and chemical reactions.

The latter application is especially germane:
several recent wet-lab experiments demonstrate the systematic engineering of custom-designed chemical reactions~\cite{zhang2011dynamic, chen2013programmable, baccouche2014dynamic, srinivas2015programming},
unfortunately in all cases having a cost that scales linearly with the number of unique chemical species (states).
(The cost can even be quadratic if certain error-tolerance mechanisms are employed~\cite{thachuk2015leakless}.)
Thus, it is imperative in implementing a molecular computational system to keep the number of distinct chemical species at a minimum.
On the other hand, it is common (and relatively cheap) for the total number of such molecules (agents) to number in the trillions in a single test tube.
It is thus important to understand the computational power enabled by a large number of agents $n$,
where each agent has only a constant number of states (each agent is a finite state machine).

A population protocol is said to require a leader if every valid initial configuration contains a single agent in a special ``leader'' state that helps to coordinate the computation.
Studying computation without a leader is important for understanding essentially distributed systems where symmetry breaking is difficult.
Further, in the chemical setting obtaining single-molecule precision in the initial configuration is difficult.
Thus, it would be highly desirable if the population protocol did not require an exquisitely tuned initial configuration.

\subsection{Introduction to the model}
A population protocol is defined by a finite set $\Lambda$ of \emph{states} that each agent may have, 
together with a \emph{transition function}\footnote{Some work allows nondeterministic transitions, in which the transition function maps to subsets of $\Lambda \times \Lambda$. Our results are independent of whether transitions are nondeterministic, and we choose a deterministic, symmetric transition function, rather than a more general relation $\delta \subseteq \Lambda^4$, merely for notational convenience.} $\delta: \Lambda^2 \to \Lambda^2$.
A \emph{configuration} is a nonzero vector $\vc\in\N^\Lambda$ describing, 
for each $\os \in \Lambda$, the \emph{count} $\vc(\os)$ of how many agents are in state $\os$.
By convention we denote the number of agents by $n = \|\vc\| = \sum_{\os\in\Lambda} \vc(\os).$
Given states $\ans{r}_1,\ans{r}_2,\ans{p}_1,\ans{p}_2 \in \Lambda$,
if $\delta(\ans{r}_1,\ans{r}_2)=(\ans{p}_1,\ans{p}_2)$ 
(denoted $\ans{r}_1,\ans{r}_2 \to \ans{p}_1,\ans{p}_2$), 
and if a pair of agents in respective states $\ans{r}_1$ and $\ans{r}_2$ interact, then their states become $\ans{p}_1$ and $\ans{p}_2$.\footnote{In the most generic model, there is no restriction on which agents are permitted to interact. If one prefers to think of the agents as existing on nodes of a graph, then it is the complete graph $K_n$ for a population of $n$ agents.}
The next pair of agents to interact is chosen uniformly at random. 
The expected (parallel) time for any event to occur is the expected number of interactions, divided by the number of agents $n$.
This measure of time is based on the natural parallel model where each agent participates in a constant number of interactions in one unit of time; 
hence $\Theta(n)$ total interactions are expected per unit time~\cite{angluin2006fast}.

The most well-studied population protocol task is computing Boolean-valued \emph{predicates}.
It is known that a protocol \emph{stably decides} a predicate $\phi:\N^k \to \{0,1\}$
(meaning computes the correct answer with probability 1; 
see Section~\ref{sec:predicates} for a formal definition)
if~\cite{angluin2006passivelymobile} and only if~\cite{AngluinAE2006semilinear} $\phi$ is semilinear.

Population protocols can also compute integer-valued \emph{functions} $f:\N^k\to\N$.
Suppose we start with $m \leq n/2$ agents in ``input'' state $\ins$ 
and the remaining agents in a ``quiescent'' state $\qs$.
Consider the protocol with a single transition rule $\ins,\qs \to \outs,\outs$.
Eventually exactly $2m$ agents are in the ``output'' state $\outs$,
so this protocol computes the function $f(m) = 2m$.
Furthermore (letting $\# \os =$ count of state $\os$), 
if $\# \qs - 2m = \Omega(n)$ initially (e.g., $\# \qs = 3m$),
then it takes $\Theta(\log n)$ expected time until $\# \outs = 2m$.
Similarly, the transition rule $\ins,\ins\to \outs,\qs$ computes the function $f(m) = \floor{m/2}$, 
but exponentially slower, in expected time $\Theta(n)$.
The transitions $\ins_1,\qs \to \outs,\qs$ and $\ins_2,\outs \to \qs,\qs$ 
compute $f(m_1,m_2)=m_1-m_2$ (assuming $m_1 \geq m_2$),
also in time $\Theta(n)$ if $m_1=m_2+O(1)$.

Formally,
we say a population protocol \emph{stably computes} a function $f:\N^k\to\N$ if,
for every ``valid'' initial configuration $\vi\in\N^\Lambda$ representing input $\vm\in\N^k$
(via counts $\vi(\ins_1),\ldots,\vi(\ins_k)$ of ``input'' states $\Sigma = \{\ins_1,\ldots,\ins_k\} \subseteq \Lambda$)
with probability 1 the system reaches from $\vi$ to $\vo$ such that $\vo(\outs)=f(\vm)$ 
($\outs \in \Lambda$ is the ``output'' state)
and $\vo'(\outs)=\vo(\outs)$ for every $\vo'$ reachable from $\vo$
(i.e., $\vo$ is \emph{stable}).
Defining what constitutes a ``valid'' initial configuration
(i.e., what non-input states can be present initially, and how many)
is nontrivial.
In this paper we focus on population protocols without a \emph{leader}---a state present in count $1$, or small count---in the initial configuration.
Here, we equate ``leaderless'' with initial configurations in which no positive state count is sublinear in the population size $n$.

It is known that a function $f:\N^k\to\N$ is stably computable by a population protocol if and only if its \emph{graph} $\{(\vm,f(\vm)) \mid \vm\in\N^{k} \} \subset \N^{k+1}$ is a semilinear set~\cite{AngluinAE2006semilinear, CheDotSolDetFuncNaCo}.
This means intuitively that it is piecewise affine, with each affine piece having rational slopes.

Despite the exact characterization of predicates and functions stably computable by population protocols, 
we still lack a full understanding of which of the stably computable (i.e., semilinear) predicates and functions are computable \emph{quickly} (say, in time polylogarithmic in $n$)
and which are only computable slowly (linear in $n$).
For positive results, much is known about time to \emph{convergence}
(time to get the correct answer).
It has been known for over a decade that with an initial leader, any semilinear predicate can be stably computed with polylogarithmic convergence time~\cite{angluin2006fast}.
Furthermore, 
it has recently been shown that all semilinear predicates can be computed \emph{without} a leader with sublinear convergence time~\cite{kosowski2018brief}.
(See Section~\ref{sec:related} for details.)

In this paper, however, 
we exclusively study time to \emph{stabilization} without a leader
(time after which the answer is \emph{guaranteed to remain correct}).
Except where explicitly marked otherwise with a variant of the word ``converge'', 
all references to time in this paper refer to time until stabilization.
Section~\ref{sec:conclusion} explains in more detail the distinction between the two.

\subsection{Contributions} \label{sec:results}

\paragraph{Undecidability of many predicates in sublinear time.}
Every semilinear predicate $\phi:\N^k\to\{0,1\}$
is stably decidable in $O(n)$ time~\cite{angluin2006fast}.
Some, such as $\phi(m_1,m_2) = 1$ iff $m_1 \geq 1$, are stably decidable in $O(\log n)$ time by a leaderless protocol, 
in this case by the transition $\ins_1,\ins_2 \to \ins_1,\ins_2$,
where $\ins_1$ ``votes'' for output $1$ and $\ins_2$ votes 0.
A predicate is \emph{eventually constant} if $\phi(\vm) = \phi(\vm')$ for all sufficiently large $\vm \neq \vm'$.
We show 
in Theorem~\ref{thm:predicates}
that unless $\phi$ is eventually constant,
any leaderless population protocol stably deciding a predicate $\phi$ requires at least linear time.
Examples of non-eventually constant predicates include
parity ($\phi(m) = 1$ iff $m$ is odd),
majority ($\phi(m_1,m_2) = 1$ iff $m_1 \geq m_2$), 
and
equality ($\phi(m_1,m_2)=1$ iff $m_1=m_2$).
It does \emph{not} include certain semilinear predicates,
such as
$\phi(m_1,m_2) = 1$ iff $m_1 \geq 1$ (decidable in $O(\log n)$ time)
or
$\phi(m_1,m_2) = 1$ iff $m_1 \geq 2$ (decidable in $O(n)$ time, and no faster protocol is known).

\paragraph{Definition of function computation and approximation.} 
We formally define computation and approximation of functions $f:\N^k\to\N$ for population protocols.
This mode of computation was discussed briefly in the first population protocols paper\cite[Section 3.4]{angluin2006passivelymobile},
which focused more on Boolean predicate computation,
and it was defined formally in the more general model of chemical reaction networks\cite{CheDotSolDetFuncNaCo, DotHajLDCRNNaCo}.
Some subtle issues arise that are unique to population protocols.
We also formally define a notion of function \emph{approximation} with population protocols.

\paragraph{Inapproximability of most linear functions with sublinear time and sublinear error.}
Recall that the transition rule $\ins,\ins \to \outs,\qs$ computes $f(m)=\floor{m/2}$ in linear time.
Consider the transitions $\ans{a},\ins \to \ans{b},\outs$ and $\ans{b},\ins \to \ans{a},\qs$,
starting with $\# \ins = m$, $\# \as = \gamma m$ for some $0 < \gamma < 1$, and $\# \outs = \# \qs = 0$
(so $n=m+\gamma m$ total agents).
Then eventually $\# \outs \in \{m/2, \ldots, m/2+\gamma m\}$ 
and $\# \ins = 0$ (stabilizing $\# \outs$),
after $O(\frac{1}{\gamma} \log n)$ expected time.
%\todo{AB: fix reference}
%(This is analyzed in more detail in Section~\ref{sec:approx-positive-results}.)
(This is analyzed in more detail in Section~\ref{sec:positive-results}.)
Thus, if we tolerate an error linear in $n$,
then $f$ can be approximated in logarithmic time.
However, Theorem~\ref{thm:main} shows this error bound to be tight:
\emph{any} leaderless population protocol that approximates $f(m)=\floor{m/2}$,
or any other linear function with a coefficient outside of $\N$
%$f(\vm) = \sum_{i=1}^k \floor{c_i \vm(i)}$ with some $c_i \not\in \N$
(such as $\floor{4 m/3}$ or $m_1-m_2$),
requires at least linear time to achieve sublinear error.

As a corollary, such functions cannot be stably computed in sublinear time
(since computing exactly is the same as approximating with zero error).
Conversely, it is simple to show that any linear function with all coefficients in $\N$  is stably computable in logarithmic time
(Observation~\ref{obs:stably-compute-positive-integer-coefficient}).
Thus we have a dichotomy theorem for the efficiency (with regard to stabilization) of computing linear functions $f$ by leaderless population protocols: 
if all of $f$'s coefficients are in $\N$, 
then it is computable in logarithmic time, 
and otherwise it requires linear time. % to compute.

\paragraph{Approximability of nonnegative rational-coefficient linear functions with logarithmic time and linear error.}
Theorem~\ref{thm:main} says that no linear function with a coefficient outside of $\N$ can be stably computed with sublinear time and sublinear error.
In a complementary positive result,
Theorem~\ref{thm:positive}, 
by relaxing the error to linear, 
and restricting the coefficients to be \emph{nonnegative} rationals (but not necessarily integers), 
 we show how to approximate any such linear function in logarithmic time.
(It is open if $m_1 - m_2$ can be approximated with linear error in logarithmic time.)

\paragraph{Uncomputability of many nonlinear functions in sublinear time.}
What about non-linear functions?
Theorem~\ref{thm:negative-result-eventually-positive-integral-linear} 
states that sublinear time computation cannot go much beyond linear functions with coefficients in $\N$:
unless $f$ is \emph{eventually $\N$-linear},
meaning linear with nonnegative integer coefficients on all sufficiently large inputs,
any protocol stably computing $f$ requires at least linear time.
Examples of non-eventually-$\N$-linear functions, 
that provably cannot be computed in sublinear time, include 
$f(m_1,m_2) = \min(m_1,m_2)$
(computable slowly via 
$\ins_1,\ins_2 \to \outs,\qs$), 
and $f(m) = m-1$
(computable slowly via 
$\ins,\ins \to \ins,\outs$).

The only remaining semilinear functions 
whose asymptotic time complexity remains unknown
are those ``piecewise linear'' functions that switch between pieces only near the boundary of  $\N^k$;
for example, $f(m) = 0$ if $m \leq 3$ and $f(m)=m$ otherwise.

Note that there is a fundamental difficulty in extending the negative results to functions and predicates that ``do something different only near the boundary of $\N^k$''.
This is because for inputs where one state is present in small count, the population protocol could in principle use that input as a ``leader state''---and no longer be leaderless.
However, this does not directly lead to a positive result for such inputs,
because it is not obvious how to use (for instance) 
state $\ins_1$ as a leader when its count is 1 while still maintaining correctness for larger counts of $\ins_1$.

Our results leave open the possibility that non-eventually constant predicates and non-eventually-$\N$-linear functions, 
which cannot be computed in sublinear time in our setting, 
\emph{could} be efficiently computed in the following ways:
\begin{enumerate}
    \item 
    \emph{With an initial leader} stabilizing to the correct answer in sublinear time,
    
    \item
    Stabilizing to an output in expected sublinear time but 
    \emph{allowing a small probability of incorrect output}
    (with or without a leader),
    or

    \item \label{open-ques:3}
    Without an initial leader but \emph{converging} to the correct output in sublinear time. 
    Recently Kosowski and Uzna\'{n}ski~\cite{kosowski2018brief} showed that this task is indeed possible.
    (See Section~\ref{sec:related}.)
\end{enumerate}

\subsection{Essential proof techniques}

Techniques developed in previous work for proving time lower bounds~\cite{LeaderElectionDIST, alistarh2017timespace} can certainly generalize beyond leader election and majority,
although it was not clear what precise category of computation they cover.
However, to extend the impossibility results to all non-eventually-$\N$-linear functions, 
we needed to develop new tools.

Compared to our previous work showing the impossibility of sublinear time leader election~\cite{LeaderElectionDIST}, 
we achieve three main advances in proof technique. 
First, the previous machinery did not give us a way to affect \emph{large}-count states predictably to change the answer,
but rather focused on using surgery to remove a single leader state.
Second, we need much additional reasoning 
to argue if a predicate is not eventually constant,
then we can find infinitely many $\alpha$-dense inputs that differ on their output but are close together.
This leads to a contradiction when we use transition manipulation arguments to show how to absorb the small extra difference between the inputs without changing the output.
Third,
we need entirely different reasoning to argue that if a semilinear function is not eventually $\N$-linear,
then we can find infinitely many $\alpha$-dense inputs $\vm$ that do not appear ``locally affine'':
pushing a small distance $\vv$ from $\vm$ changes the function $f(\vm)$ to $f(\vm+\vv) = f(\vm) + \epsilon$,
but pushing by the same distance again changes it a different amount,
i.e., $f(\vm + 2\vv) = f(\vm) + \epsilon + \delta$, 
where $\epsilon \neq \delta$.
This leads to a contradiction when we use transition manipulation arguments to show how,
from input $\vm + 2\vv$,
to stabilize the count of the output to the incorrect value $f(\vm) + 2\epsilon$.\footnote{
    These arguments are easier to understand for the special case when we can assume $f$ is linear.
    Thus Section~\ref{sec:negative-results} concentrates on this special case,
    obtaining an exact characterization of the efficiently computable linear functions.
    Section~\ref{sec:nonlinear-functions} reasons about the more difficult case of arbitrary semilinear functions.
}

Both in prior and current work, the high level intuition of the proof technique is as follows.
The overall argument is a proof by contradiction: 
if sublinear time computation is possible, 
then we find a nefarious execution sequence that stabilizes to an incorrect output.
In more detail, sublinear time computation requires avoiding ``bottlenecks''---having to go through a transition in which both states are present in small count 
(constant independent of the number of agents $n$).
Traversing even a single such transition requires linear time.
Technical lemmas show that bottleneck-free execution sequences from $\alpha$-dense initial configurations 
(i.e., where every state that is present is present in at least $\alpha n$ count) are amenable to predictable ``surgery''~\cite{LeaderElectionDIST, alistarh2017timespace}.
At the high level, the surgery lemmas show how states that are present in ``low'' count when the population protocol stabilizes, 
can be manipulated (added or removed) such that only ``high'' count other states are affected.
Since it can also be shown that changing high count states in a stable configuration does not affect its stability,
this means that the population protocol cannot ``notice'' the surgery, 
and remains stabilized to the previous output.
For leader election, 
the surgery allows one to remove an additional leader state 
(leaving us with no leaders).
For majority computation~\cite{alistarh2017timespace},
the minority input must be present in low count (or absent) at the end.
This allows one to add enough of the minority input to turn it into the majority, 
while the protocol continues to output the wrong answer.

However, applying the previously developed surgery lemmas to fool a \emph{function} computing population protocol is more difficult.
The surgery to consume additional input states affects the count of the output state, 
which could be present in ``large count'' at the end. 
How do we know that the effect of the surgery on the output is not consistent with the desired output of the function?
In order to arrive at a contradiction we develop two new techniques, both of which are necessary to cover all cases.
The first involves showing that the slope of the change in the count of the output state as a function of the input states is inconsistent.
The second involves exposing the semilinear structure of the graph of the function being computed, and forcing it to enter the ``wrong piece'' (i.e., periodic coset).

\subsection{Related work} \label{sec:related}
\paragraph{Positive results.}
Angluin, Aspnes, Diamadi, Fischer, and Peralta~\cite{angluin2006passivelymobile} 
showed that any semilinear predicate can be decided in expected parallel time $O(n \log n)$, 
later improved to $O(n)$ by Angluin, Aspnes, and Eisenstat~\cite{angluin2006fast}.
More strikingly, the latter paper showed that if an initial leader is present
(a state assigned to only a single agent in every valid initial configuration),
then there is a protocol for $\phi$ that \emph{converges} to the correct answer in expected time $O(\log^5 n)$.
However, this protocol's expected time to \emph{stabilize} is still provably $\Omega(n)$.
Section~\ref{sec:conclusion} explains this distinction in more detail.
Chen, Doty, and Soloveichik~\cite{CheDotSolDetFuncNaCo} showed in the related model of chemical reaction networks
(borrowing techniques from the related predicate results~\cite{angluin2006passivelymobile, AngluinAE2006semilinear})
that any semilinear \emph{function}
(integer-output $f:\N^k\to\N$)
can similarly be computed with expected convergence time $O(\log^5 n)$ 
if an initial leader is present,
but again with much slower stabilization time $O(n \log n)$.
Doty and Hajiaghayi~\cite{DotHajLDCRNNaCo} showed that any semilinear function can be computed by a chemical reaction network without a leader with expected convergence and stabilization time $O(n)$.
Although the chemical reaction network model is more general,
these results hold for 
%our definition of function computation with 
population protocols.

Kosowski and Uzna\'{n}ski~\cite{kosowski2018brief} show that all semilinear predicates can be computed without an initial leader,
converging in $O(\polylog~n)$ time if a small probability of error is allowed,
and converging in $O(n^\epsilon)$ time with probability 1, 
where $\epsilon$ can be made arbitrarily close to 0 by changing the protocol.
They also showed leader election protocols
(which can be thought of as computing the constant function $f(m) = 1$)
with the same properties.

Since efficient computation seems to be helped by a leader, the computational task of leader election has received significant recent attention.
In particular, Alistarh and Gelashvili~\cite{polylogleaderICALP} showed that in a variant of the model allowing the number of states $\lambda_n$ to grow with the population size $n$, 
a protocol with $\lambda_n = O(\log^3 n)$ states can elect a leader with high probability in $O(\log^3 n)$ expected time.
Alistarh, Aspnes, Eisenstat, Gelashvili, and Rivest~\cite{alistarh2017timespace} later showed how to reduce the number of states to $\lambda_n = O(\log^2 n)$,
at the cost of increasing the expected time to $O(\log^{5.3} n \log \log n)$.
Gasieniec and Stachowiak~\cite{gasieniec2018fast} showed that there is a protocol with $O(\log \log n)$ states electing a leader in $O(\log^2 n)$ time in expectation and with high probability, recently improved to $O(\log n \log \log n)$ time by Gasieniec, Stachowiak, and Uzna\'{n}ski~\cite{gasieniec2018almost}.
This asymptotically matches the $\Omega(\log \log n)$ states provably required for sublinear time leader election
(see negative results below).

%\todoi{DD: see if there is any connection with this paper (possibly with dense configs): \url{https://arxiv.org/abs/1604.07187}}

\paragraph{Negative results.}
The first attempt to show the limitations of sublinear time population protocols,
using the more general model of chemical reaction networks,
was made by Chen, Cummings, Doty, and Soloveichik~\cite{SpeedFaultsDIST}.
They studied a variant of the problem in which negative results are easier to prove,
an ``adversarial worst-case'' notion of sublinear time: 
the protocol is required to be sublinear time not only from the initial configuration, 
but also from any reachable configuration.
They showed that the predicates computable in this manner are 
precisely those whose output depends only on the presence or absence of states (and not on their exact positive counts).
% Refining these techniques for use in the standard model of time complexity (measuring expected time only from the initial configuration),
Doty and Soloveichik~\cite{LeaderElectionDIST} showed the first $\Omega(n)$ lower bound on expected time from valid initial configurations, 
proving that any protocol electing a leader with probability 1 takes $\Omega(n)$ time.

These techniques were improved by 
Alistarh, Aspnes, Eisenstat, Gelashvili, and Rivest~\cite{alistarh2017timespace},
who showed that even with up to $\lambda_n = O(\log \log n)$ states, 
any protocol electing a leader with probability 1 requires nearly linear time: $\Omega(n/\polylog\ n)$.
They used these tools to prove time lower bounds for another important computational task:
majority (detecting whether state $\ans{x}_1$ or $\ans{x}_2$ is more numerous in the initial population, 
by stabilizing on a configuration in which the state with the larger initial count occupies the whole population).
Alistarh, Aspnes, and Gelashvili~\cite{alistarh2018space} strengthened the state lower bound,
showing that $\Omega(\log n)$ states are required to compute majority in $O(n^{1-c})$ time for some $c>0$, 
when a certain ``natural'' condition is imposed on the protocol that holds for all known protocols.

% They showed that there exist $c \in (0,1)$ and $K \geq 1$ so that any protocol deciding majority,
% with initial gap $|\# \ans{x}_1 - \# \ans{x}_2| \geq \epsilon n$
% and $\lambda_n \leq c \log \log n$ states,
% must take $\Omega(n / (K^{\lambda_n} + \epsilon n)^2)$ expected time.
% Thus, with a constant number of states and initial gap $|\# \ans{x}_1 - \# \ans{x}_2| = O(1)$, 
% any protocol requires $\Omega(n)$ time.

In contrast to these previous results on the specific tasks of leader election and majority,
we obtain time lower bounds for a broad class of functions and predicates,
showing ``most'' of those computable at all by population protocols,
cannot be computed in sublinear time.
Since they all \emph{can} be computed in linear time,
this settles their asymptotic population protocol time complexity.
%\opt{sub}{Section~\ref{sec:open-questions} contains a discussion of several relevant open questions related to the current work.}

Informally, one explanation for our result could be that some computation requires electing ``leaders'' as part of the computation, and other computation does not.
Since leader election itself requires linear time as shown in~\cite{LeaderElectionDIST}, the computation that requires it is necessarily inefficient.
It is not clear, however, how to define the notion of a predicate or function computation requiring electing a leader somewhere in the computation,
but recent work by Michail and Spirakis helps to clarify the picture~\cite{michail2016many}.

\subsection{Organization of this paper}
Section~\ref{sec:prelim} defines population protocol model and notation.
Section~\ref{subsec:technical-tools} proves the technical lemmas that are used in all the time lower bound proofs.
Section~\ref{sec:predicates} shows that a wide class of predicates requires $\Omega(n)$ time to compute.
Section~\ref{sec:function-computation-defn} explains our definitions of function computation and approximation.
Section~\ref{sec:negative-results} shows that linear functions with either a negative 
(e.g., $m_1-m_2$)
or non-integer 
(e.g., $\floor{m/2}$)
coefficient cannot be stably approximated with $o(n)$ error in $o(n)$ time.
Section~\ref{sec:positive-results} shows our positive result, 
Theorem~\ref{thm:positive},
that linear functions with all nonnegative rational coefficients
(e.g., $\floor{2m_1/3} + 3m_2$)
\emph{can} be stably approximated with $O(n)$ error in $O(\log n)$ time.
% It is an open question whether a function with integer negative coefficients
% (e.g., $m_1-m_2$)
% can be stably approximated with sublinear error in $o(n)$ time.
Section~\ref{sec:nonlinear-functions} studies non-linear functions,
showing that a large class of those computable by population protocols require $\Omega(n)$ time to compute.
Section~\ref{sec:conclusion} states conclusions and open questions.

\section{Preliminaries}
\label{sec:prelim}

% \todo{Since we use the following concept repeatedly, it might be worth defining:
% say a sequence $\vx_0,\vx_1,\ldots \in \N^\Lambda$ is \emph{strongly increasing} 
% if for all $\os \in \Lambda$, and all $i\in\N$, 
% $\vx_i(\os) < \vx_{i+1}(\os)$, 
% i.e., all coordinates increase each time $i$ increases.}

If $\Lambda$ is a finite set 
(in this paper, of \emph{states}, 
which will be denoted as lowercase Roman letters with an overbar such as $\os$), 
we write $\N^\Lambda$ to denote the set of functions $\vc:\Lambda \to \N$.
Equivalently, we view an element $\vc\in\N^\Lambda$ as a vector of $|\Lambda|$ nonnegative integers, 
with each coordinate ``labeled'' by an element of $\Lambda$.
(By assuming some canonical ordering $\os_1,\ldots,\os_{k}$ of $\Lambda$, 
we also interpret $\vc\in\N^\Lambda$ as a vector $\vc \in \N^k$.)
Given $\os \in \Lambda$ and $\vc \in \N^\Lambda$, we refer to $\vc(\os)$ as the \emph{count of $\os$ in $\vc$}.
Let $\|\vc\| = \| \vc \|_1 = \sum_{\os\in\Lambda} \vc(\os)$. % denote the total number of agents.
We write $\vc \leq \vc'$ to denote that $\vc(\os) \leq \vc'(\os)$ for all $\os \in \Lambda$. 
Since we view vectors $\vc\in\N^\Lambda$ equivalently as multisets of elements from $\Lambda$, if $\vc \leq \vc'$ we say $\vc$ is a \emph{subset} of $\vc'$.
For $\alpha>0$, we say that $\vc\in\N^k$ is \emph{$\alpha$-dense} if, 
for all $i\in\{1,\ldots,k\}$, if $\vc(i) > 0$, then $\vc(i) \geq \alpha \|\vc\|$.

It is sometimes convenient to use multiset notation to denote vectors, e.g., $\{\ins,\ins,\outs\}$ and $\{2\ins,\outs\}$ both denote the vector $\vc$ defined by $\vc(\ins)=2$, $\vc(\outs)=1$, and $\vc(\os)=0$ for all $\os \not\in \{\ins,\outs\}$.
Given $\vc,\vc' \in \N^\Lambda$, we define the vector component-wise operations of addition $\vc+\vc'$, subtraction $\vc-\vc'$, and scalar multiplication $m \vc$ for $m \in \N$.
For a set $\Delta \subset \Lambda$, we view a vector $\vc \in \N^\Delta$ equivalently as a vector $\vc \in \N^\Lambda$ by assuming $\vc(\os)=0$ for all $\os \in \Lambda \setminus \Delta.$
Write $\vc \rest \Delta$ to denote the vector $\vd \in \N^\Delta$ such that $\vc(\os)=\vd(\os)$ for all $\os \in \Delta$.
%\todo{DD: we only use this in the proof of Claim~\ref{claim:firstfixing}; maybe move it to there.}
%Given $s_1,\ldots,s_k \in \Lambda$, $\vc\in\N^\Lambda$, and $n_1,\ldots,n_k\in\Z$,we write $\vc + \{n_1 s_1,\ldots,n_k s_k\}$ to denote vector addition of $\vc$ with the vector $\vv\in\Z^{\{s_1,\ldots,s_k\}}$ defined by $\vv(s_i)=n_i$.
For any vector or matrix $\vc$, let $\amax(\vc)$ denote the largest absolute value of any component of $\vc$.
Also, given $b\in\N$ and $\vm\in\N^k$, $\max(b,\vm)$ is a shorthand for $\max \left( \{b\} \cup \bigcup_{i=1}^k \{\vm(i)\} \right)$,
and similar for $\amax(b,\vm)$.

In this paper, the floor function $\floor{\cdot}:\R\to\Z$ is defined to be the integer \emph{closest to 0} that is  distance $< 1$ from the input, e.g., $\floor{-3.4} = -3$ and $\floor{3.4} = 3$.
For an (infinite) set/sequence of configurations $C$, let 
$\bdd(C) = \{\os \in \Lambda \mid (\exists b\in\N)(\forall \vc \in C)\ \vc(\os) < b\}$
be the set of states whose counts are bounded by a constant in $C$.
Let $\unbdd(C) = \Lambda \setminus \bdd(C)$.
For $m \in \N$, let $\N^k_{\geq m_0} = \{ \vm \in \N^k \mid (\forall i \in \{1,\ldots,k\}\ \vm(i) \geq m_0 \}$,
denote the set of vectors in which each coordinate is at least $m_0$.
%and let $\N^k_{\geq m_0} = \{\vm \in \N^k \mid (\forall i \in \{1,\ldots,k\}\ \vm(i) = 0 \text{ or } \vm(i) \geq m \}$
%denote the set of vectors in which each coordinate is either 0 or at least $m_0$.\todo{DS: I assume the last one should be taken out?}

%In other words, any state present in $\vc$ occupies at least an $\alpha$ fraction of the total number of agents.

\subsection{Population Protocols}
\label{subsec:pp-defn}

A \emph{population protocol} is a pair $\calP=(\Lambda,\delta)$,
where $\Lambda$ is a finite set of \emph{states}
and $\delta:\Lambda^2 \to \Lambda^2$ is the (symmetric) \emph{transition function}. 
%, $\Sigma = \{\ins_1,\ldots,\ins_k\} \subset \Lambda$ is the set of \emph{input symbols}, and $\outs \in \Lambda \setminus \Sigma$ is the \emph{output symbol}.
A \emph{configuration} of a population protocol is a vector $\vc \in \N^\Lambda$, with the interpretation that $\vc(\os)$ agents are in state $\os\in\Lambda$.
If there is some ``current'' configuration $\vc$ understood from context, we write $\# \os$ to denote $\vc(\os)$.
By convention, the value $n\in\Z_{\geq 1}$ represents the total number of agents $\|\vc\|$.
A \emph{transition} is a 4-tuple $\tau = (\ans{r}_1,\ans{r}_2,\ans{p}_1,\ans{p}_2) \in \Lambda^4$,
written $\tau: \ans{r}_1,\ans{r}_2 \to \ans{p}_1,\ans{p}_2$,
such that $\delta(\ans{r}_1,\ans{r}_2)=(\ans{p}_1,\ans{p}_2)$.
If an agent in state $\ans{r}_1$ interacts with an agent in state $\ans{r}_2$, then they change states to $\ans{p}_1$ and $\ans{p}_2$.
This paper typically defines a protocol by a list of transitions, with $\delta$ implicit.
There is a \emph{null} transition $\delta(\ans{r}_1,\ans{r}_2)=(\ans{r}_1,\ans{r}_2)$ if a different output for $\delta(\ans{r}_1,\ans{r}_2)$ is not specified.

Given $\vc\in\N^\Lambda$ and transition $\tau: \ans{r}_1,\ans{r}_2 \to \ans{p}_1,\ans{p}_2$, we say that $\tau$ is \emph{applicable} to $\vc$ if $\vc \geq \{\ans{r}_1,\ans{r}_2\}$, i.e., $\vc$ contains 2 agents, one in state $\ans{r}_1$ and one in state $\ans{r}_2$.
If $\tau$ is applicable to $\vc$, then write $\tau(\vc)$ to denote the configuration $\vc - \{\ans{r}_1,\ans{r}_2\} + \{\ans{p}_1,\ans{p}_2\}$ (i.e., that results from applying $\tau$ to $\vc$); otherwise $\tau(\vc)$ is undefined.
A finite or infinite sequence of transitions $(\tau_i)$ is a \emph{transition sequence}.
Given a $\vc_0\in\N^\Lambda$ and a transition sequence $(\tau_i)$, 
the induced \emph{execution sequence} (or \emph{path}) 
is a finite or infinite sequence of configurations $(\vc_0, \vc_1, \ldots)$ such that, for all $i \geq 1$, 
$\vc_i = \tau_{i-1}(\vc_{i-1})$.\footnote{When the initial configuration to which a transition sequence is applied is clear from context, we may overload terminology and refer to a transition sequence and an execution sequence interchangeably.}
If a finite execution sequence, with associated transition sequence $\qs$, starts with $\vc$ and ends with $\vc'$, we write $\vc \reach_q \vc'$.
We write $\vc \reach_\calP \vc'$ (or $\vc \reach \vc'$ when $\calP$ is clear from context) if such a path exists (i.e., it is possible to reach from $\vc$ to $\vc'$) and we say that $\vc'$ is \emph{reachable} from $\vc$.
Let $\post_\calP(\vc) = \{ \vc' \mid \vc \reach_\calP \vc' \}$ to denote the set of all configurations reachable from $\vc$, writing $\post(\vc)$ when $\calP$ is clear from context.
If it is understood from context what is the initial configuration $\vi$, then say $\vc$ is simply \emph{reachable} if $\vi \reach \vc$.
%Note that this notation omits mention of $\calP$; we always deal with a single protocol at a time, so it is clear from context which protocol is defining the transitions.
If a transition $\tau: \ans{r}_1,\ans{r}_2 \to \ans{p}_1,\ans{p}_2$ has the property that for $i\in\{1,2\}$, $\ans{r}_i\not\in\{\ans{p}_1,\ans{p}_2\}$, or if ($\ans{r}_1=\ans{r}_2$ and ($\ans{r}_i \neq \ans{p}_1$ or $\ans{r}_i \neq \ans{p}_2$)), then we say that $\tau$ \emph{consumes} $\ans{r}_i$;
i.e., applying $\tau$ reduces the count of $\ans{r}_i$.
We say $\tau$ \emph{produces} $\ans{p}_i$ if it increases the count of $\ans{p}_i$.

%We use the following shorthand notations, borrowed from the theory of chemical reaction networks.
%We write $\ins \to \outs,z$ to denote that for all states $\os \in \Lambda$, there is a transition $\ins,\os \to \outs,z$.
%We write $\ins,\outs \to z$ (respectively, $\ins,\outs \to \emptyset$ to denote that there is a state $w$ such that $\ins,\outs \to z,w$ (respectively, $\ins,\outs \to w,w$), such that $w$ is a ``null'' state: it is not an input state for any non-null transitions.

\subsection{Time Complexity}
\label{subsec:time-defn}

The model used to analyze time complexity is a discrete-time Markov process, whose states correspond to configurations of the population protocol.
In any configuration the next interaction is chosen by selecting a pair of agents uniformly at random and applying transition function $\delta$ to determine the next configuration.
Since a transition may be null, self-loops are allowed.
To measure time we count the expected total number of interactions (including null),
and divide by the number of agents $n$.
(In the population protocols literature, this is often called ``parallel time''; i.e. $n$ interactions among a population of $n$ agents corresponds to one unit of time).
Let $\vc\in\N^\Lambda$ and $C \subseteq \N^\Lambda$.
Denote the probability that the protocol reaches from $\vc$ to some configuration $\vc'\in C$ by $\Pr[\vc \reach C]$.
%We want to talk about the expected time to reach from a configuration $\vc$ to another $\vc'$, but it may be the case that we never reach configuration $\vc'$.
%However, suppose there is a set of configurations $C$, such that from configuration $\vc$, the protocol has probability 1 to eventually reach some $\vc' \in C$.
If $\Pr[\vc \reach C]=1$,\footnote{Since population protocols have a finite reachable configuration space, this is equivalent to requiring that for all $\vx\in\post(\vc)$, there is a $\vc' \in C \cap \post(\vx)$.}
define the \emph{expected time to reach from  $\vc$ to $C$}, denoted $\time{\vc}{C}$, to be the expected number of interactions to reach from $\vc$ to some $\vc' \in C$, divided by the number of agents $n=\|\vc\|$.
If $\Pr[\vc \reach C]<1$ then $\time{\vc}{C} = \infty$.

\section{Technical tools}
\label{subsec:technical-tools}

In this section we explain some technical results that are used in proving the time lower bounds of
Theorems~\ref{thm:predicates},
\ref{thm:subtraction-slow}, 
\ref{thm:division-slow},
\ref{thm:affine-not-sublinear-computable}, 
and
\ref{thm:negative-result-eventually-positive-integral-linear}.
In some cases the main ideas are present in previous papers, 
but several must be adapted significantly to the current problem.
Throughout Section~\ref{subsec:technical-tools}, let $\calP = (\Lambda,\delta)$ be a population protocol.

Although other results from this section are used in this paper,
the key technical result of this section is Corollary~\ref{cor:push-Delta-simplified}.
It gives a generic method to start with an initial configuration $\vi$ reaching in sublinear time to a  configuration $\vo$
(in all our uses $\vo$ is a stable configuration, but this is not required by the corollary),
and starting from \emph{two} copies of $\vi$,
to manipulate the transitions leading from $2\vi$ to $2\vo$ while having a predictable effect on the counts of certain states,
possibly also starting with a ``small'' number of extra states,
denoted $\vd^\Delta$ in Corollary~\ref{cor:push-Delta-simplified}.
This leads to a contradiction when the effect on the counts of the states representing the output can be shown to be incorrect for the given input $2\vi + \vd^\Delta$.

We often deal with infinite sequences of configurations.\footnote{In general these will not be \emph{execution} sequences. Typically none of the configurations are reachable from any others because they are configurations with increasing numbers of agents.}
The following lemma, used frequently in reasoning about population protocols, shows that we can always take a nondecreasing subsequence.

\begin{lem}[Dickson's Lemma~\cite{dickson}] \label{lem:dickson}
    Any infinite sequence $\vc_0,\vc_1,\ldots \in \N^k$ has an infinite nondecreasing subsequence $\vc_{i_0} \leq \vc_{i_1} \leq \ldots$, where $i_0 < i_1 < \ldots \in \N$. 
\end{lem}

\subsection{Bottleneck transitions take linear time}

Let $b\in\N$.
A transition $\ans{r}_1,\ans{r}_2 \to \ans{p}_1,\ans{p}_2$ is a \emph{$b$-bottleneck} for configuration $\vc$ if $\vc(\ans{r}_1) \leq b$ and $\vc(\ans{r}_2) \leq b$.

    The next observation, proved in~\cite{LeaderElectionDIST}, states that, 
    if to get from a configuration $\vx \in \N^\Lambda$ to some configuration in a set $S \subseteq \N^\Lambda$, 
    it is necessary to execute a transition $\ans{r}_1,\ans{r}_2 \to \ans{p}_1,\ans{p}_2$ in which the counts of $\ans{r}_1$ and $\ans{r}_2$ are both at most some number $b$, then the expected time to reach from $\vx$ to some configuration in $S$ is $\Omega(n/b^2)$.
    
    \begin{obs}[\cite{LeaderElectionDIST}] \label{obs:bottleneck-linear-time}
        Let $b\in\N$, $\vx \in \N^\Lambda$, and $S \subseteq \N^\Lambda$ such that $\Pr[\vx \reach S]=1$.
        If every path taking $\vx$ to a configuration $\vo \in S$ has a $b$-bottleneck, then $\time{\vx}{S} \geq \frac{n-1}{2(b |\Lambda|)^2} > \frac{n}{3 b^2 \|\Lambda\|^2}$.
    \end{obs}
    
    The next corollary is useful.

\begin{obs}[\cite{LeaderElectionDIST}] \label{cor:bottleneck-linear-time}
    Let $\gamma > 0$, $b\in\N$, $\vc\in\N^\Lambda$, and $X,S \subseteq \N^\Lambda$ such that $\Pr[\vc \reach X] \geq \gamma$, $\Pr[\vc \reach S] = 1$, and every path from every $\vx \in X$ to some $\vo \in S$ has a $b$-bottleneck.
    Then $\time{\vc}{S} > \gamma \frac{n}{3 b^2 |\Lambda|^2}.$
\end{obs}

\subsection{Transition ordering lemma}

The following lemma was originally proved in~\cite{SpeedFaultsDIST} 
and was restated in the language of population protocols as Lemma 4.5 in~\cite{LeaderElectionDIST}.
Intuitively, the lemma states that a ``fast'' transition sequence (meaning one without a bottleneck transition) that decreases certain states from large counts to small counts must contain transitions of a certain restricted form.
In particular the form is as follows: if $\Delta$ is the set of states whose counts decrease from large to small, then we can write the states in $\Delta$ in some order $\ds_1,\ds_2,\ldots,\ds_d$, such that for each $1 \leq i \leq d$, there is a transition $\tau_i$ that consumes $\ds_i$, and every other state involved in $\tau_i$ is either not in $\Delta$, or comes later in the ordering.
These transitions will later be used to do controlled ``surgery'' on fast transition sequences, because they give a way to alter the count of $\ds_i$, by inserting or removing the transitions $\tau_i$, knowing that this will not affect the counts of $\ds_1,\ldots,\ds_{i-1}$.

Let $\Delta \subseteq \Lambda$, with $d=|\Delta|$.
We say that $\calP$ is \emph{$\Delta$-ordered (via $\tau_1,\ldots,\tau_d$)} 
if there is an order on $\Delta$, so that we may write $\Delta = \{\ds_1,\ldots,\ds_d\}$, 
such that, for all $i \in \{1,\ldots,d\}$, 
there is a transition $\tau_i: \ds_i,\os_i \to \ans{o}_i,\ans{o}'_i$, 
such that $\os_i,\ans{o}_i,\ans{o}'_i \not \in \{\ds_{1},\ldots,\ds_i\}$.
In other words, for each $i$ there is a transition consuming exactly one $\ds_i$ without affecting $\ds_1,\ldots,\ds_{i-1}$.

\begin{lem}[adapted from~\cite{SpeedFaultsDIST}] \label{lem:ordering}
  Let $b_1,b_2 \in \N$ such that $b_2 > |\Lambda| \cdot b_1$.
  Let $\vx,\vo \in \N^\Lambda$ such that $\vx \reach_p \vo$ via a path $p$ without a $b_2$-bottleneck.
  Define
  $\Delta = \{ \ds \in \Lambda \mid \vx(\ds) \geq b_2$ and $\vo(\ds) \leq b_1 \}.$
  Then $\calP$ is $\Delta$-ordered via $\tau_1,\ldots,\tau_d$, 
  and each $\tau_i$ occurs at least $(b_2 - |\Lambda| \cdot b_1)/|\Lambda|^2$ times in $p$.
\end{lem}

\subsection{Sublinear time from dense configuration implies bottleneck free path from dense configuration with every state present}

Say that $\vc \in \N^\Lambda$ is \emph{full} if $(\forall \os\in\Lambda)\ \vc(\os)>0$, i.e., every state is present.
The following theorem states that with high probability, a population protocol will reach from an $\alpha$-dense configuration to a configuration in which all states are present (full) in ``large'' count ($\beta$-dense, for some $0 < \beta < \alpha$).\footnote{With the same probability, this happens in time $O(1)$, although this fact is not needed in this paper.}
It was proven in~\cite{DotyTCRN2014} in the more general model of chemical reaction networks, for a subclass of such networks that includes all population protocols.

\begin{thm}[adapted from~\cite{DotyTCRN2014}]\label{thm:large-counts}
  Let $\alpha > 0$.
  Then there are constants $\epsilon,\beta > 0$ such that,
  letting
  $X_\beta = \{\ \vx\in\N^\Lambda \ |\ \vx$ is full and $\beta$-dense $\}$,
  for all sufficiently large $\alpha$-dense $\vi\in\N^\Lambda$,
  $\Pr[\vi \reach X_\beta] \geq 1 - 2^{-\epsilon \|\vi\|}.$
\end{thm}

The following was originally proved as Lemma 4.4 in~\cite{LeaderElectionDIST}. 
The result was stated with $S$ being the set of what was called ``$Q$-stable configs,''
but we have adapted it to make the statement more general and quantitatively relate the bound $b(n)$ to the expected time $t(n)$.
It states that if a protocol goes from an $\alpha$-dense configuration to a set of states $S$ in expected time $\leq t(n)$,
then there is a full $\beta$-dense (for $0 < \beta < \alpha$) 
reachable configuration $\vx$ 
and a path from $\vx$ to a state in $S$ with no $b(n)$-bottleneck transition, 
where $b(n) = O\left( \sqrt{ \frac{n}{t(n)} } \right)$.
If $t(n) = o(n)$,
then $b(n) = \omega(1)$,
which suffices for our subsequent results.

\begin{lem}[adapted from~\cite{LeaderElectionDIST}]
    \label{lem:pos-prob-states-sublinear-expected-time-implies-bottleneck-free}
    For all $\alpha > 0$, there is a $\beta > 0$ such that the following holds.
    Suppose that for some $t:\N\to\N$, some set $S\subseteq\N^\Lambda$ and some set of $\alpha$-dense initial configurations $I$, 
    for all $\vi\in I$, 
    $\time{\vi}{S} \leq t(n)$.
    Define $b(n) = \frac{1}{|\Lambda|} \sqrt{\frac{n}{6 t(n)}}$.
    There is an $n_0\in\N$ such that
    for all $\vi \in I$ with $\|\vi\| = n \geq n_0$,
    there is $\vx \in \post(\vi)$ and path $p$ such that:
    \begin{enumerate}
    %(1) 
    \item
    $\vx(\os) \geq \beta n$ for all $\os \in \Lambda$,
    %(2) 
    \item
    $\vx \reach_p \vo$, where $\vo \in S$, and
    %(3) 
    \item
    $p$ has no $b(n)$-bottleneck transition.
    \end{enumerate}
\end{lem}

\begin{proof}
    Intuitively, the lemma follows from the fact that state $\vx$ is reached with high probability by Theorem~\ref{thm:large-counts}, and if no paths such as $p$ existed, 
    then all paths from $\vx$ to a stable configuration would have a bottleneck and require more than the stated time by Observation~\ref{cor:bottleneck-linear-time}. 
    Since $\vx$ is reached with high probability, 
    this would imply the entire expected time is linear.
    
    For any configuration $\vx$ reachable from some configuration $\vi \in I$, 
    there is a transition sequence $p$ satisfying condition (2) by the fact that $\Pr[\vi \reach S] = 1$.
    It remains to show we can find $\vx$ and $p$ satisfying conditions (1) and (3).
    
    By Theorem~\ref{thm:large-counts} there exist $\epsilon,\beta$ (which depend only on $\calA$ and $\alpha$) such that, 
    starting in any sufficiently large initial configuration $\vi$,
    with probability at least $1 - 2^{-\epsilon n}$,
    $\calA$ reaches a configuration $\vx$ where all states $\os \in \Lambda$ have count at least $\beta n$, where $n = \|\vi\|$.
    For all $\vi$, let $X_{\vi} = \post(\vi) \cap X_\beta = \{\vx\; |\; \vi \reach \vx \text{ and } (\forall \os \in \Lambda) \, \vx(\os) \geq \beta \|\vi\|\}$.
    %Given any $m \in \N$, 
    Let $n_0$ be a lower bound on $n$ such that Theorem~\ref{thm:large-counts} applies for all $n \geq n_0$
    and
    $1 - 2^{-\epsilon n_0} \geq \frac{1}{2}$. %, and further $n_0 \geq m / \beta$.
    Then for all $\vi \in I$ such that $\|\vi\| = n \geq n_0$,
    $\Pr[\vi \reach X_{\vi}] \geq \frac{1}{2}$.
    Choose any $n \geq n_0$ for which there is $\vi \in I$ with $\|\vi\|=n$.
    Then any $\vx \in X_{\vi}$ satisfies condition (1): $\vx(\os) \geq \beta n$ for all $\os \in \Lambda$.
    We now show that by choosing $\vx$ from $X_{\vi}$ for a large enough $n$, 
    we can find a corresponding $p$ satisfying condition (3) as well.
    
    Suppose for the sake of contradiction that, 
    we cannot satisfy condition (3) when choosing $\vx$ as above, no matter how large we make $n$.
    This means that for infinitely many $\vi \in I$, 
    (and therefore infinitely many population sizes $n=\|\vi\|$), 
    all transition sequences from $X_{\vi}$ to $S$ have a $b(n)$-bottleneck.
    Applying Observation~\ref{cor:bottleneck-linear-time}, 
    letting $\vc=\vi$, $\gamma=\frac{1}{2}$, $X = X_{\vi}$, 
    tells us that 
    $t(n) = \time{\vi}{S} > \frac{1}{2} \frac{n}{3 b(n)^2 |\Lambda|^2}$, 
    so $b(n) > \frac{1}{|\Lambda|} \sqrt{\frac{n}{6 t(n)}}$, a contradiction.
    \end{proof}

In the following lemma, note that the indexing is over a subset $N \subseteq \N$;
for example, the sequence might be indexed $\vi_3, \vi_6, \vi_7, \vi_{12}, \ldots$ if $N=\{3,6,7,12,\ldots\}$,
allowing us to retain the convention that the population size $\|\vi_n\|$ is represented by $n$.
Lemma~\ref{lem:cor:pos-prob-states-sublinear-expected-time-implies-bottleneck-free} 
essentially states that,
if infinitely many configurations $\vi$ satisfy the hypothesis of 
Lemma~\ref{lem:pos-prob-states-sublinear-expected-time-implies-bottleneck-free},
then we can find three infinite sequences satisfying the conclusion of
Lemma~\ref{lem:pos-prob-states-sublinear-expected-time-implies-bottleneck-free}:
initial configurations $\vi_n$, 
intermediate full configurations $\vx_n$,
and 
``final'' configurations $\vo_n$ 
(in our applications all $\vo_n$ will be stable),
which by Dickson's lemma can all be assumed nondecreasing.

\begin{lem}
    \label{lem:cor:pos-prob-states-sublinear-expected-time-implies-bottleneck-free}
    For all $\alpha > 0$, there is a $\beta > 0$ such that the following holds.
    Suppose that for some set $S\subseteq\N^\Lambda$ and infinite set of $\alpha$-dense initial configurations $I$, 
    for all $\vi\in I$, 
    $\time{\vi}{S} \leq t(n)$.
    Define $b(n) = \frac{1}{4 |\Lambda|} \sqrt{\frac{n}{t(n)}}$.
    There is an infinite set $N \subseteq \N$ and infinite sequences of configurations
    $(\vi_n \in I)_{n\in N}$, 
    $(\vx_n \in \N^\Lambda)_{n\in N}$, 
    $(\vo_n \in \N^\Lambda)_{n\in N}$, 
    where $(\vx_n)_{n\in N}$ and $(\vo_n)_{n\in N}$ are nondecreasing,
    and an infinite sequence of paths $(p_n)_{n\in N}$
    such that, for all $n\in N$,
    \begin{enumerate}
    %(1) 
    \item
    $\|\vi_n\|=\|\vx_n\|=\|\vo_n\|=n$,
    %(2) 
    \item
    $\vi_n \reach \vx_n$,
    %(3) 
    \item
    $\vx_n(\os) \geq \beta n$ for all $\os \in \Lambda$,
    %(4) 
    \item
    $\vx_n \reach_{p_n} \vo_n$, where $\vo_n \in S$, and
    %(5) 
    \item
    $p_n$ has no $b(n)$-bottleneck transition.
    \end{enumerate}
\end{lem}

\begin{proof}
    Since $I$ is infinite, the set $I_{n_0} = \{\vi \in I \mid \|\vi_n\| \geq n_0 \}$ is infinite.
    Pick an infinite sequence $(\vi_n)$ from $I_{n_0}$, where $\|\vi_n\|=n$ ($n$ may range over a subset of $\N$ here, but for each $n\in\N$, at most one configuration in the sequence has size $n$).
    For each $\vi_n$ in the sequence, pick $\vx_n$, $p_n$ and $\vo_n$ for $\vi_n$ as in Lemma~\ref{lem:pos-prob-states-sublinear-expected-time-implies-bottleneck-free}.
    By Dickson's Lemma (Lemma~\ref{lem:dickson}) there is an infinite subset $N \subseteq \N$ such that $(\vx_n)_{n\in N}$ and $(\vo_n)_{n \in N}$ are nondecreasing on the respective subsequences of $(\vx_n)_{n \in \N}$ and $(\vo_n)_{n \in \N}$ corresponding to $N$.
    Lemma~\ref{lem:pos-prob-states-sublinear-expected-time-implies-bottleneck-free} ensures that properties (1)-(5) are satisfied.
\end{proof}

The conclusion of Lemma~\ref{lem:cor:pos-prob-states-sublinear-expected-time-implies-bottleneck-free},
with its various infinite sequences, is quite complex.
The hypothesis of Lemma~\ref{lem:push-Delta} is equally complex; 
they are used in tandem to prove Lemma~\ref{lem:push-Delta-simplified}
and Corollary~\ref{cor:push-Delta-simplified},
the latter being our main technical tool for proving the time lower bounds 
of Theorems~\ref{thm:predicates},
\ref{thm:subtraction-slow}, 
\ref{thm:division-slow},
\ref{thm:affine-not-sublinear-computable}, 
and
\ref{thm:negative-result-eventually-positive-integral-linear}.

The idea of Lemma~\ref{lem:push-Delta-simplified} 
is to start with a protocol satisfying the hypothesis of Lemma~\ref{lem:cor:pos-prob-states-sublinear-expected-time-implies-bottleneck-free},
which reaches in sublinear time from some set $I$ of $\alpha$-dense initial configurations to some set $S$
(in all applications, $S$ is the set of stable configurations reachable from $I$).
Then, invoke Lemma~\ref{lem:push-Delta} to show that it is possible from certain initial configurations to drive some states in the set $\Delta = \bdd((\vo_n))$ to 0.

The reason that the statement of Lemma~\ref{lem:push-Delta-simplified} is also fairly complex,
and references some of these infinite sequences,
is that the set $\Delta$ appearing in the conclusion of Lemma~\ref{lem:push-Delta} depends on the particular infinite sequence $(\vo_n)$ defined in the conclusion of Lemma~\ref{lem:cor:pos-prob-states-sublinear-expected-time-implies-bottleneck-free}.
Several infinite sequences, each with their own $\Delta$, 
could satisfy the hypothesis of Lemma~\ref{lem:push-Delta}, and it matters which one we pick.
Thus, in applying these results, before reaching the conclusion of Lemma~\ref{lem:push-Delta},
we must explicitly define these infinite sequences to know the particular $\Delta$ to which the conclusion of Lemma~\ref{lem:push-Delta} applies.

\subsection{Path manipulation}

This is the most technically dense subsection,
with many intermediate technical lemmas that culminate in our primary technical tool for proving time lower bounds,
Corollary~\ref{cor:push-Delta-simplified}.
Each lemma statement is complex and involves many interacting variables.
The first three lemmas are accompanied by an example and figures to help trace through the intuition.

The next two lemmas,
Lemmas~\ref{lem:eliminate-Delta} and~\ref{lem:produce-e},
apply to population protocols that have transitions as described in Lemma~\ref{lem:ordering}.
Both use these transitions in order to manipulate a configuration 
(by manipulating a ``fast'' path leading to it from another configuration) 
until it has prescribed counts of states in $\Delta$ from Lemma~\ref{lem:ordering}.

Lemmas~\ref{lem:eliminate-Delta} and~\ref{lem:produce-e} are based on statements first proven as ``Claim 1'' and ``Claim 2'' in~\cite{LeaderElectionDIST}.
Since their statements in that paper were not self-contained 
(being claims as part of a larger proof), 
we have rephrased them as self-contained Lemmas~\ref{lem:eliminate-Delta} and~\ref{lem:produce-e}, 
and we give self-contained proofs.
Furthermore, we have significantly adapted both the statements and proofs to make them more generally useful for proving negative results, 
in particular stating the minimum conditions required to apply the lemmas,
in addition to quantitatively accounting for the precise effect that the path manipulation has on the underlying configurations.
%}

%\opt{normal}{\explanationLemmasClaims}

We use linear algebra to describe changes in counts of states.
It is beneficial to fix some notational conventions first.
Recall $\Lambda$ is the set of all states, 
$\Delta \subset \Lambda$ where $\Delta = \{\ds_1,\ldots,\ds_d\}$, 
and $\Gamma = \Lambda \setminus \Delta$ where $\Gamma = \{\gs_1,\ldots,\gs_g\}$.
A matrix $\tvC \in \Z^{\Gamma\times\Delta}$ is an integer-valued matrix with $g$ rows and $d$ columns, with row $j$ corresponding to state $\gs_j$ and column $i$ corresponding to state $\ds_i$.
Given a vector $\vc^\Delta \in \N^{\Delta}$ representing counts of states in $\Delta$, 
then $\tvC.\vc^\Delta = \vd^\Gamma$ is a vector $\vd^\Gamma\in\Z^\Gamma$ representing changes in counts of states in $\Gamma$.

    Our notation for indexing these matrices will generally follow our usual vector convention of using the name of the state itself,
    rather than an integer index,
    so for example, $\tvC(\gs_i,\ds_j)$ refers to the entry 
    in the column corresponding to $\ds_j$
    and the row corresponding to $\gs_i$.
    If necessary to identify the position, this will correspond to the $i$'th row and $j$'th column.
    Where convenient, 
    we also use the traditional notation $\tvC(i,j)$ as well: 
    for instance, a protocol being $\Delta$-ordered  
    implies a 1-1 correspondence between transitions $\tau_1,\ldots,\tau_d$
    and $\Delta = \{\ds_1,\ldots,\ds_d\}$, 
    which can both be indexed by $i \in \{1,\ldots,d\}$.
%}

Similarly, when convenient we will abuse notation and consider a vector $\vv \in \N^k$,
for a predicate or function with $k$ inputs,
to equivalently represent a configuration or subconfiguration in $\N^\Sigma$,
where $\Sigma \subseteq \Lambda$ is the set of $k$ input states of the population protocol.

%\opt{normal}{\explanationLinearNotation}

The next lemma says that for any amount $\vc^\Delta \in \N^\Delta$ of states in $\Delta$, 
there exists an amount $\ve \in \N^\Lambda$ of states that,
if present in addition to $\vc^\Delta$, 
can be used to remove $\vc^\Delta$ and $\ve \rest \Delta$ 
(the states of $\ve$ that are in $\Delta$), 
resulting in a configuration $\vz^\Gamma \in \N^\Gamma$ with no states in $\Delta$.
Furthermore, both $\ve$ and $\vz^\Gamma$ are linear functions of $\vc^\Delta$.

So when we employ Lemma~\ref{lem:eliminate-Delta} later, 
where will these extra agents $\ve$ come from?
Although we talk about them as if they are somehow physically added, 
in actuality, we'll start with a larger initial configuration and ``guide'' some of the agents to the desired states that make up $\ve$;
this is the work of Lemma~\ref{lem:produce-e}.

\begin{lem}[adapted from~\cite{LeaderElectionDIST}]   \label{lem:eliminate-Delta}
    Let $\Delta \subseteq \Lambda$ such that $\calP$ is $\Delta$-ordered,
    $d=|\Delta|$,
    and let $\Gamma = \Lambda \setminus \Delta$.
    Then there are matrices 
    $\vC_1 \in \N^{\Gamma \times \Delta}$ 
    and $\vC_2 \in \N^{\Lambda \times \Delta}$,
    with $\max(\vC_1) < 2^{d+1}$, $\max(\vC_2) < 2^d$,
    such that,
    for all $\vc^\Delta \in\N^\Delta$, 
    setting $\ve = \vC_2.\vc^\Delta \in \N^\Lambda$
    and $\vz^\Gamma = \vC_1.\vc^\Delta \in \N^\Gamma$,
    then $\vc^\Delta + \ve \reach \vz^\Gamma$.
\end{lem}

\begin{proof}
    Intuitively, the proof works as follows.
    Since $\calP$ is $\Delta$-ordered, for each $i\in\{1,\ldots,d\}$,
    there is a transition $\tau_i: \ds_i,\ans{s}_i \to \ans{o}_i,\ans{o}'_i$,
    such that for all $i$, $\ds_i \in \Delta$ and $o_i,o'_i \not\in \{\ds_1,\ldots,\ds_i\}$.
    We will construct the path $p$ such that $\vc^\Delta + \ve \reach_p \vz$ as follows.
    A na\"{i}ve approach would simply consume states in $\vc^\Delta$,
    by adding $\vc^\Delta(i)$ copies of $\tau_i$ to the path $p$, 
    and $\vc^\Delta(i)$ copies of the other input $\os_i$ to $\ve$.
    Since $\calP$ is $\Delta$-ordered this would indeed result in count 0 of $\ds_1$.
    However, although for $i \in \{2,\ldots,d\}$,
    this consumes $\vc^\Delta(\ds_i)$ copies of $\ds_i$,
    it might produce additional copies of $\ds_i$ 
    if it appears as an output state of some transitions $\tau_1,\ldots,\tau_{i-1}$ that were added.
    Let $\vc^\Delta_1$ denote these counts.
    Since $\vc^\Delta_1(\ds_1)=0$, we won't need to add any more $\tau_1$.
    Repeat the na\"{i}ve approach a second time to consume $\vc^\Delta_1$, which will result in $\vc^\Delta_2$,
    where $\vc^\Delta_2(\ds_1)=\vc^\Delta_2(\ds_2)=0$.
    Repeating this $d$ times consumes all of $\Delta$.
    
    We now formally define matrices that will help to account for the exact changes in state counts that result from executing this path.
    First, we define a matrix $\vT \in \N^{d \times d}$.
    Intuitively, if $\vc^\Delta \in \N^\Delta$ represents counts of states in $\Delta$, 
    then the vector $\vt \in \N^d$ defined by $\vt = \vT.\vc^\Delta$
    represents counts of transitions $\tau_1,\ldots,\tau_d$ in the path $p$
    such that $\vc^\Delta + \ve \reach_p \vz^\Gamma$.
    In particular, $\vt(\ds_i)$ will represent the total number of transitions $\tau_i$ that we add to path $p$,
    in order to consume all copies of $\ds_i$,
    not only the $\vc^\Delta(\ds_i)$ present initially, 
    but also any added because of transitions $\tau_j$ for $j<i$ appearing previously in $p$, 
    if one of the outputs of $\tau_j$ is $\ds_i$.
    
    Define the $d \times d$ matrix $\vT_1$ as follows.
    Intuitively, $\vT_1$ is a matrix such that, 
    if $\vt \in \N^d$ represents ``counts of transition executions'',
    i.e., $\vt(j)$ means ``execute transition $\tau_j$ $\vt(j)$ times'',
    then $\vT_1.\vt \in \N^d$ (equivalently, $\N^\Delta$) 
    represents the total count of output states \emph{in $\mathit{\Delta}$}
    that would be \emph{produced as outputs} of these transitions.
    It does not account for the number of input states consumed, 
    nor the number of output states in $\Gamma$ produced.
    
    Formally, $\vT_1$ is a strictly lower diagonal matrix ($0$'s on and above the diagonal).
    Column $j$ is $0$'s, other than potentially up to two positive entries, described below.
    \begin{itemize}
        \item If $\tau_j$ has output states $\ds_{k},\ds_{k'}$ where $j<k \neq k'$, 
            then $\vT(k,j)=\vT(k',j)=1$. 
            %and $\vT(k',j)=0$ for all $k' \not\in \{k_1,k_2\}$.
        \item If $\tau_j$ has output states $\ds_k,\ds_k$ where $j<k$,
            then $\vT(k,j)=2$. 
            %and $\vT(k',j)=0$ for all $k' \neq k$.
        \item If $\tau_j$ has output states $\vd_k,\ans{g}$, where $j<k$ and $\ans{g} \in \Gamma$, 
            then $\vT(k,j)=1$. 
            % and $\vT(k',j)=0$ for all $k' \neq k$.
        %\item If $\tau_j$ has output states $\ans{g},\ans{g}' \in \Gamma$, then $\vT(k',j)=0$ for all $k' \in \{1,\ldots,d\}$.
    \end{itemize}
    By the fact that $\calP$ is $\Delta$-ordered via $\tau_1,\ldots,\tau_d$,
    there are no other forms the transitions can take.
    For example, if we have transitions
    \[
    \begin{array}{rll}
        \tau_1: & \ds_1,\ds_3     &\to \ds_2,\ds_4 \\
        \tau_2: & \ds_2,\ans{g}_1 &\to \ds_3,\ds_4 \\
        \tau_3: & \ds_3,\ds_6     &\to \ds_5,\ans{g}_1 \\
        \tau_4: & \ds_4,\ds_6     &\to \ans{g}_1,\ans{g}_2 \\
        \tau_5: & \ds_5,\ans{g}_2 &\to \ds_6,\ds_6 \\
        \tau_6: & \ds_6,\ans{g}_1 &\to \ans{g}_1,\ans{g}_2
    \end{array}
    \]
    where $\ans{g}_1,\ans{g}_2 \in \Gamma$, then
    \[
        \vT_1 = \left[\begin{array}{cccccc}
            0 & 0 & 0 & 0 & 0 & 0 \\
            1 & 0 & 0 & 0 & 0 & 0 \\
            0 & 1 & 0 & 0 & 0 & 0 \\
            1 & 1 & 0 & 0 & 0 & 0 \\
            0 & 0 & 1 & 0 & 0 & 0 \\
            0 & 0 & 0 & 0 & 2 & 0
        \end{array}\right]
    \]
    
    We define $\vT$ based on $\vT_1$.
    Na\"{i}vely, to consume states in $\vc^\Delta$,
    for each $i \in \{1,\ldots,d\}$ one would add 
    $\vc^\Delta(i)$ copies of $\tau_i$.
    Since $\calP$ is $\Delta$-ordered this would indeed result in count 0 of $\ds_1$.
    However, although for $i \in \{2,\ldots,d\}$,
    this consumes $\vc^\Delta(\ds_i)$ copies of $\ds_i$,
    it also produces $(\vT_1.\vc^\Delta)(i)$ copies of $\ds_i$,
    which is positive if $\ds_i$ is an output of some transition in $\tau_1,\ldots,\tau_{i-1}$.
    
    Applying the na\"{i}ve idea a second time, 
    to consume the states that were produced on the first step,
    for each $i \in \{1,\ldots,d\}$
    we add $(\vT_1.\vc^\Delta)(\ds_i)$ copies of $\tau_i$.
    (Note that $(\vT_1.\vc^\Delta)(\ds_1)=0$ so this second step adds no additional copies of $\tau_1$.)
    Thus, this results in count $0$ of $\ds_2$, 
    and although it consumes the copies of $\ds_3,\ldots,\ds_d$ that remained after the first step,
    it also produces $\vT_1.(\vT_1\vc^\Delta)(\ds_i)$ additional copies of $\ds_i$.
    The number of transitions after two steps is then described by summing steps 1 and 2:
    $\vc^\Delta + \vT_1.\vc^\Delta$.
    We iterate this procedure a total of $d$ steps, 
    where the transitions added in step $i$ are described by the vector $\vT_1^{i-1}.\vc^\Delta$.
    
    Since the $i$'th step results in getting to count $0$ of $\ds_1,\ldots,\ds_i$,
    all $\ds_1,\ldots,\ds_d$ will have count 0 after $d$ steps.
    The total number of transitions applied over all steps is then described by the vector obtained by summing the $d$ vectors indicating transition counts for each step 1 through $d$:
    $\vc^\Delta + \vT_1.\vc^\Delta + \vT_1^2.\vc^\Delta + \vT_1^3.\vc^\Delta + \ldots + \vT_1^{d-1}.\vc^\Delta$.
    Thus, taking $\vT_1^0$ to be the $d\times d$ identity matrix,
    we can define the matrix $\vT = \sum_{i=0}^{d-1} \vT_1^i$.
    Since each column of $\vT_1$ is either all $0$, has one or two $1$'s, or has a single $2$,
    a simple induction shows that for all $i\in\{1,\ldots,d-1\}$,
    $\max(\vT_1^i) \leq 2^i$.
    Thus $\max(\vT) \leq \sum_{i=0}^{d-1} \max(\vT_1^i) \leq \sum_{i=0}^{d-1} 2^i < 2^d$.
    (This bound is nearly tight; e.g., transitions $\tau_i: \ds_i,\os \to \ds_{i+1},\ds_{i+1}$ result in $\vT(d,1) = 2^{d-1}$.)
    
    Now that we have defined $\vT$, 
    which tells us that we will have $(\vT.\vc^\Delta)(i)$ copies of $\tau_i$ in path $p$, 
    it is easy to define $\vC_1$ and $\vC_2$ based on $\vT$.
    For each copy of $\tau_i:\ds_i,\os \to \ans{o},\ans{o}'$, we add a copy of $\os$ to $\ve$.
    Thus, define the matrix $\vS \in \N^{\Lambda\times\Delta}$ so that,
    for all $i\in\{1,\ldots,d\}$, $\vS(\os,\ds_i) = 1$ if transition $\tau_i$,
    which by definition has one input state $\ds_i$, has $\os$ as its other input state.
    All other entries of $\vS$ are $0$.
    Then $\vC_2 = \vS.\vT$,
    and $\max(\vC_2) = \max(\vT) < 2^d$.
    
    It remains to define $\vC_1$, so that $\vC_1.\vc^\Delta$ describes 
    the vector $\vz^\Gamma$ of states in $\Gamma$ produced by path $p$
    First, define the matrix $\vG\in\N^{\Gamma\times\Delta}$ as follows,
    which intuitively maps a count vector of transitions in $\tau_1,\ldots,\tau_d$ 
    to a total count of states in $\Gamma$ produced as output by the transitions.
    Let $j\in\{1,\ldots,d\}$ and let $\tau_j: \ds_j,\os \to \ans{o},\ans{o}'$.
    If $\ans{o},\ans{o}' \in \Delta$, then the $j$'th column of $\vG$ is all $0$.
    If exactly one (w.l.o.g.) $\ans{o} \in \Gamma$, 
    then $\vG(\ans{o},\ds_j)=1$ and the remainder of the $j$'th column of $\vG$ is all $0$.
    If both $\ans{o},\ans{o}' \in \Gamma$, and $\ans{o}=\ans{o}'$, 
    then $\vG(\ans{o},\ds_j)=2$ and the remainder of the $j$'th column of $\vG$ is all $0$.
    If both $\ans{o},\ans{o}' \in \Gamma$, and $\ans{o} \neq \ans{o}'$, 
    then $\vG(\ans{o},\ds_j)=\vG(\ans{o}',\ds_j)=1$ and the remainder of the $j$'th column of $\vG$ is all $0$.
    Then $\vC_1 = \vG.\vT$,
    and $\max(\vC_1) \leq 2 \cdot \max(\vT) < 2^{d+1}$.
\end{proof}

We demonstrate Lemma~\ref{lem:eliminate-Delta} with a concrete example. Consider a Population Protocol $\calP$ defined by transitions
\[
\begin{array}{rll}
    \tau_1: & \ds_1,\ds_3     &\to \gs_1,\ds_2 \\
    \tau_2: & \gs_1,\ds_2     &\to \gs_1,\ds_3 \\
    \tau_3: & \gs_1,\ds_3     &\to \gs_1,\gs_1 \\
    \tau_4: & \gs_1,\gs_2     &\to \gs_1, \gs_1 \\
    \tau_5: & \gs_1,\gs_2     &\to \gs_2, \gs_2 
\end{array}
\]

% \begin{figure}[ht]
% \includegraphics[width= 5cm]{transition_rules.pdf}
% \caption{Color-coded transition rules for $\mathcal{P}$}
% \label{fig:transition_rules}
% \end{figure}

Let $\Delta = \{\ds_1, \ds_2, \ds_3\}$. 
Then $\calP$ is $\Delta$-ordered via $\tau_1, \tau_2, \tau_3$, 
because $\tau_2$ does not reference $\ds_1$, 
and $\tau_3$ does not reference $\ds_2$ or $\ds_3$,
and for all $i\in\{1,2,3\}$, $\tau_i$ contains exactly one reference to $\ds_i$ as an input.
By the terminology of Lemma~\ref{lem:eliminate-Delta}, $\Gamma = \{\gs_1,\gs_2\}$.

For a given $\vc^\Delta$, we can design a configuration $\ve \in \N^\Delta$ such that $\vc^\Delta + \ve \reach \vz^\Gamma$.
We will remove agents in $\Delta$ using the transitions given to us by $\mathcal{P}$ being $\Delta$-ordered. 
For example, to remove $n$ agents of $\ds_1$ we will need transition $\tau_1$ to occur $n$ times. 
Similarly, to remove $m$ agents of $\ds_2$, we will need $\tau_2$ to occur at least $m$ times, and we may need more if $\tau_1$ generated additional copies of $\ds_2$.

% If these go before statement of lemmas, then should explain in more detail (or not use terminology?)
Let $\vT$ be defined as in the proof of Lemma~\ref{lem:eliminate-Delta} and let
%Let $\vc^\Delta = (5,1,2)^T$
\[
        \vc^\Delta= \left[\begin{array}{c}
            5  \\
            1 \\
            2 \\
        \end{array}\right]
    \]

\begin{figure}[ht]
\centering
\includegraphics[width=\textwidth]{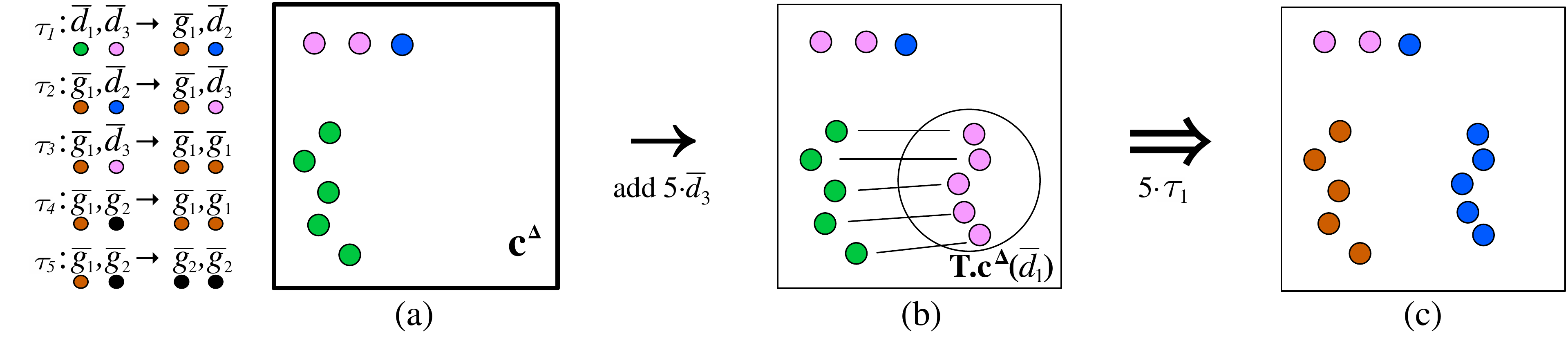}
\caption{First step of the surgery detailed in Lemma~\ref{lem:eliminate-Delta}. 
{\bf (a)} Our initial configuration $\vc^\Delta$.
{\bf (b)} Add 5 $\ds_3$ agents which will react with 5 $\ds_1$ agents.
{\bf (c)} All 5 copies of $\ds_1$ have been removed.}
\label{fig:claim1_1}
\end{figure}

In order to remove 5 copies of $\ds_1$, we need enough copies of $\ds_3$ for $\tau_1$ to occur 5 times. 
As such, we add $\vT.\vc^\Delta(\ds_1) = 5$ $\ds_3$ agents to $\ve$ allowing $\tau_1$ to occur 5 times, as shown in Figure~\ref{fig:claim1_1}. This effectively removes all copies of $\ds_1$ and produces 5 extra copies of both $\gs_1$ and $\ds_2$. 
Since $\mathcal{P}$ is $\Delta$-ordered, we know that the only states created will either be in $\Gamma$ or they will be in $\Delta$ but further in the ordering - allowing us to remove the extra agents at a later time.  
    
%\todoi{DD: replace straight arrow on left below with straight-plus-curvy arrow from bottom (showing addition of new agents)}
    
\begin{figure}[ht]
\includegraphics[width=\textwidth]{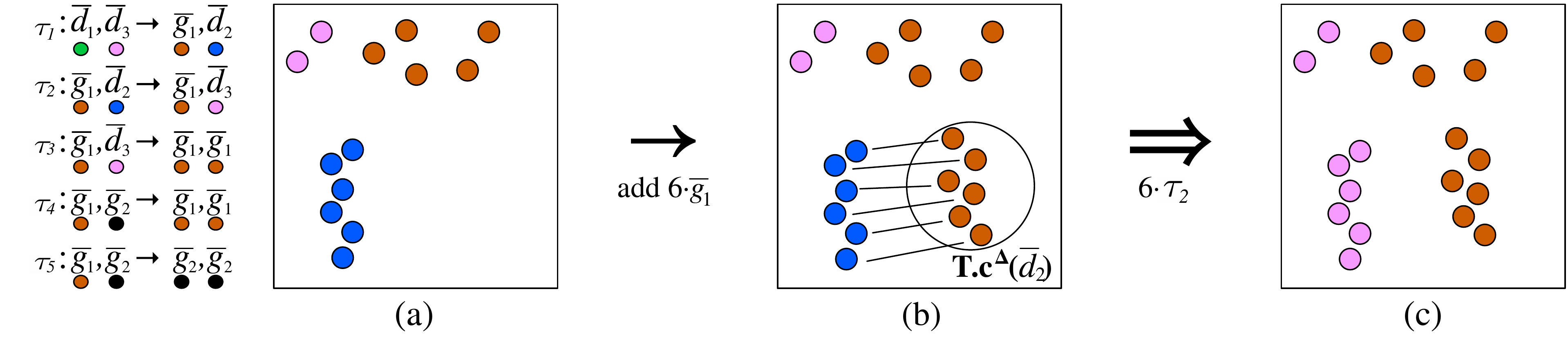}
\caption{Second step of the surgery detailed in Lemma~\ref{lem:eliminate-Delta}.
{\bf (a)} The same configuration as in panel (c) of Figure~\ref{fig:claim1_1}.
{\bf (b)} Add 6 $\gs_1$ agents which will react with 6 $\ds_2$ agents.
{\bf (c)} All 6 copies of $\ds_2$ have been removed.
}
\label{fig:claim1_2}
\end{figure}

Now, we must remove 1 original copy of $\ds_2$ plus 5 newly created copies created via $\tau_1$ in the previous step. To do so, we can use $\tau_2$, which requires we add $\vT.\vc^\Delta(\ds_2) = 6$ additional copies of $\gs_1$ to $\ve$. 
Via $\tau_2$, an additional 6 copies of both $\ds_3$ and $\gs_1$ are created. 
This process is illustrated in Figure~\ref{fig:claim1_2}
Again, since $\ds_3$ comes after $\ds_2$ in the ordering, we can still remove all copies of $\ds_3$ at a later step. 

\begin{figure}[ht]
\includegraphics[width=\textwidth]{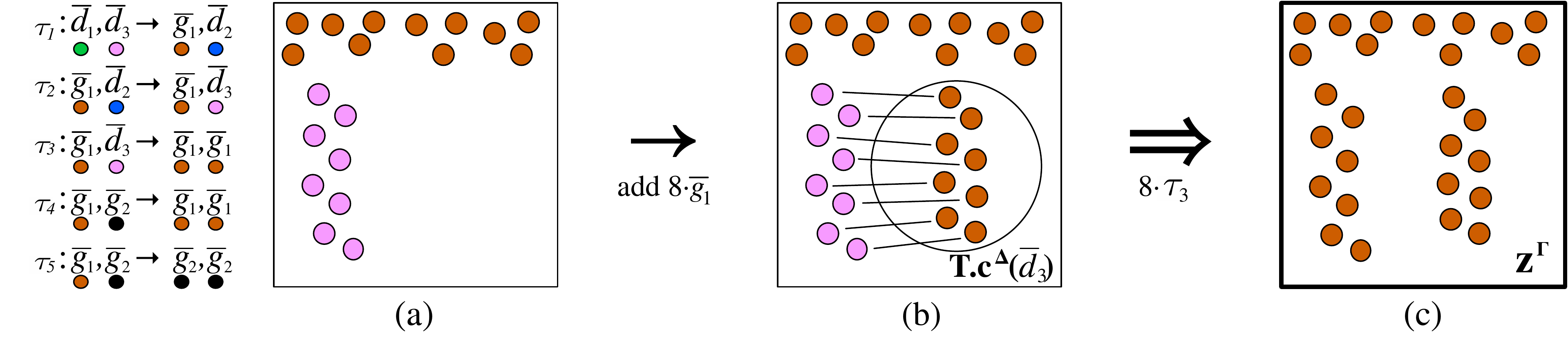}
\caption{Third step of the surgery detailed in Lemma~\ref{lem:eliminate-Delta}.
{\bf (a)} The same configuration as in panel (c) of Figure~\ref{fig:claim1_2}.
{\bf (b)} Add 8 $\gs_1$ agents which will react with 8 $\ds_3$ agents.
{\bf (c)} All 8 copies of $\ds_3$ have been removed. The resulting configuration contains only states in $\Gamma$.}
\label{fig:claim1_3}
\end{figure}

There are now 8 copies of $\ds_3$ to remove due to additional agents being produced in the previous steps. We will add $\vT.\vc^\Delta(\ds_3) = 8$ copies of $\gs_1$ to $\ve$ to allow 8 instances of $\tau_3$ to take place.
This will transition all instances of $\ds_3$ into instances of $\gs_1$ and leave us with a configuration $\vz^\Gamma$ of states only in $\Gamma$ as shown in Figure~\ref{fig:claim1_3}

\begin{figure}[ht]
\includegraphics[width=\textwidth]{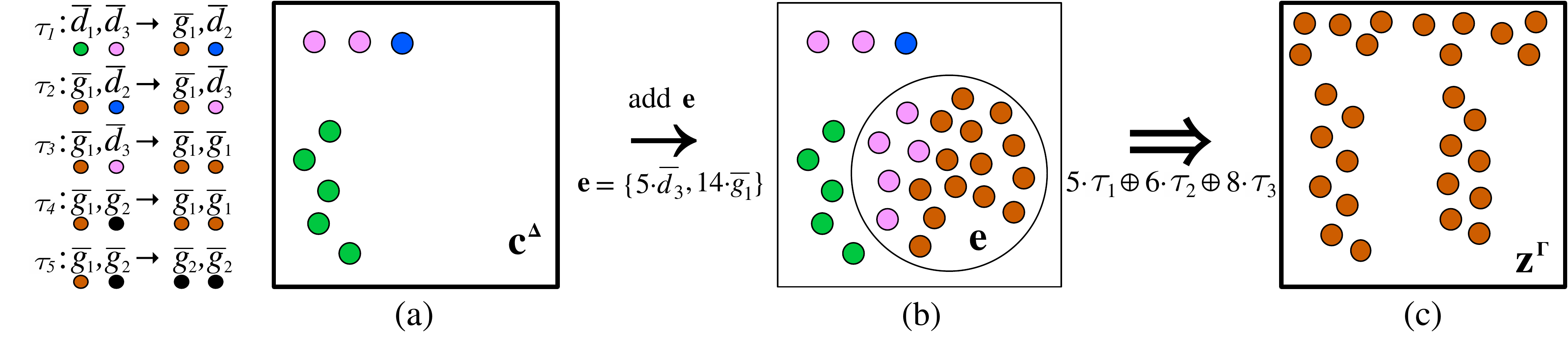}
\caption{Illustration of the whole path surgery technique detailed in Lemma~\ref{lem:eliminate-Delta}. We combine the steps shown in Figures~\ref{fig:claim1_1}, \ref{fig:claim1_2}, and \ref{fig:claim1_3} to form a single transition sequence leading from $\vc^\Delta$ to $\vz^\Gamma$. 
{\bf (a)} The same initial configuration $\vc^\Delta$ as in panel (a) of Figure~\ref{fig:claim1_1}.
{\bf (b)} Add $\ve = \{ 5 \ds_3\} + \{ 6 \gs_1 \} + \{ 8 \gs_1 \} = \{ 5 \ds_3, 14 \gs_1 \} $ to $\vc^\Delta$.  
{\bf (c)} All states in $\Delta$ are removed and we reach a configuration $\vz^\Gamma$.}
\label{fig:claim1_combo}
\end{figure}

While the process can be presented as several separate steps as above, $\ve$ will be the sum of all the agents added in the prior steps. In Figure~\ref{fig:claim1_combo}, we add $\ve$ to $\vc^\Delta$, which will then transition to $\vz^\Gamma$. We have removed all agents in $\Delta$ and arrived at a configuration $\vz^\Gamma$ as desired.

%%%%%%%%%%%%%%%%%%%%%%%%%%%%%%%%%%%%%%%
%%%%% See Amanda's diagrams above %%%%%
%%% Lemma~\ref{lem:eliminate-Delta} %%%
%%%%%%%%%%%%%%%%%%%%%%%%%%%%%%%%%%%%%%%

The next lemma works toward generating the vector of states $\ve$ needed to apply Lemma~\ref{lem:eliminate-Delta}.
The ``cost'' for Lemma~\ref{lem:produce-e} is that the path must be taken ``in the context'' of additional agents in states captured by $\vp$.
The intuitive reason $\vp$ is needed is this:
In Lemma~\ref{lem:eliminate-Delta}, we add new transitions and add enough new states ($\ve$) to supply the inputs for all these transitions.
Thus the resulting counts of states in $\Gamma$ can only be larger than the the original path.
However, the manipulation of Lemma~\ref{lem:produce-e} does not add new states,
so inputs of added transitions may have lower count.
Also, unlike Lemma~\ref{lem:produce-e}, 
the manipulation also involves \emph{removing} transitions;
thus the \emph{outputs} of those transitions may have their counts lowered.
The states in $\vp$ are used as a ``buffer'' to keep any state count from becoming negative,
which could happen if the manipulation were applied to the path starting from just the original configuration $\vx$.
Because of this, we do not specify where precisely in the transition sequence certain transitions are added, 
nor which specific occurrences are removed.
These could be chosen anywhere along the sequence,
and the buffer $\vp$ ensures that the entire sequence remains applicable.

Importantly, the net effect of the path preserves $\vp$,
which will give a way to ``interleave'' Lemmas~\ref{lem:eliminate-Delta} and \ref{lem:produce-e}, in order to start from a configuration with large counts of all states and reach a configuration with count $0$ of all states in $\Delta$.
Note that unlike matrices $\tvC_1$ and $\tvC_2$ of Lemma~\ref{lem:eliminate-Delta}, 
the matrix $\tvC_3$ may have negative entries,
since the resulting configuration is described as a difference from the configuration $\vo$,
and some states in $\Gamma$ may have lower count in the new configuration than in $\vo$.
%certain prescribed state counts in $\Delta$ (described by $\ve^\Delta$).

% \todo{DD: explain that the transitions can be added/removed anywhere along the path in the next lemma, since $\vec{p}$ buffers enough to keep all counts positive no matter where the surgery happens}

\begin{lem}[adapted from~\cite{LeaderElectionDIST}]   \label{lem:produce-e}
    Let $b_1,b_2 \in \N$.
    Let $\vx,\vo \in \N^\Lambda$ such that $\vx \reach_q \vo$ via path $q$ that does not contain a $b_2$-bottleneck.
    Define
    $\Delta = \{\ans{d} \in \Lambda \mid  \vo(\ans{d}) \leq b_1 \}$,
    $d=|\Delta|$,
    and $\Gamma = \Lambda \setminus \Delta$.
    Let $\vo^\Delta=\vo\rest\Delta$ and $\vo^\Gamma=\vo\rest\Gamma$.
    Then there is a matrix $\tvC_3 \in \Z^{\Gamma \times \Delta}$
    with $\amax(\tvC_3) < 2^{d+1}$,
    such that
    for all $\ve^\Delta \in \N^\Delta$,
    if $b_2 \geq |\Lambda| \cdot b_1 + d 2^d \cdot \max(b_1, \ve^\Delta) \cdot |\Lambda|^2$,
    and if $\vx(\os) \geq b_2$ for all $\os\in\Lambda$ and $\vo(\os) \geq b_2$ for all $\os \in \Gamma$,
    letting $\vp \in \N^\Lambda$ be defined 
    $\vp(\os) = d 2^{d+1} \cdot \max(b_1, \ve^\Delta)$ for all $\os \in \Lambda$,
    then $\vp + \vx \reach \vp + \vo^\Gamma + \tvC_3.\vo^\Delta - \tvC_3.\ve^\Delta + \ve^\Delta$.

\end{lem}

Note that we consider configurations in which all counts in $\vo^\Gamma$ are arbitrarily large
(see Lemma~\ref{lem:push-Delta}), 
whereas counts in $\vo^\Delta$ and $\ve^\Delta$,
as well as the entries of $\tvC_3$, 
are bounded.
Thus, for sufficiently large starting configurations, 
$\vo^\Gamma \geq \tvC_3.\vo^\Delta - \tvC_3.\ve^\Delta + \ve^\Delta$,
justifying our earlier claim that in this lemma, we ``get $\vp$ back at the end.''

\begin{proof}
    By Lemma~\ref{lem:ordering},
    $\calP$ is $\Delta$-ordered via $\tau_1,\ldots,\tau_d$, 
    and each $\tau_i$ occurs at least $(b_2 - |\Lambda| \cdot b_1)/|\Lambda|^2$ times in $q$.
    
    Intuitively, this proof is similar to that of Lemma~\ref{lem:eliminate-Delta},
    except that instead of targeting count $0$ of all states in $\Delta$, we target counts given by $\ve^\Delta$.
    Also, rather than constructing a new path consisting solely of transitions of type $\tau_1,\ldots,\tau_d$,
    we alter the path $q$, which may contain other types of transitions 
    (although we will only modify transitions of type $\tau_1,\ldots,\tau_d$).
    Since $\vx \reach_p \vo$,
    we can think of our ``starting value'' for counts in $\Delta$ as being $\vo^\Delta$,
    and thus the total change in counts that we want to make is described by the vector $\vo^\Delta - \ve^\Delta$.
    In particular, since we may have $\vo^\Delta(\ds_i) < \ve^\Delta(\ds_i)$ for some $\ds \in \Delta$,
    this may require \emph{removing} transitions from $q$ as well as adding them.
    % This is why the statement of the lemma, unlike that of Lemma~\ref{lem:eliminate-Delta},
    % references the path $q$ such that $\vx \reach_p \vo$, 
    % since it is this path from which transitions may need to be removed.
    Furthermore, since we have no $\ve$ at the start as in Lemma~\ref{lem:eliminate-Delta},
    when adding or removing transition $\ds_i,\os \to \ans{o},\ans{o}'$ to alter the count of $\ds_i$,
    we must account not only for the effect this has on the output states $\ans{o},\ans{o}'$,
    but also the effect on the other input state $\os$.
    This may result in a path that is not valid, 
    in the sense that some counts may be negative after the modification.
    The extra states in $\vp$ have the purpose of keeping the entire path valid.
    The bound on $\vp$ will then be derived from the bound on the size of the changes to $q$ that we make.
    
    First, we define a matrix $\tvT \in \Z^{d \times d}$.
    Intuitively, if $\tvc^\Delta \in \Z^\Delta$ represents changes in counts of states in $\Delta$ 
    that we wish to achieve through addition and removal of transitions $\tau_1,\ldots,\tau_d$ from the path $q$,
    then the vector $\tvt \in \Z^d$ defined by $\tvt = \tvT.\tvc^\Delta$
    represents changes in counts of transitions $\tau_1,\ldots,\tau_d$ in the path $q$
    to achieve this.
    More precisely, the counts in $\vo$ are given by $\vo^\Delta$,
    but we wish them to be $\ve^\Delta$ instead.
    Letting $\tvc^\Delta = \vo^\Delta - \ve^\Delta$, 
    then $\tvt = \tvT.\tvc^\Delta$ describes how many transitions of each type to add or remove from $q$.
    
    Define the $d \times d$ matrix $\tvT_1$ as follows.
    Intuitively, $\tvT_1$ is a matrix such that, 
    if $\tvt \in \Z^d$ represents (possibly negative) ``counts of transition executions'',
    i.e., $\tvt(j)$ means ``execute transition $\tau_j$ an additional $\tvt(j)$ times'' 
    (where executing a transition an additional negative number of times means removing it from $q$),
    then $\tvT_1.\tvt \in \Z^d$ (equivalently, $\Z^\Delta$) 
    represents the total count of states \emph{in $\mathit{\Delta}$}
    that would be \emph{produced as outputs} of these transitions
    or \emph{consumed as the second input}.
    Here, the ``second'' input means, 
    for transition $\tau_i:\ds_i,\os \to \ans{o},\ans{o}'$,
    the input $\os$ that is \emph{not} $\ds_i$.
    It does not account for the number of input states in $\Gamma$ consumed, 
    nor the number of output states in $\Gamma$ produced.
    
    Formally, $\tvT_1$ is a strictly lower diagonal matrix ($0$'s on and above the diagonal).
    Column $j$ is $0$'s, other than potentially up to two nonzero entries, described below.
    \begin{itemize}
        \item If $\tau_j$ has output states $\ds_{k},\ds_{k'}$ where $j<k \neq k'$, 
            then $\tvT(k,j)=\tvT(k',j)=1$.
        \item If $\tau_j$ has output states $\ds_k,\ds_k$ where $j<k$,
            then $\tvT(k,j)=2$.
        \item If $\tau_j$ has output states $\ds_k,\ans{g}$, where $j<k$ and $\ans{g} \in \Gamma$, 
            then $\tvT(k,j)=1$.
        \item If $\tau_j$ has second input state $\ds_k$ where $j<k$, 
            then $\tvT(k,j)=-1$.
    \end{itemize}
    By the fact that $\calP$ is $\Delta$-ordered via $\tau_1,\ldots,\tau_d$,
    there are no other forms the transitions can take.
    For example, if we have transitions
    \[
    \begin{array}{rll}
        \tau_1: & \ds_1,\ds_3     &\to \ds_2,\ds_4 \\
        \tau_2: & \ds_2,\ans{g}_1 &\to \ds_3,\ds_4 \\
        \tau_3: & \ds_3,\ds_6     &\to \ds_5,\ans{g}_1 \\
        \tau_4: & \ds_4,\ds_6     &\to \ans{g}_1,\ans{g}_2 \\
        \tau_5: & \ds_5,\ans{g}_2 &\to \ds_6,\ds_6 \\
        \tau_6: & \ds_6,\ans{g}_1 &\to \ans{g}_1,\ans{g}_2
    \end{array}
    \]
    
    where $\ans{g}_1,\ans{g}_2 \in \Gamma$, then
    \[
        \tvT_1 = \left[\begin{array}{rrrrrr}
            0  & 0  & 0  & 0  & 0  & 0 \\
            1  & 0  & 0  & 0  & 0  & 0 \\
            -1 & 1  & 0  & 0  & 0  & 0 \\
            1  & 1  & 0  & 0  & 0  & 0 \\
            0  & 0  & 1  & 0  & 0  & 0 \\
            0  & 0  & -1 & -1 & 2  & 0
        \end{array}\right]
    \]
    
    We define $\tvT$ based on $\tvT_1$.
    Na\"{i}vely, to consume (respectively, produce) counts of states as described in $\tvc^\Delta$,
    for each $i \in \{1,\ldots,d\}$ one would add
    $\tvc^\Delta(i)$ copies of $\tau_i$ (where adding a negative amount means removing from path $q$).
    Since $\calP$ is $\Delta$-ordered this would indeed result in altering the count of $\ds_1$ by $\tvc^\Delta(\ds_1)$.
    However, although for $i \in \{2,\ldots,d\}$,
    this consumes (resp. produces) $\vc^\Delta(\ds_i)$ copies of $\ds_i$,
    it also produces $(\tvT_1.\tvc^\Delta)(i)$ copies of $\ds_i$ 
    (where ``producing'' a negative number corresponds to consuming copies of $\ds_i$).
    Note that $(\tvT_1.\tvc^\Delta)(1) = 0$ since $\ds_1$ does not appear in any transition other than $\tau_1$.
    
    We take the same approach as in the proof of Lemma~\ref{lem:eliminate-Delta},
    in which we take the vector $\tvT_1.\tvc^\Delta$,
    which represents the difference between the count of states in $\Delta$, compared to our target
    \emph{after} doing step 1 above.
    Step 2 consists of adding transitions as described by the vector $\tvT_1.\tvc^\Delta$,
    resulting in $\tvT_1^2.\tvc^\Delta$,
    in which $\tvT_1^2.\tvc^\Delta(1)=\tvT_1^2.\tvc^\Delta(2)=0$.
    The $i$'th step involves adding transitions according to the vector $\tvT_1^i.\tvc^\Delta$.
    We define $\tvT = \sum_{i=0}^{d-1} \tvT_1^i$.

    Thus, if transition $\tau_i$ appears $a_i$ times in $q$,
    then in the altered path $q'$,
    it appears $a_i + (\tvT.\tvc^\Delta)(i)$ times.
    Thus, so long as for each $i$, $a_i \geq -(\tvT.\tvc^\Delta)(i)$,
    there are sufficiently many transitions of each type in $q$ to potentially remove.
    Since each column of $\tvT_1$ has either a single $2$, or two $1$'s, and at most one $-1$,
    a simple induction shows that for each $i$,
    $\amax(\tvT_1^i) \leq 2^i$.
    Thus $\amax(\tvT) \leq \sum_{i=0}^{d-1} \tvT_1^i \leq \sum_{i=0}^{d-1} 2^i < 2^d$,
    which implies that $\max(-(\tvT.\tvc^\Delta)) < d 2^d \cdot \amax(\tvc^\Delta)$.
    Thus, it suffices if $a_i \geq d 2^d \cdot \amax(\tvc^\Delta)$.
    By Lemma~\ref{lem:ordering},
    $a_i \geq (b_2 - |\Lambda| \cdot b_1)/|\Lambda|^2$.
    Thus it suffices to show that 
    $d 2^d \cdot \amax(\tvc^\Delta) \leq (b_2 - |\Lambda| \cdot b_1)/|\Lambda|^2$.
    Note that $b_1 \geq \max(\vo^\Delta)$ by the defintion of $\Delta$ in the statement of the lemma.
    Then we have
    $d 2^d \cdot \amax(\tvc^\Delta)
    = d 2^d \cdot \amax(\vo^\Delta-\ve^\Delta)
    \leq d 2^d \cdot \max(\vo^\Delta, \ve^\Delta)
    \leq d 2^d \cdot \max(b_1, \ve^\Delta)
    \leq (b_2 - |\Lambda| \cdot b_1)/|\Lambda|^2$
    where the last inequality follows from the assumption that
    $b_2 \geq |\Lambda| \cdot b_1 + d 2^d \cdot \max(b_1, \ve^\Delta) \cdot |\Lambda|^2$.
    
    It remains to define $\tvC_3$, so that 
    the vector $\tvC_3.\tvc^\Delta = \tvC_3.\vo^\Delta - \tvC_3.\ve^\Delta$ 
    describes relative counts of states in $\Gamma$ compared to $\vo^\Gamma$.
    First, define the matrix $\vG\in\N^{\Gamma\times\Delta}$ as follows,
    which intuitively maps a count vector of transitions in $\tau_1,\ldots,\tau_d$ 
    to a total count of states in $\Gamma$ 
    either produced as output or consumed as a second input by the transitions.
    Let $j\in\{1,\ldots,d\}$ and let $\tau_j: \ds_j,\os \to \ans{o},\ans{o}'$.
    Let $i \in \{1,\ldots,g\}$ (recall $g=|\Gamma|$).
    Let $\vG(i,j) \in \{-1,0,1,2\}$ denote the net number of $\ans{g}_i$ produced by $\tau_j$;
    e.g., $-1$ if $\os=\ans{g}_i \neq \ans{o},\ans{o}'$,
    $0$ if $\ans{g}_i \neq \os,\ans{o}, \ans{o}'$ or if $\ans{g}_i = \os = \ans{o} \neq \ans{o}'$,
    etc.
    Then $\tvC_3 = \vG.\vT$,
    and $\amax(\tvC_3) \leq 2 \cdot \amax(\tvT) < 2^{d+1}$.
    
    Finally, recall that the new path may not be valid since, although the final configuration
    $\vo^\Gamma + \tvC_3.\vo^\Delta - \tvC_3.\ve^\Delta + \ve^\Delta \in \N^\Lambda$ is nonnegative,
    some intermediate configurations between $\vx$ and the final configuration could be negative.
    Define $\vG' \in \N^{\Lambda \times \Delta}$ similarly to $\vG$ above, 
    but reflecting the effect of a transition $\tau_1,\ldots,\tau_d$ on \emph{every} state $\os \in \Lambda$
    (not just those in $\Gamma$).
    Then letting $\tvC'_3 = \vG'.\vT$,
    $\amax(\tvC'_3.\tvc^\Delta) < d 2^{d+1} \amax(\tvc^\Delta)$ is an upper bound on how much any individual state count can change.
    Thus, for all $\os \in \Lambda$,
    letting $\vp(\os)
    = d 2^{d+1} \cdot \amax(\tvc^\Delta)
    = d 2^{d+1} \cdot \amax(\vo^\Delta-\ve^\Delta)
    \leq d 2^{d+1} \cdot \max(\vo^\Delta,\ve^\Delta)
    \leq d 2^{d+1} \cdot \max(b_1,\ve^\Delta)$
    suffices to ensure that the whole path
    $\vp + \vx \reach \vp + \vo^\Gamma + \tvC_3.\vo^\Delta - \tvC_3.\ve^\Delta + \ve^\Delta$ is valid.
\end{proof}

The following example provides intuition for Lemma~\ref{lem:produce-e}. Let $\Delta$, $\Gamma$, $\vx$, $\vo$, $\vp$ be defined as in Lemma~\ref{lem:produce-e}.
For all $\ve^\Delta$, we can alter a transition sequence $p$ --- either by adding or removing transitions --- to ensure we finish with only states in $\Gamma$ plus exactly $\ve^\Delta$ from $\Delta$.
In other words, given that we have additional agents in $\vp$ to use, we can manipulate a population protocol to have the exact amount of agents in $\Delta$ that we desire. 
Depending on the desired $\ve^\Delta$, doing so will effect the counts of states in $\Gamma$ in a predictable way. 

Let us continue with the above example using $\mathcal{P}$. 
In order to transition $\vc^\Delta$ to $\vz^\Gamma$, we needed to add $\ve$, which contained 14 copies of $\gs_1$ and 5 copies of $\ds_3$.
So, for this example, we will show how to produce $\ve^\Delta$ such that 
%$\ve^Delta(\ds_3) = 5$
\[
        \ve^\Delta= \left[\begin{array}{c}
            0  \\
            0 \\
            5 \\
        \end{array}\right]
    \]

Let $\vx \reach_p \vo$ by transition sequence $p$ and let 
\[
        \vo^\Delta= \left[\begin{array}{c}
            3  \\
            1 \\
            2 \\
        \end{array}\right]
    \]

% Given the counts of states in $\vo$ we will need to alter $p$ to remove 3 copies of $\ds_1$ and 1 copy of $\ds_2$ and add 3 copies of $\ds_3$, taking into account that the adding and removing of transitions within the transition sequence may affect the counts of other states in $\Delta$. 

 So, 
\[
        \tvc^\Delta = \vo^\Delta - \ve^\Delta=  \left[\begin{array}{c}
            3  \\
            1 \\
            -3 \\
        \end{array}\right]
    \]
representing the number of agents of each state in $\Delta$ that must be removed. The proof of Lemma~\ref{lem:produce-e} provides further detail on how to create matrices to determine exactly how many instances of each transition are added or removed based on $\tvc^\Delta$.
We will need to remove two instances of $\tau_3$ and add 3 instances of $\tau_1$ and 4 instances of $\tau_2$.

Additional transition executions can be added to the end of the transition sequence without otherwise affecting the original transition sequence; 
however, you can only remove transition executions where they take place. Thus, to remove two instance of $\tau_3$, we need to alter $p$ in the middle of the sequence.

Removing transitions has an affect on the overall counts of other states. 
We have extra states in $\vp$ to account for this. 
If removing a transition in the middle of the sequence causes the count of a state to become zero later in the sequence, and if a transition requiring that state was meant to take place after that point, the extra agents in $\vp$ can be used instead. 

Our example population protocol $\mathcal{P}$ includes $\tau_4$ and $\tau_5$ to demonstrate how $\vp$ may be used.
While these transition rules may not seem particularly useful in practice, they allow us to show a simple situation in which the count of one state dips to zero. 

\begin{sidewaysfigure}[p]
\centering
\includegraphics[width=\textwidth]{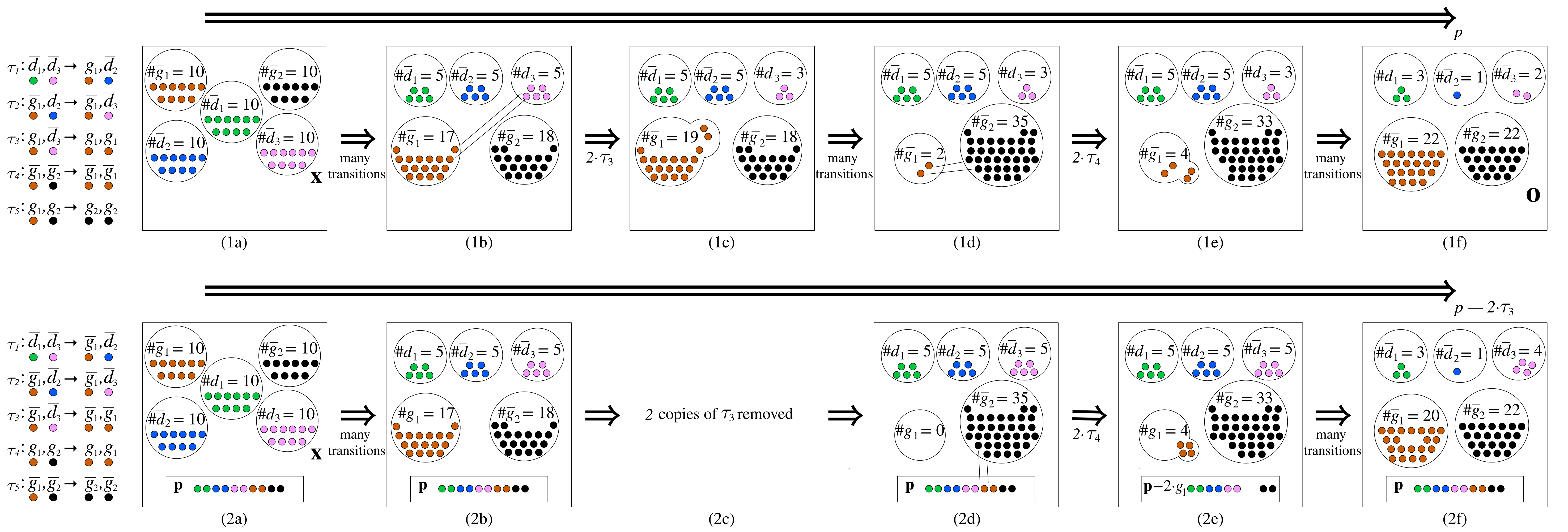}
\caption[LoF entry]{
Surgery as described in Lemma~\ref{lem:produce-e} to produce two extra copies of $\ds_3$. 
This figure shows the difference between the original transition sequence and the sequence after surgery.
The top diagram shows $\vx \reach_p \vo$ without any interference. 
{\bf (1a)} The initial configuration $\vx$. 
{\bf (1b)} an intermediate configuration.
{\bf (1c)} 2 instances of $\tau_3$ take place. 
{\bf (1d)} A configuration after many instances of $\tau_5$ have caused the counts of $\gs_1$ to drop to 2. 
{\bf (1e)} 2 instances of $\tau_4$ take place bringing the count of $\gs_1$ to 4. 
{\bf (1f)} Finally, the transition sequence ends in $\vo$.
Consistent with the definition of $\Delta$ and $\Gamma$, the counts of states in $\Delta$ are low, and counts of both $\gs_1$ and $\gs_2$ are ``large''.

{\bf (2a)} The same initial configuration $\vx$ from (1a).
{\bf (2b)} the same intermediate configuration from (1b).
{\bf (2c)} Remove 2 instances of $\tau_3$. 
{\bf (2d)} Due to the removal of 2 instances of $\tau_3$, there are 2 less copies of $\gs_1$ and 2 extra copies of $\ds_3$ compared to (1d). The transition sequence reaches a point where there are not enough agents outside of $\vp$ for $\tau_4$ to execute.
{\bf (2e)} We must use the additional agents in $\vp$ to continue with the remainder of the transition sequence $p - 2 \tau_3$.
{\bf (2f)} We are left with 2 additional copies of $\ds_3$ and 2 less copies of $\gs_1$ due to removing 2 instances of $\tau_3$.
We have created some of the copies of $\ds_3$ required for our desired $\ve^{\Delta}$.
Note that $\vp$ can be ``restored'' by taking copies of $\gs_1$ from $\vo^\Gamma$ since it has ``large'' counts.
%When employed in Lemma~\ref{lem:push-Delta}, entries of $\vo^\Gamma$ (in this example, representing counts of $\gs_1$ and $\gs_2$) grow without bound, so for sufficiently large starting configurations $\vx$, this is always possible.
}
\label{fig:claim2_12}
\end{sidewaysfigure}

In $\mathcal{P}$, removing two instances of $\tau_3$ will increase the count of $\ds_3$ by 2, at the same time reducing the count of $\gs_1$ by 2. 
At some point in $p$, the count of $\gs_1$ dips to 2. With 2 fewer $\gs_1$ agents in our modified transition sequence, we instead arrive at a configuration where there are zero copies of $\gs_1$. 
In order for the next $\tau_4$ transition to take place, we must use the extra agents provided by $\vp$. 
Thus, the presence of additional agents allows the rest of the transition sequence to remain unchanged after removing transitions from the transition sequence. 
This is demonstrated in Figure~\ref{fig:claim2_12}.

\begin{figure}[ht]
\centering
\includegraphics[height=4cm]{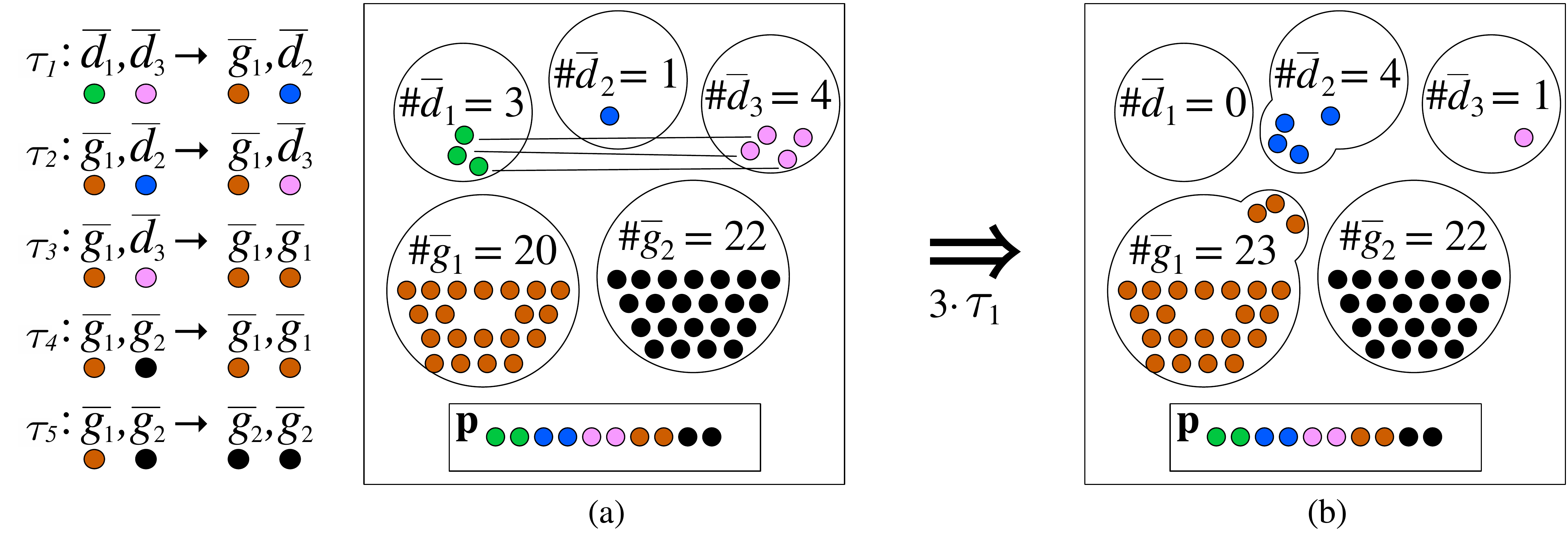}
\caption{Removing all instances of $\ds_1$ using techniques detailed in Lemma~\ref{lem:produce-e}. 
Since we are adding transitions (not removing them as in Figure~\ref{fig:claim2_12}), 
this is similar to the surgery shown in Figures~\ref{fig:claim1_1},~\ref{fig:claim1_2}, and~\ref{fig:claim1_3}.
{\bf (a)} The same configuration as seen in panel (2f) of Figure~\ref{fig:claim2_12}.
{\bf (b)}  Add 3 instances of $\tau_1$ to remove all copies of $\ds_1$ as $\ve^{\Delta}$ contains no copies of $\ds_1$} 
\label{fig:claim2_step3}
\end{figure}

\begin{figure}[ht]
\centering
\includegraphics[height=4cm]{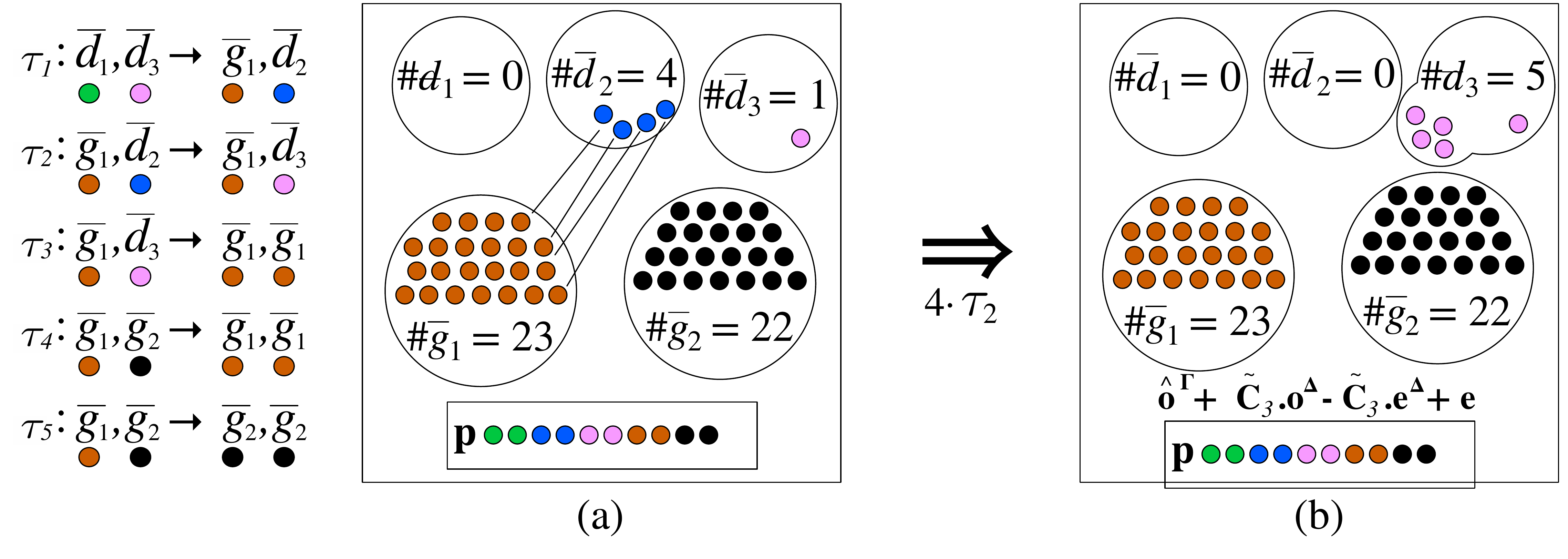}
\caption{Removing all instances of $\ds_2$ using techniques detailed in Lemma~\ref{lem:produce-e}. 
This is similar to the surgery shown in Figures~\ref{fig:claim1_1},~\ref{fig:claim1_2}, and~\ref{fig:claim1_3}.
{\bf (a)} The same configuration as seen in panel (b) of Figure~\ref{fig:claim2_step3}.
{\bf (b)} Add 4 instances of $\tau_2$ to remove all copies of $\ds_2$ as  $\ve^{\Delta}$ contains no copies of $\ds_2$
{\bf (c)} The only agents from $\Delta$ that remain are exactly those needed in $\ve^\Delta = \{ 5 \ds_3\}$.}
\label{fig:claim2_step4}
\end{figure}

    % Letting $\tvc^\Delta = \vo^\Delta - \ve^\Delta$, 
    % then $\tvt = \tvT.\tvc^\Delta$ describes how many transitions of each type to add or remove from $q$.  
    
We can then add the transitions necessary to remove $\ds_1$ and $\ds_2$ from the resulting configuration, since our final configuration $\ve^{\Delta}$ contains no agents from these states. 
This process is similar to the one detailed above to describe Lemma~\ref{lem:eliminate-Delta}. 
In Figure~\ref{fig:claim2_step3} we add 3 instances of $\tau_1$ and in Figure~\ref{fig:claim2_step4} we add 4 instances of $\tau_3$, which completes the adjusted transition sequence.

In the final configuration of Figure~\ref{fig:claim2_step4}, counts of states in $\Gamma$ are altered and counts of states in $\Delta$ are exactly equal to those in $\ve^\Delta$ as desired. 
This demonstrates how we can create $\ve$ as required by Lemma~\ref{lem:eliminate-Delta}.
This concludes the description of the example showing Lemma~\ref{lem:produce-e}.

%%%%%%%%%%%%%%%%%%%%%%%%%%%%%%%%%%%%%%%
%%%%% See Amanda's diagrams above %%%%%
%%%%%% Lemma~\ref{lem:produce-e} %%%%%%
%%%%%%%%%%%%%%%%%%%%%%%%%%%%%%%%%%%%%%%

The following is a general lemma that  
uses Lemmas~\ref{lem:eliminate-Delta} and~\ref{lem:produce-e}
to ``steer'' certain configurations 
(each of which is expressed as 
``twice a configuration $\vx_n$ where all counts are large ($\geq b_2$), 
plus a few more states described by $\vd^\Delta$'')
to a configuration with a ``target'' $\vt^\Delta$ amount of states in $\Delta$.

Intuitively,
the proof goes like this:
Letting $\vc^\Delta$ in Lemma~\ref{lem:eliminate-Delta} equal $\vo_n \rest \Delta + \vd^\Delta$,
the states in $\Delta$ in the final configuration $\vo_n$,
plus the states $\vd^\Delta$ that we have added at the start,
which we want to eliminate,
use Lemma~\ref{lem:eliminate-Delta} to determine what states $\ve$ need to be added to apply the extra transitions of Lemma~\ref{lem:eliminate-Delta} and eliminate all of $\vd^\Delta + \vo_n \rest \Delta$.
Then, apply Lemma~\ref{lem:produce-e} to produce these states $\ve$ from $\vx_n$.
However, Lemma~\ref{lem:produce-e} requires the presence of an extra ``buffer'' $\vp$ of states to enable $\ve$ to be produced.
A second copy of $\vx_n$ serves as this extra buffer 
(for large enough $n$ $\vx_n \geq \vp$, 
since $\vx_n(\os) \geq \beta n$ for all state counts in $\vx_n$, 
due to $\vx_n$, when we employ Lemma~\ref{lem:push-Delta},
being the sequence of configurations from Lemma~\ref{lem:cor:pos-prob-states-sublinear-expected-time-implies-bottleneck-free} in which all state counts are at least $\beta n$ for a fixed $\beta > 0$).
Finally,
we can generalize to not only eliminate all of $\vd^\Delta + \vo_n \rest \Delta$
(i.e., set final counts of states in $\Delta$ all to $0$),
but to target other \emph{positive} counts of states in $\Delta$,
represented by the vector $\vt^\Delta$.
The states in $\vt^\Delta$ can simply be generated by Lemma~\ref{lem:produce-e} alongside the states in $\ve$.
In this paper we use only Corollary~\ref{cor:push-Delta-simplified},
which sets $\vt^\Delta = \vec{0}$,
but it may be useful in other applications to be able to choose a nonzero $\vt^\Delta$.

\begin{lem} \label{lem:push-Delta}
    Let $b_2 \in \N$.
    Let $N \subseteq \N$ be infinite.
    Let $(\vx_n)_{n\in N}$ and $(\vo_n)_{n\in N}$
    be nondecreasing sequences of configurations. 
    Let $(p_n)_{n\in N}$ be a sequence of paths
    such that, for all $n\in N$,
    \begin{enumerate}
        \item \label{lem:push-Delta:cond-size-n}
        $\|\vx_n\|=\|\vo_n\|=n$,
        
        \item \label{lem:push-Delta:cond-xn-big}
        $\vx_n(\os) \geq b_2$ for all $\os \in \Lambda$,
        
        \item \label{lem:push-Delta:cond-reach}
        $\vx_n \reach_{p_n} \vo_n$, and
    
        \item \label{lem:push-Delta:cond-no-bottleneck}
        $p_n$ has no $b_2$-bottleneck transition.
    \end{enumerate}
    Let $\Delta = \bdd((\vo_n)_{n\in N})$,
    $d=|\Delta|$,
    and $\Gamma = \Lambda \setminus \Delta$.
    Then there are matrices 
    $\tvD_1 \in \Z^{\Gamma \times \Delta}$ 
    and 
    $\tvD_2 \in \Z^{\Gamma \times \Delta}$
    with $\amax(\tvD_1),\amax(\tvD_2) \leq d2^{2d+2}$ 
    and the following holds.
    For each $n\in N$, let $\vo_n^\Delta=\vo_n\rest\Delta$ and $\vo_n^\Gamma=\vo_n\rest\Gamma$.
    Let $b_1 = \max\limits_{n \in N,\ans{d} \in \Delta} \vo_n^\Delta(\ans{d})$.
    For all $b\in\N$,
    there is $n_b \in N$ such that for all $n \in N$ with $n \geq n_b$ 
    and all $\vd^{\Delta},\vt^\Delta \in \N^{\Delta}$ 
    with 
    $\vd^\Delta \leq b$,
    $\vt^\Delta \leq d 2^d (b_1 + b)$,
    and 
    $b_2 \geq 
    \max( 
        |\Lambda| \cdot b_1 + d 2^d \cdot \max(b_1, d 2^{d+1} (b_1+b)) \cdot |\Lambda|^2, 
        d^2 2^{2d+2} (b_1+b)
    )$,
    we have
    $2 \vx_n + \vd^\Delta \reach 2 \vo_n^\Gamma + \tvD_1.\vo_n^\Delta + \tvD_2.\vd^\Delta + \vt^\Delta$.
\end{lem}

\begin{proof}
    Choose $n\in N$ sufficiently large to satisfy the conditions below as needed.
    The bound on $b_2$ in the hypothesis is a max of two different bounds,
    each needed for its own purpose:
    \begin{enumerate}
        \item 
        $b_2 \geq |\Lambda| \cdot b_1 + d 2^d \cdot \max(b_1, d 2^{d+1} (b_1+b)) \cdot |\Lambda|^2$
        is necessary to apply Lemma~\ref{lem:produce-e}.
        
        \item
        $b_2 \geq d^2 2^{2d+2} (b_1+b)$
        is necessary to ensure that $\vx_n$,
        which has $\vx_n(\os) \geq b_2$ for all $\os \in \Lambda$ by hypothesis,
        obeys $\vx_n \geq \vp$ as defined below,
        where $\vp$ is used as in Lemma~\ref{lem:produce-e}.
    \end{enumerate}
    
    Choose $\tvC_3$ for $\Delta$, $\Gamma$, $\vx_n$, $\vo_n$, and $p_n$ as in Lemma~\ref{lem:produce-e}.
    By Lemma~\ref{lem:eliminate-Delta}, 
    letting $\vc^\Delta = \vd^\Delta + \vo_n^\Delta$,
    there are $\vC_1\in\N^{\Gamma\times\Delta}$ and $\vC_2\in\N^{\Lambda\times\Delta}$,
    so that, setting $\ve = \vC_2.\vc^\Delta$,
    there is a transition sequence $q_1$ such that
    \begin{align*}
        \vd^\Delta + \vo_n^\Delta + \ve 
    &\reach_{q_1} 
        \vC_1.(\vd^\Delta + \vo_n^\Delta) 
    \\&= 
        \vC_1.\vd^\Delta + \vC_1.\vo_n^\Delta.
    \end{align*}
    Since 
    $\max(C_1) < 2^{d+1}$, 
    $\max(C_2) < 2^d$,
    $\max(\vo_n^\Delta) \leq b_1$,
    and 
    $\max(\vd^\Delta) \leq b$,
    we have
    $\max(\vC_1.(\vd^\Delta + \vo_n^\Delta) < d 2^{d+1} (b_1+b) $
    and
    $\max(\ve) = \max(\vC_2.(\vd^\Delta + \vo_n^\Delta)) < d 2^d (b_1+b).$
    
    Let $\ve^\Delta = \ve\rest\Delta$ and $\ve^\Gamma = \ve\rest\Gamma$.
    Let $\ve_2^\Delta = \ve^\Delta + \vt^\Delta$.
    Since $\max(\ve) < d 2^d (b_1+b)$ and $\max\vt^\Delta \leq d 2^d (b_1+b)$,
    we have that
    $\max(\ve_2^\Delta) \leq d 2^{d+1} (b_1+b)$.
    
    Apply Lemma~\ref{lem:produce-e} on $\ve_2^\Delta$, which says that,
    letting $\vp \in \N^\Lambda$ be defined 
    \begin{align*}
        \vp(\os) = d 2^{d+1} \cdot \max(b_1, \ve_2^\Delta)
    &\leq 
        d 2^{d+1} \cdot \max(b_1, d 2^{d+1} (b_1+b))
    \\&= 
        d 2^{d+1} \cdot d 2^{d+1} (b_1+b) 
    \\&= 
        d^2 2^{2d+2} (b_1+b)
    \end{align*}
    for all $\os \in \Lambda$,
    there is a transition sequence $q_2$ such that
    \begin{align*}
        \vp + \vx_n 
    &\reach_{q_2}
        \vp + \vo_n^\Gamma + \tvC_3.(\vo_n^\Delta) - \tvC_3.\ve_2^\Delta + \ve_2^\Delta 
    \\&= 
        \vp + \vo_n^\Gamma + \tvC_3.(\vo_n^\Delta) - \tvC_3.\ve_2^\Delta + \ve^\Delta + \vt^\Delta.
    \end{align*}
    
    Since 
    $b_2 \geq \max( |\Lambda| \cdot b_1 + d 2^d \cdot \max(b_1, \ve_2^\Delta) \cdot |\Lambda|^2, (d^2+1) 2^{2d+1} (b_1+b))$,
    we have
    $b_2 \geq (d^2+1) 2^{2d+1} (b_1+b)$.
    Since
    $\vx_n(\os) \geq b_2$ for all $\os \in \Lambda$,
    $\vx_n \geq \vp$, 
    so by additivity
    \[
        2 \vx_n
    \reach_{q_2} 
        \vx_n 
        + \vo_n^\Gamma 
        + \tvC_3.(\vo_n^\Delta) 
        - \tvC_3.\ve^\Delta 
        + \ve^\Delta 
        + \vt^\Delta.
    \]

    For large enough $n$, 
    for all $\os \in \Gamma$,
    $\vo_n^\Gamma(\os) \geq \ve^\Gamma(\os).$
    Define $\hat{\vo}_n^\Gamma = \vo_n^\Gamma - \ve^\Gamma$.
    Then $\vo_n^\Gamma + \ve^\Delta = \hat{\vo}_n^\Gamma + \ve$.
    Therefore 
    $2 \vx_n \reach_{q_2} \vx_n + \hat{\vo}_n^\Gamma + \tvC_3.\vo_n^\Delta - \tvC_3.\ve^\Delta + \ve^\Delta + \vt^\Delta$.
    Recall that the transition sequence $p_n$ takes $\vx_n \reach_{p_n} \vo_n$.
    By additivity we have
    (the relevant parts of the configuration needed for the subsequence transitions are \underline{underlined})
    \begin{align*}
        \underline{2 \vx_n}
        + \vd^\Delta 
        &\reach_{q_2}
        \underline{\vx_n}
        + \hat{\vo}_n^\Gamma 
        + \tvC_3.\vo_n^\Delta 
        - \tvC_3.\ve^\Delta 
        + \ve 
        + \vt^\Delta
        + \vd^\Delta 
    \\&\reach_{p_n} 
        \underline{\vo_n}
        + \hat{\vo}_n^\Gamma 
        + \tvC_3.\vo_n^\Delta 
        - \tvC_3.\ve^\Delta 
        + \vd^\Delta 
        + \ve 
        + \vt^\Delta
    \\&= 
        \vo_n^\Gamma
        + \hat{\vo}_n^\Gamma 
        + \tvC_3.\vo_n^\Delta 
        - \tvC_3.\ve^\Delta 
        + \underline{\vo_n^\Delta
        + \vd^\Delta
        + \ve}
        + \vt^\Delta
    \\&\reach_{q_1} 
        \vo_n^\Gamma 
        + \underline{\hat{\vo}_n^\Gamma}
        + \tvC_3.\vo_n^\Delta 
        - \tvC_3.\ve^\Delta 
        + \vC_1.\vd^\Delta 
        + \vC_1.\vo_n^\Delta 
        + \vt^\Delta
    \\&=
        2 \vo_n^\Gamma 
        + \tvC_3.\vo_n^\Delta 
        - \tvC_3.\ve^\Delta 
        + \vC_1.\vd^\Delta 
        + \vC_1.\vo_n^\Delta 
        - \ve^\Gamma 
        + \vt^\Delta
    \end{align*}
    Recall that $\ve = \vC_2.(\vd^\Delta + \vo_n^\Delta)$,
    so $\ve^\Gamma = \vC_2.(\vd^\Delta + \vo_n^\Delta) \rest \Gamma$.
    Thus, 
    the last configuration above is
    \begin{align*}
        2 \vo_n^\Gamma 
        + \tvC_3.\vo_n^\Delta 
        - \tvC_3.(\vC_2.(\vd^\Delta + \vo_n^\Delta)\rest\Delta) 
        + \vC_1.\vd^\Delta 
        + \vC_1.\vo_n^\Delta 
        - \vC_2.(\vd^\Delta + \vo_n^\Delta)\rest\Gamma
        + \vt^\Delta
    \end{align*}
    Let $\vC_2^\Gamma$ be $\vC_2$ with the rows corresponding to $\Delta$ removed
    (so that for any $\vv\in\N^{\Delta}$, $\vC_2^\Gamma.\vv \in \Z^\Gamma$; 
    this has the same effect as restricting the output vector to $\Gamma$, 
    as above with $\vC_2.(\vd^\Delta + \vo_n^\Delta)\rest\Gamma$).
    Similarly, let $\vC_2^\Delta$ be $\vC_2$ with the rows corresponding to $\Gamma$ removed,
    and let $\tvC_4\in\Z^{\Gamma\times\Delta}$ be $\tvC_3.\vC_2^\Delta$,
    Then the above configuration is
    \begin{align*}
        &
        2 \vo_n^\Gamma 
        + \tvC_3.\vo_n^\Delta 
        - \tvC_4.(\vd^\Delta + \vo_n^\Delta) 
        + \vC_1.\vd^\Delta 
        + \vC_1.\vo_n^\Delta 
        - \vC_2^\Gamma.(\vd^\Delta + \vo_n^\Delta)
        + \vt^\Delta
        \\=\ &
        2 \vo_n^\Gamma 
        + \tvC_3.\vo_n^\Delta 
        - \tvC_4.\vo_n^\Delta 
        + \vC_1.\vo_n^\Delta 
        - \vC_2^\Gamma.\vo_n^\Delta 
        - \tvC_4.\vd^\Delta 
        + \vC_1.\vd^\Delta 
        - \vC_2^\Gamma.\vd^\Delta
        + \vt^\Delta.
    \end{align*}
    Letting $\tvD_1 = \tvC_3 - \tvC_4 + \vC_1 - \vC_2^\Gamma$
    and $\tvD_2 = - \tvC_4 + \vC_1 - \vC_2^\Gamma$,
    the above is 
    $2 \vo_n^\Gamma + \tvD_1.\vo_n^\Delta + \tvD_2.\vd^\Delta + \vt^\Delta.$
    Since $\max(\vC_2) < 2^d$, and $\amax(\tvC_3) < 2^{d+1}$, we can conclude that $\amax(\tvC_4) < d2^{2d+1}$. 
    Since $\max(\vC_1) < 2^{d+1}$ and $\vC_1$ and $\vC_2$ are both nonnegative, we know that $\amax(\vC_1 - \vC_2^{\Gamma}) < 2^{d+1}$.
    Thus, $\amax(\tvD_1) < 2^{d+1} + d2^{2d+1} + 2^{d+1} < d2^{2d+2}$, 
    and similarly for $\amax(\tvD_2)$.
\end{proof}

We will use the same population protocol used in the previous two examples, 
$\mathcal{P}$, to demonstrate how Lemma~\ref{lem:eliminate-Delta} and Lemma~\ref{lem:produce-e} can be used in conjunction with each other.
We choose the special case of target $\vt^\Delta = \vec{0}$.
We will show that if a transition sequence arrives at a configuration without any $b_2$-bottleneck transitions, then we can also bring all counts of states in $\Delta$ to zero. 
In the statement of Lemma~\ref{lem:push-Delta}, we define $\Gamma$ and $\Delta$ over an infinite sequence. 
Here, will we be looking at a single population protocol with configuration $\vx_n$ that satisfies the constraints of Lemma~\ref{lem:push-Delta}. 
Let $\vC_1$ be defined as in the proof of Lemma~\ref{lem:push-Delta}

We will begin with two copies of $\vx_n$ plus $\vd^\Delta$ which contains additional agents from $\Delta$ that need to be removed. 

\[
    \vx_n= \left[\begin{array}{c}
        10  \\
        10 \\
        10 \\
        10 \\
        10 \\
    \end{array}\right]
\]

\[
    \vd^\Delta= \left[\begin{array}{c}
        2  \\
        0 \\
        0 \\
    \end{array}\right]
\]

The second copy of $\vx_n$ will serve the same purpose as $\vp$ in Lemma~\ref{lem:produce-e}.

\begin{figure}[ht]
\includegraphics[width=\textwidth]{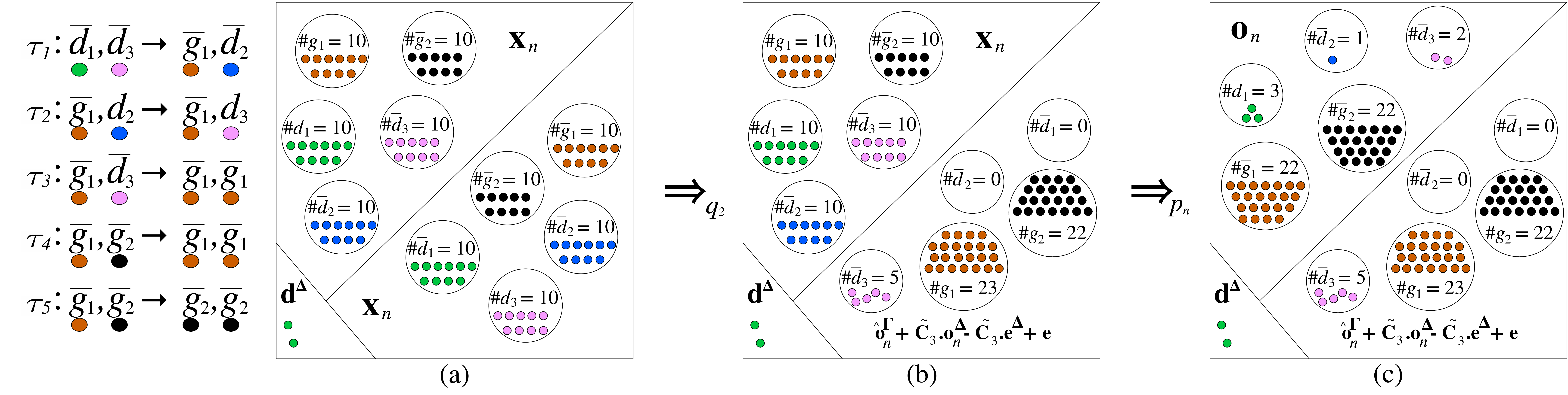}
\caption{The first steps of Lemma~\ref{lem:push-Delta} use techniques from Lemma~\ref{lem:produce-e} to push one copy of $\vx_n$ to the desired configuration and produce $\ve$.
{\bf (a)} 
We start with two copies of $\vx_n$ plus $\vd^\Delta$.
{\bf (b)} 
Via transition sequence $q_2$ based on the techniques of  Lemma~\ref{lem:produce-e}, we produce the $\vec{e} = \{ 5 \ds_3, 14 \gs_1 \}$ needed for Lemma~\ref{lem:eliminate-Delta}.
{\bf (c)} Via transition sequence $p_n$, the second copy of $\vx_n$ transitions without interference to $\vo_n$. 
We end in the configuration
$\vd^\Delta + \vo_n + \hat{\vo}_n^\Gamma + \tvC_3.\vo_n^\Delta - \tvC_3.\ve^\Delta + \ve$.
Figure~\ref{fig:claim3_2} shows how to use Lemma 3.7 to use the $\ve$ produced in (b) to remove states in $\Delta$ in $\vo_n$ in (c).}
\label{fig:claim3_1}
\end{figure}

Using Lemma~\ref{lem:produce-e}, one copy of $\vx_n$ will transition via transition sequence $q_2$ to $\hat{\vo}_n^\Gamma + \tvC_3.\vo_n^\Delta - \tvC_3.\ve^\Delta + \ve$.
We use the same techniques here to build the desired $\ve^\Delta$ as in Lemma~\ref{lem:produce-e}.
Then, the second copy of $\vx_n$ will transition ``normally'' via $p_n$ to $\vo_n$.
The process is shown in Figure~\ref{fig:claim3_1}.

\[
    \vo_n^\Delta= \left[\begin{array}{c}
        3  \\
        1 \\
        2 \\
    \end{array}\right]
\]

\begin{figure}[ht]
\includegraphics[width=\textwidth]{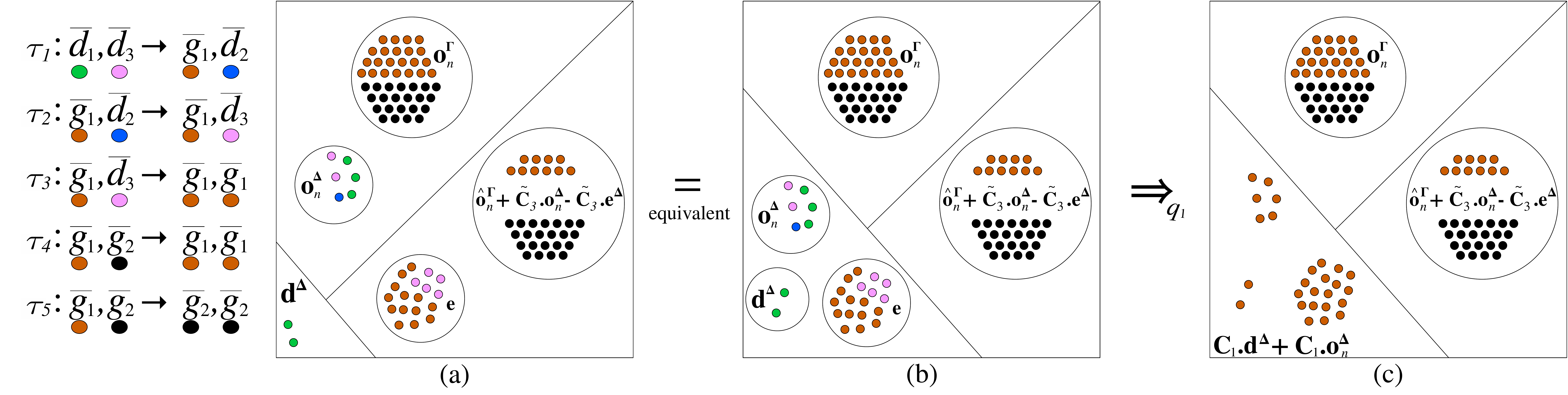}
\caption{The second steps of Lemma~\ref{lem:push-Delta} use techniques from Lemma~\ref{lem:eliminate-Delta} to remove all remaining states from $\Delta$.
{\bf (a)}
This is the same configuration as in panel (c) in Figure~\ref{fig:claim3_1}.
{\bf (b)} We visually separate $\ve$ from $\hat{\vo}_n^\Gamma + \tvC_3.\vo_n^\Delta - \tvC_3.\ve^\Delta + \ve$ and $\vo_n^\Delta$ from $\vo_n^\Gamma$.
{\bf (c)} Via transition sequence $q_1$ based on the techniques of Lemma~\ref{lem:eliminate-Delta}, $\vd^\Delta + \vo_n^\Delta + \ve \reach_{q_1} \vC_1.\vd^\Delta + \vC_1.\vo_n^\Delta$. The agents in $\ve$ react with agents from $\vo_n$ and $\vd^\Delta$ to remove all remaining states in $\Delta$. The final configuration has only states from $\Gamma$, as desired.}
\label{fig:claim3_2}
\end{figure}

At this point, our goal is to remove all agents from $\vd^\Delta$ and $\vo_n^\Delta$. 
Using the techniques from Lemma~\ref{lem:eliminate-Delta} along with $\ve$ created previously using the techniques of  Lemma~\ref{lem:produce-e}, $\ve$ has exactly the counts of agents from $\Delta$ needed to remove all of $\vd^\Delta + \vo_n^\Delta$.
And so, $\vd^\Delta + \vo_n^\Delta + \ve \reach_{q_1} \vC_1.(\vd^\Delta + \vo_n^\Delta) = \vC_1.\vd^\Delta + \vC_1.\vo_n^\Delta$
The resulting final configuration contains only agents in $\Gamma$. 
As shown in Figure~\ref{fig:claim3_2}, all agents in $\Delta$ have been removed, as desired. 

%%%%%%%%%%%%%%%%%%%%%%%%%%%%%%%%%%%%%%%
%%%%% See Amanda's diagrams Below %%%%%
%%%%%% Lemma~\ref{lem:push-Delta} %%%%%
%%%%%%%%%%%%%%%%%%%%%%%%%%%%%%%%%%%%%%%

% \clearpage

Finally,
we combine 
Lemmas~\ref{lem:cor:pos-prob-states-sublinear-expected-time-implies-bottleneck-free}
and~\ref{lem:push-Delta}
into a single lemma,
which is the main technical result of this subsection,
used 
(via Corollary~\ref{cor:push-Delta-simplified}, 
which sets $\vt^\Delta = \vec{0}$)
for proving time lower bounds in 
Theorems~\ref{thm:predicates},
\ref{thm:subtraction-slow}, 
\ref{thm:division-slow},
\ref{thm:affine-not-sublinear-computable}, 
and
\ref{thm:negative-result-eventually-positive-integral-linear}.

\begin{lem}
    \label{lem:push-Delta-simplified}
    Let $\alpha > 0$.
    Suppose that for some set $S\subseteq\N^\Lambda$ and infinite set $I$ of $\alpha$-dense configurations, 
    for all $\vi \in I$,
    letting $n = \|\vi\|$,
    $\time{\vi}{S} = o(n)$.
    
    Then there are matrices 
    $\tvD_1 \in \Z^{\Gamma \times \Delta}$ 
    and 
    $\tvD_2 \in \Z^{\Gamma \times \Delta}$,
    % with $\amax(\tvD_1)$, $\amax(\tvD_2) \leq d2^{2d+2}$,
    an infinite set $N \subseteq \N$, 
    and infinite nondecreasing sequences of configurations
    $(\vi_n)_{n\in N}$
    and
    $(\vo_n)_{n\in N}$
    such that the following holds.
    Let $\Delta = \bdd((\vo_n)_{n\in N})$,
    $d=|\Delta|$,
    and $\Gamma = \Lambda \setminus \Delta$.
    For each $n\in N$, let
    $\vo_n^\Delta=\vo_n\rest\Delta$ 
    and 
    $\vo_n^\Gamma=\vo_n\rest\Gamma$.
    Then $\amax(\tvD_1)$, $\amax(\tvD_2) \leq d2^{2d+2}$ and 
    \begin{enumerate}
        \item
        For all $n\in N$, 
        $\vi_n \in I$,
        $\vo_n \in S$,
        $\|\vi_n\| = \|\vo_n\| = n$,
        and
        $\vi_n \reach \vo_n$.
        
        \item %\label{lem:push-Delta-simplified:cond:reach}
        Let $b_1 = \max\limits_{n\in N,\ans{d}\in\Delta} \vo_n^\Delta(\ans{d})$.
        For all 
        $b \in \N$,
        there is $n_b \in N$ such that,
        for all $n \in N$ such that $n \geq n_b$
        and all $\vd^{\Delta},\vt^\Delta \in \N^{\Delta}$ such that 
        $\vd^\Delta \leq b$
        and
        $\vt^\Delta \leq d 2^d (b_1 + b)$,
        we have that
        $2 \vi_n + \vd^\Delta \reach 2 \vo_n^\Gamma + \tvD_1.\vo_n^\Delta + \tvD_2.\vd^\Delta + \vt^\Delta$.
    \end{enumerate}
\end{lem}

\begin{proof}
    Let $t:\N\to\N$ be such that 
    $\time{\vi_n}{S} \leq t(n)$.
    Define $b(n) = \frac{1}{4 |\Lambda|} \sqrt{\frac{n}{t(n)}}$; 
    note that $b(n) = \omega(1)$ since $t(n) = o(n)$ by hypothesis.
    By Lemma~\ref{lem:cor:pos-prob-states-sublinear-expected-time-implies-bottleneck-free}
    there is $\beta > 0$ such that the following holds.
    There is an infinite set $N \subseteq \N$ and infinite sequences of configurations
    $(\vi_n \in I)_{n\in N}$, 
    $(\vx_n \in \N^\Lambda)_{n\in N}$, 
    $(\vo_n \in \N^\Lambda)_{n\in N}$, 
    where $(\vx_n)_{n\in N}$ and $(\vo_n)_{n\in N}$ are nondecreasing,
    and an infinite sequence of paths $(p_n)_{n\in N}$
    such that, for all $n\in N$,
    \begin{enumerate}
        \item 
        $\|\vi_n\|=\|\vx_n\|=\|\vo_n\|=n$,
        
        \item
        $\vi_n \reach \vx_n$,
        
        \item
        $\vx_n(\os) \geq \beta n$ for all $\os \in \Lambda$,
        
        \item
        $\vx_n \reach_{p_n} \vo_n$, where $\vo_n \in S$, and
        
        \item
        $p_n$ has no $b(n)$-bottleneck transition.
    \end{enumerate}
    Conditions~\eqref{lem:push-Delta:cond-size-n}
    and~\eqref{lem:push-Delta:cond-reach}
    of Lemma~\ref{lem:push-Delta} are two of the above.
    Let $b_1 = \max\limits_{n\in N,\ans{d}\in\Delta} \vo_n(\ans{d})$.
    % Let $\vd^\Delta,\vt^\Delta \in \N^\Delta$.
    Let $b_2
        = 
        \max( 
            |\Lambda| \cdot b_1 + d 2^d \cdot \max(b_1, d 2^{d+1} (b_1+b)) \cdot |\Lambda|^2, 
            d^2 2^{2d+2} (b_1+b)
        )$.
    %|\Lambda| \cdot b_1 + d 2^{2d}\cdot (b_1 + \amax(\vd^\Delta - \vt^\Delta)) \cdot |\Lambda|^2$.
    For sufficiently large $n$, $\beta n > b_2$,
    satisfying 
    condition~\eqref{lem:push-Delta:cond-xn-big} 
    of Lemma~\ref{lem:push-Delta},
    and $b(n) \geq b_2$, 
    satisfying condition~\eqref{lem:push-Delta:cond-no-bottleneck}
    of Lemma~\ref{lem:push-Delta}.
    
    Thus Lemma~\ref{lem:push-Delta} 
    tells us that there are 
    $\tvD_1,\tvD_2 \in \Z^{\Gamma \times \Delta}$
    such that for all $b \in \N$,
    there exists $n_b\in N$
    such that 
    for all $n \geq n_b$ such that $n \in N$
    and all $\vd^{\Delta},\vt^\Delta \in \N^{\Delta}$ such that 
    $\vd^\Delta \leq b$
    and
    $\vt^\Delta \leq d 2^d (b_1 + b)$,
    letting $\vo_n^\Delta = \vo_n \rest\Delta$ and $\vo_n^\Gamma = \vo_n \rest \Gamma$,
    $2 \vx_n + \vd^\Delta \reach 2 \vo_n^\Gamma + \tvD_1.\vo_n^\Delta + \tvD_2.\vd^\Delta + \vt^\Delta$.
    
    Since $\vi_n \reach \vx_n$,
    by additivity
    $2 \vi_{n} + \vd^\Delta \reach 2 \vo_n^\Gamma + \tvD_1.\vo_n^\Delta + \tvD_2.\vd^\Delta + \vt^\Delta$.
    Lemma~\ref{lem:push-Delta} also gives that 
    $\amax(\tvD_1),\amax(\tvD_2) \leq d2^{2d+2}$.
\end{proof}

The following corollary,
which sets $\vt^\Delta = \vec{0}$,
is the only result of this subsection employed for our time lower bounds.
% TODO: I'd like the following simplification, but one or two proofs actually use the linear algebra exact counts.
% It also simplifies the statement by removing the matrices $\tvD_1$ and $\tvD_2$,
% simply concluding that the change in counts of states in $\Gamma$ in $2 \vo_n$
% (represented by $\tvD_1.\vo_n^\Delta$ and $\tvD_2.\vd^\Delta$)
% is a constant independent of $n$,
% represented by $\vc^\Gamma$ in the statement of the corollary.

\newcommand{\corPushDeltaSimplifiedConclusion} {
    Then there are matrices 
    $\tvD_1 \in \Z^{\Gamma \times \Delta}$ 
    and 
    $\tvD_2 \in \Z^{\Gamma \times \Delta}$,
    an infinite set $N \subseteq \N$, 
    and infinite nondecreasing sequences of configurations
    $(\vi_n)_{n\in N}$
    and
    $(\vo_n)_{n\in N}$
    such that the following holds.
    Let $\Delta = \bdd((\vo_n)_{n\in N})$,
    $d=|\Delta|$,
    and $\Gamma = \Lambda \setminus \Delta$.
    % Let $b_1 = \max\limits_{n\in N,\ans{d}\in\Delta} \vo_n(\ans{d})$.
    For each $n\in N$, let $\vo_n^\Delta=\vo_n\rest\Delta$ and $\vo_n^\Gamma=\vo_n\rest\Gamma$.
    Then $\amax(\tvD_1), \amax(\tvD_2) \leq d2^{2d+2}$ and
    \begin{enumerate}
        \item
        For all $n\in N$, 
        $\vi_n \in I$,
        $\vo_n \in S$,
        $\|\vi_n\| = \|\vo_n\| = n$,
        and
        $\vi_n \reach \vo_n$.
        
        \item %\label{lem:push-Delta-simplified:cond:reach}
        For all 
        $b \in \N$,
        there is $n_b \in N$ 
        % and $\vc \in \N^\Gamma$ 
        such that,
        for all $n \in N$ such that $n \geq n_b$
        and all $\vd^{\Delta} \in \N^{\Delta}$ such that 
        $\max(\vd^\Delta) \leq b$, 
        we have
        $2 \vi_n + \vd^\Delta \reach 2 \vo_n^\Gamma + \tvD_1.\vo_n^\Delta + \tvD_2.\vd^\Delta$.
    \end{enumerate}
}

\begin{cor}
    \label{cor:push-Delta-simplified}
    Let $\alpha > 0$.
    Suppose that for some set $S\subseteq\N^\Lambda$ and infinite set $I$ of $\alpha$-dense configurations, 
    for all $\vi \in I$,
    letting $n = \|\vi\|$,
    $\time{\vi}{S} = o(n)$.
    
    \corPushDeltaSimplifiedConclusion
\end{cor}

For all $n \in N$,
define $\vv_n^\Gamma = 2 \vo_n^\Gamma + \tvD_1.\vo_n^\Delta + \tvD_2.\vd^\Delta$ 
to be the configuration reached in part~$(2)$ of the conclusion.
We note two important properties of $\vv_n^\Gamma$:
\begin{enumerate}
    \item
    $\vv_n^\Gamma \in \N^\Gamma$ (i.e., it has no states in $\Delta$).\footnote{
        Compare this to Lemma~\ref{lem:push-Delta-simplified},
        which allows one to ``target'' a particular nonzero count $\vt^\Delta$
        of $\Delta$ states to be present in the final configuration.
    }
    This is crucial for the final statement of Observation~\ref{obs:unbdd-state-configs-stable} below arguing that 
    $\vv_n^\Gamma$ is stable.
    
    \item
    There is a constant $c \in \N$ 
    such that for all $\os \in \Gamma$ and $n\in N$,
    $| \vv_n^\Gamma(\os) - 2 \vo_n(\os) | \leq c$,
    i.e., the counts of $\Gamma$ states in $\vv_n^\Gamma$ are within a constant $c$ of those in $2 \vo_n$.
    The constant $c$ depends on the sequence $(\vo_n)$, 
    which defines $\max(\vo_n^\Delta)$,
    and $c$ also depends on the bound $b$ on $\max(\vd^\Delta)$,
    as well as the values of entries of $\tvD_1$ and $\tvD_2$,
    but crucially $c$ does \emph{not} depend on $n$.
    % In particular,
    % for sufficiently large $n$,
    % $2 \vo_n^\Gamma + \tvD_1.\vo_n^\Delta + \tvD_2.\vd^\Delta$ has all values greater than 0, 
    % so is a well-defined configuration.
\end{enumerate}

\subsection{Stable configurations and unbounded states}
\label{sec:StableConfigsInputStatesAndUnboundedStates}

In this subsection, let $\calC$ be a function-computing or function-approximating population protocol with input states $\Sigma$ and output state $\outs$, taking ``stable'' in the next two observations to mean stable with respect to $\outs$.
The set $\Delta$ (and its complement $\Gamma$) referenced frequently in previous sections will play a key role the main proof as the set of states with bounded counts in some infinite sequence of reachable stable configurations.

The next observation states that if we have a stable configuration $\vo$, 
and we modify it by reducing the counts of states that are already ``small'' (contained in $\Delta$)
and changing in either direction the counts of states that are ``large'' (contained in $\Gamma$),
then the resulting configuration $\vv$ is also stable.

\begin{obs} \label{obs:unbdd-state-configs-stable}
    If there is an infinite nondecreasing sequence $(\vo_n)$ of stable configurations such that 
    $\Gamma = \unbdd((\vo_n))$ and $\Delta = \bdd((\vo_n))$,
    for every $\vv \in \N^\Lambda$ such that $\vv\rest\Delta \leq \vo_n \rest \Delta$ for some $n\in\N$, $\vv$ is stable.
    In particular, any $\vv \in \N^\Gamma$ is stable.
\end{obs}

This follows since for sufficiently large $n$, $\vv \leq \vo_n$, and stability is closed downward.
The following corollary is useful,
which states that adding any amount $\vu$ of states in $\Gamma$ to a stable configuration,
as well as removing any amount $\vw$ of states (whether in $\Gamma$ or not), 
keeps it stable.

\begin{corollary} \label{cor:unbdd-state-configs-stable}
    If there is an infinite nondecreasing sequence $(\vo_n)$ of stable configurations such that 
    $\Gamma = \unbdd((\vo_n))$,
    for every $\vo_n$ and every $\vu \in \N^\Gamma$ and $\vw \in \N^\Lambda$,
    $\vo_n + \vu - \vw$ is stable.
\end{corollary}

% \todoi{DD: I think the next observation is false. I think it can be handled in each of the individual cases, based on memories of discussions from a long time ago, but right now I forget how to handle it in each case.}

% Next we argue that $\Delta$ must contain the input states.
% Intuitively, this follows because otherwise Observation~\ref{obs:unbdd-state-configs-stable} applies to such a state $\ins_i$, implying that adding more agents in state $\ins_i$ to a stable configuration $\vo$ preserves its stability.
% But for sufficiently large counts of $\ins_i$, this would be a configuration reachable from an initial configuration almost identical to the original, 
% but with more $\ins_i$, which should have a different output value, a contradiction.

% \begin{obs} \label{obs:input-states-in-Delta}
%     If $(\vo_n)$ is an infinite nondecreasing sequence of stable configurations,
%     then $\Sigma \subseteq \bdd((\vo_n))$.
% \end{obs}

\section{Predicate computation}
\label{sec:predicates}

In this section we show that a wide class of Boolean predicates cannot be stably computed in sublinear time by population protocols
(without a leader).
This is the class of predicates $\phi:\N^k\to\{0,1\}$
that are not \emph{eventually constant} (see definition below):
for all $m \in\N$,
there are two inputs $\vm_0,\vm_1 \in \N^k_{\geq m}$
such that $\phi(\vm_0) \neq \phi(\vm_1)$.

\subsection{Definition of predicate computation}

    Computation of Boolean predicates $\phi:\N^k\to\{0,1\}$ was the first type of computational problem studied in population protocols~\cite{angluin2006passivelymobile, angluin2006fast, AngluinAE2006semilinear, angluin2007computational}.
    Compared to function computation
    (defined formally in Section~\ref{sec:function-computation-defn}),
    it is a bit more complex to define output,
    since we require a convention for converting several integer counts to a single Boolean value.
    However, the definition is also simpler because there is no need for initial configurations to contain quiescent states (see Section~\ref{sec:function-computation-defn}):
    whatever predicates are computable by population protocols,
    are computable from initial configurations containing \emph{only} the input states~\cite{angluin2007computational}.
    Thus we have a 1-1 correspondence between inputs to $\phi$ and valid initial configurations.

    It is worth mentioning that, using the output convention from the foundational work on predicate computation with population protocols~\cite{angluin2006passivelymobile, angluin2006fast, AngluinAE2006semilinear, angluin2007computational},
    we cannot merely consider predicates a special case of functions with integer outputs in $\{0,1\}$.
    If this were the case, then the results of this section would follow trivially from Theorem~\ref{thm:negative-result-eventually-positive-integral-linear}.
    The reason this does not work is that
    the output convention requires not merely to produce a single $\outs$ if and only if the answer is yes;
    instead it requires \emph{all} agents to vote unanimously on a ``yes'' or ``no'' output.\footnote{
        The reason this issue is not trivially resolved by converting the ``$0/1$-function'' output convention to the ``whole population votes'' convention is two-fold:
        1) If $\outs$ is absent,
        there is no straightforward way to detect this in order to ensure that 0-voters are produced.
        It turns out that this output convention is equivalent if time complexity is not an issue, although this is not straightforward to prove~\cite{decocsccrn}.
        But this leads to the second issue:
        2) Even with a symmetric convention where a single $\ans{n}$ state is present to represent output $0$,
        and a single $\outs$ state to represent output $1$,
        it takes at least linear time to convert all other voters by a standard scheme where the single agent representing output directly interacts with all other agents.
    }

Formally, a \emph{predicate-deciding leaderless population protocol} is a tuple $\calD = (\Lambda,\delta,\Sigma,\Upsilon_1)$,
where $(\Lambda,\delta)$ is a population protocol,
$\Sigma \subseteq \Lambda$ is the set of \emph{input states},
and $\Upsilon_1 \subseteq \Lambda$ is the set of \emph{1-voters}.
By convention, we define $\Upsilon_0 = \Lambda \setminus \Upsilon_1$ to be the set of \emph{0-voters}.
The \emph{output} $\Phi(\vc)$ of a configuration $\vc \in\N^\Lambda$
is $b\in\{0,1\}$ if $\vc(\os)=0$ for all $\os\in\Upsilon_{1-b}$ (i.e., if the vote is unanimously $b$);
the output is undefined if voters of both types are present.
We say $\vo\in\N^\Lambda$ is \emph{stable} if $\Phi(\vo)$ is defined
and for all $\vo'\in\post(\vo)$, $\Phi(\vo')=\Phi(\vo)$.
For all $\vm\in\N^k$,
define initial configuration $\vi_\vm \in \N^\Lambda$ by
$\vi_\vm\rest\Sigma=\vm$ and
$\vi_\vm\rest(\Lambda\setminus\Sigma)=\vec{0}$.
Call such an initial configuration \emph{valid}.
For any valid initial configuration $\vi_\vm\in\N^\Lambda$
and predicate $\phi:\N^k\to\{0,1\}$,
let $S_{\vi_\vm,\phi} = \{\vo\in\N^\Lambda \mid \vi_\vm \reach \vo, \vo$ is stable, and $\Phi(\vo)=\phi(\vm)\}$.
A population protocol \emph{stably decides}\footnote{
    The original definition~\cite{angluin2006passivelymobile} used the term stably \emph{compute}, which we reserve for integer-valued function computation.
}
a predicate $\phi:\N^k\to\{0,1\}$ if,
for any valid initial configuration $\vi_\vm\in\N^\Lambda$,
$\Pr[\vi_\vm \reach S_{\vi_\vm,\phi}] = 1$.
This is equivalent to requiring that for all $\vc \in \post(\vi_\vm)$,
there is $\vo\in\post(\vc)$ such that $\vo$ is stable and $\Phi(\vo) = \phi(\vm)$.

For example, the protocol defined by transitions
\begin{align*}
    \ins_1,\ins_2 &\to \qs_1,\qs_2
    \\
    \ins_1,\qs_2 &\to \ins_1,\qs_1
    \\
    \ins_2,\qs_1 &\to \ins_2,\qs_2
    \\
    \qs_1,\qs_2 &\to \qs_1,\qs_1
\end{align*}
if $\Upsilon_1 = \{\ins_1,\qs_1\}$ and $\Upsilon_0 = \{\ins_2,\qs_2\}$,
decides whether $m_1 = \vi(\ins_1) \geq m_2 = \vi(\ins_2)$.
The first transition stops once the less numerous input state is gone.
If $\ins_1$ (resp. $\ins_2$) is left over,
then the second (resp. third) transition converts $\qs_i$ states to its vote.
If neither is left over (i.e., if $m_1=m_2$, requiring output $1$),
the fourth transition converts all $\qs_2$ states to $\qs_1$.

\paragraph{Relation to prior work.}
Alistarh, Aspnes, Eisenstat, Gelashvili, and Rivest~\cite{alistarh2017timespace} showed a linear-time lower bound on any leaderless population protocol deciding the majority predicate.
Their technique is based on
showing that after adding enough of the input in the minority to change it to the majority,
the effect of this addition can be effectively nullified by surgery of the transition sequence,
yielding a stable configuration with the original (now incorrect) answer.
The technique can be extended easily to show various other specific predicates,
such as equality and parity,
also require linear time.
We use the same technique of finding pairs of inputs with opposite correct answers
and apply a similar transition sequence surgery.
The main difficulty in showing Theorem~\ref{thm:predicates}, which covers the class of \emph{all} predicates that are semilinear but not eventually constant (see below),
is to identify a common characteristic,
derived from the semilinear structure of the predicate computed,
that can be exploited to find an infinite sequence of pairs of inputs that are all $\alpha$-dense for some fixed $\alpha>0$.
Subsection~\ref{subsec:eventually-constant-predicates} shows how this structure can be used.

\subsection{Eventually constant predicates} \label{subsec:eventually-constant-predicates}
Let $\phi:\N^k\to\{0,1\}$, and for $b\in\{0,1\}$,
define $\phi^{-1}(b) = \{ \vm\in\N^k \mid \phi(\vm)=b \}$
to be the set of inputs on which $\phi$ outputs $b$.
We say $\phi$ is \emph{eventually constant}
if there is $m_0 \in \N$ such that
$\phi$ is constant on
%$\N^k_{\geq m_0} = \{\vm\in\N^k \mid (\forall i \in \{1,\ldots,k\})\ \vm(i)=0 \text{ or } \vm(i) \geq m_0 \}$,
$\N^k_{\geq m_0} = \{\vm\in\N^k \mid (\forall i \in \{1,\ldots,k\})\ \vm(i) \geq m_0 \}$,
i.e., either
$\phi^{-1}(0) \cap \N^k_{\geq m_0} = \emptyset$
or
$\phi^{-1}(1) \cap \N^k_{\geq m_0} = \emptyset$.
In other words,
although $\phi$ may have an infinite number of each output,
``sufficiently far from the boundary''
(where all coordinates exceed $m_0$),
only one output appears.

The main result of this section, Theorem~\ref{thm:predicates},
concerns eventually constant predicates as defined above.
However, our proof technique requires reasoning about infinitely many inputs $\vm \in \N^k$ that are $\alpha$-dense for some $\alpha>0$.
A predicate $\phi:\N^k\to\{0,1\}$ can be not eventually constant,
yet for any fixed $\alpha > 0$,
have all but finitely many $\alpha$-dense inputs map to a single output.
For example, the predicate $\phi(\vm) = 1 \iff \vm(1) = \vm(2)^2$ is not eventually constant,
yet for any fixed $\alpha > 0$,
all but finitely many $\alpha$-dense inputs $\vm$ have $\phi(\vm) = 0$.
The rest of this section shows that for \emph{semilinear} predicates $\phi$,
if $\phi$ is not eventually constant,
then we \emph{can} find infinitely many $\alpha$-dense inputs mapping to each output.
The actual requirements we need to prove Theorem~\ref{thm:predicates} are a bit more technical and are captured in Corollary~\ref{cor:infinitely-many-m-doubled-different-outputs}.

Given $D \subseteq \N^k$,
we say $\phi$ is \emph{almost constant on $D$} if either
$|\phi^{-1}(0) \cap D| < \infty$
or
$|\phi^{-1}(1) \cap D| < \infty$.
In other words,
$\phi$ is constant on $D$ except for a finite number of counterexamples in $D$.\footnote{
    Note that almost constant is a stricter requirement than eventually constant,
    since the latter allows infinitely many counterexamples so long as at least one component is ``small''.
}
For $\alpha > 0$ and $k\in\N$,
let $D_\alpha^k = \{ \vm\in\N^k \mid \vm$ is $\alpha$-dense $\}$.
Say that a predicate $\phi:\N^k\to\{0,1\}$ is
\emph{$\alpha$-densely almost constant}
if $\phi$ is almost constant on $D_\alpha^k$.
Say that $\phi$ is \emph{densely almost constant}
if for all $\alpha>0$, $\phi$ is $\alpha$-densely almost constant.

The following proof uses the definition of semilinear given in Section~\ref{sec:nonlinear-functions}
in terms of finite unions of periodic cosets.
% known~\cite{ginsburg1966} to be equivalent to the ``threshold and mod'' characterization given earlier.
% A set $A \subseteq \N^k$ is \emph{linear}
% if there exist vectors $\vec{b},\vec{p}_1,\ldots,\vec{p}_p \in\N^k$ such that
% $A=\{\vb + n_1 \vec{p}_1 + \ldots + n_l \vec{p}_l \mid n_1,\ldots,n_l\in\N \}.$
% A set $A$ is \emph{semilinear} if it is a finite union of linear sets.

\begin{lemma}\label{lem:finitely-densely-supported-predicates-property}
    If $\phi:\N^k\to\{0,1\}$ is semilinear and densely almost constant,
    then $\phi$ is eventually constant.
\end{lemma}

\begin{proof}
    We prove this by contrapositive.
    Assume $\phi$ is semilinear and not eventually constant.
    We will show $\phi$ is not densely almost constant.

    For $b\in\{0,1\}$,
    let $I_b = \phi^{-1}(b)$ be the set of inputs mapping to output $b$.
    Since $\phi$ is not eventually constant,
    for all $m_0 \in \N$,
    for both $b\in\{0,1\}$,
    $|I_b \cap \N^k_{\geq m_0}| \neq \emptyset$.
    If for some $m_0$,
    $|I_b \cap \N^k_{\geq m_0}| < \infty$,
    then for sufficiently large $m'_0$,
    we would have $|I_b \cap \N^k_{\geq m'_0}| = \emptyset$,
    a contradiction.
    So in fact, for all $m_0 \in \N$,
    for both $b\in\{0,1\}$,
    $|I_b \cap \N^k_{\geq m_0}| = \infty$.

    Since $\phi$ is semilinear, so is $I_b$, so expressible as a finite union
    $I_b = \bigcup_{i=1}^p P_i$,
    of $p$ periodic cosets $P_1,\ldots,P_p$.
    Since $|I_b \cap \N^k_{\geq m_0}|=\infty$ for all $m_0\in\N$,
    without loss of generality,
    $|P_1 \cap \N^k_{\geq m_0}|=\infty$ for all $m_0\in\N$ as well.
    Let $\vb,\vp_1,\ldots,\vp_l \in\N^k$ be such that
    $P_1 = \{\vb + n_1 \vec{p}_1 + \ldots + n_l \vec{p}_l \mid n_1,\ldots,n_l\in\N \}$.

    Let $\vp = \vp_1 + \ldots + \vp_l$.
    Then $\vp(i) > 0$ for all $i\in\{1,\ldots,k\}$.
    (Otherwise, we would have $(\exists i)(\forall j) \vp_j(i)=0$ and $P_1$ could not be arbitrarily large on component $i$, so it could not intersect $\N^k_{\geq m_0}$ for all $m_0$.)
    Letting $\alpha' = \min(\vp) / \|\vp\|$, we have that for all $r\in\N$, $r \vp$ is $\alpha'$-dense.
    Let $\alpha = \alpha'/2$.
    Then for sufficiently large $r$,
    $\vb + r \vp$ is $\alpha$-dense.
    Since $(\vb + r \vp) \in P_1 \subseteq I_b$,
    this shows that $I_b$ has infinitely many $\alpha$-dense points.
    Since $b\in\{0,1\}$ was arbitrary, $\phi$ is not $\alpha$-densely almost constant,
    hence not densely almost constant.
\end{proof}

For $i \in \{1,\ldots,k\}$,
let $\vu_i \in \N^k$ be the unit vector such that $\vu_i(i)=1$ and $\vu_i(j)=0$ for all $j\neq i$.
% For each point $\vm\in\N^k$ and $d\in\N$,
% define the set $C^d_{\vm} = \{\vm + \sum_{i=1}^k c_i \vu_i \mid c_1,\ldots,c_k \in \{0,1,\ldots,d\} \}$,
% the ``cone of length $d$ above $\vm$'',
% which constitutes the $(d+1)^k$ points obtainable by adding up to $d$ to each of the $k$ coordinates of $\vm$.

\begin{lemma} \label{lem:infinitely-many-m-doubled-different-outputs}
    Let $\phi:\N^k\to\{0,1\}$.
    If $\phi$ is not densely almost constant,
    then there is $\alpha>0$
    and an infinite subset $D \subseteq D^k_\alpha$
    so that one of the following two conditions holds.
    \begin{enumerate}
        \item \label{lem:cond:not-densely-almost-constant-1}
        For all $\vm \in D$,
        $\phi(\vm) \neq \phi(2\vm)$.

        \item \label{lem:cond:not-densely-almost-constant-2}
        There is $i \in \{1,\ldots,k\}$
        such that for all $\vm \in D$,
        $\phi(\vm) \neq \phi(\vm+\vu_i)$
        and
        $\phi(\vm) \neq \phi(2(\vm + \vu_i))$.
    \end{enumerate}
\end{lemma}

\begin{proof}
    Since $\phi$ is not densely almost constant,
    for some $\alpha>0$, %and $E\subseteq\{1,\ldots,k\}$,
    $|\phi^{-1}(0) \cap D^k_\alpha| = |\phi^{-1}(1) \cap D^k_\alpha| = \infty$.
    So for infinitely many $\vm \in D^k_\alpha$,
    there is $i_{\vm} \in \{ 1,\ldots,k \}$
    such that
    $\phi({\vm}) \neq \phi({\vm} + \vu_{i_{\vm}})$.
    In other words,
    since each output $b\in\{0,1\}$ is supported on infinitely many points in $D^k_\alpha$,
    there must be infinitely many pairs of adjacent points with opposite output
    (``adjacent'' meaning that the points differ on exactly one coordinate, and that difference is 1).
    By the pigeonhole principle there is $i \in \{1,\ldots,k\}$
    so that some infinite subset of these $\vm$ use the same coordinate $i_\vm = i$,
    so consider only this infinite subset $D' \subseteq D^k_\alpha$.

    If $\phi(\vm) \neq \phi(2\vm)$
    for infinitely many $\vm \in D'$,
    take $D$ to be this infinite subset,
    and we are done.

    Otherwise,
    $\phi(\vm) = \phi(2\vm)$ for all but finitely many $\vm \in D'$.
    If $\phi(2\vm) \neq \phi(2(\vm+\vu_i))$ for infinitely many of these $\vm$,
    then $\phi(\vm) \neq \phi(2(\vm+\vu_i))$ and we are done.
    In the remaining case,
    we have that $\phi(2(\vm+\vu_i)) = \phi(2\vm) \neq \phi(\vm+\vu_i)$ for infinitely many $\vm \in D'$.
    In this case,
    replace each such $\vm$ in $D'$ with $\vm'=\vm+\vu_i$,
    calling the resulting set $D$,
    noting that $\vm'$ satisfies condition~\eqref{lem:cond:not-densely-almost-constant-1} of the lemma
    since $\phi(\vm') \neq \phi(2\vm')$.
\end{proof}

Combining Lemma~\ref{lem:infinitely-many-m-doubled-different-outputs} and the contrapositive of Lemma~\ref{lem:finitely-densely-supported-predicates-property} gives the following corollary.

\begin{cor} \label{cor:infinitely-many-m-doubled-different-outputs}
    Let $\phi:\N^k\to\{0,1\}$ be semilinear and not eventually constant.
    Then there is $\alpha>0$
    and an infinite subset $D \subseteq D^k_\alpha$
    so that one of the following two conditions holds.
    \begin{enumerate}
        \item \label{cond:not-densely-almost-constant-1}
        For all $\vm \in D$,
        $\phi(\vm) \neq \phi(2\vm)$.

        \item \label{cond:not-densely-almost-constant-2}
        There is $i \in \{1,\ldots,k\}$
        such that for all $\vm \in D$,
        $\phi(\vm) \neq \phi(\vm+\vu_i)$
        and
        $\phi(\vm) \neq \phi(2(\vm + \vu_i))$.
    \end{enumerate}
\end{cor}

\subsection{Time lower bound for non-eventually-constant predicates}

The following theorem shows that unless a predicate is eventually constant,
it cannot be stably decided in sublinear time by a leaderless population protocol.

\begin{theorem}\label{thm:predicates}
    Let $\phi:\N^k\to\{0,1\}$ and $\calD$ be a predicate-deciding leaderless population protocol that stably decides $\phi$.
    If $\phi$ is not eventually constant,
    then $\calD$ takes expected time $\Omega(n)$.
\end{theorem}

The high level intuition behind our proof technique is as follows.
Sublinear time computation requires avoiding ``bottlenecks''---having to go through a transition in which both states are present in small count (constant independent of the number of agents $n$).
Traversing even a single such transition requires linear time.
Corollary~\ref{cor:push-Delta-simplified}
shows that bottleneck-free execution sequences from $\alpha$-dense initial configurations (i.e., initial configurations where every state that is present is present in at least $\alpha n$ count) are amenable to predictable ``surgery''.
Using Corollary~\ref{cor:push-Delta-simplified},
we show how to consume additional input states but still drive the system to the same output stable answer, and thus fool the population protocol into giving the wrong answer.
Using Corollary~\ref{cor:push-Delta-simplified} is rather technical
and requires finding infinitely many candidate execution sequences and respective $\alpha$-dense initial configurations that are ``close'' to other initial configurations on which the computed predicate is supposed to evaluate to the opposite answer.
The reason that Theorem~\ref{thm:predicates} holds only for not eventually constant predicates is that the initial configurations susceptible to surgery need to be $\alpha$-dense, and thus we can only fool the population protocol if the predicate evaluates to both $0$ and $1$ ``far away'' from the boundaries of $\N^k$.

A bit of care is needed in picking the pairs of inputs that give a different answer.
In particular, we need to ensure that any input states $\ins_i$ that we add in $\vd^\Delta$
when applying Corollary~\ref{cor:push-Delta-simplified} are actually contained in $\Delta$,
the set of states with bounded counts in all the output stable configurations in the sequence.
Not all input states are contained in $\Delta$ in some cases.
For instance, consider the majority-computing protocol
\begin{eqnarray*}
    \ins_1,\ins_2 &\to& \ans{y},\ans{y}
    \\
    \ins_1,\ans{n} &\to& \ins_1,\ans{y}
    \\
    \ins_2,\ans{y} &\to& \ins_2,\ans{n}
    \\
    \ans{y},\ans{n} &\to& \ans{y},\ans{y},
\end{eqnarray*}
which decides whether $\# \ins_1 \geq \# \ins_2$ initially,
if $\phi(\ins_1) = \phi(y) = 1$ and $\phi(\ins_2) = \phi(n) = 0$.
If the sequence of inputs picked were such that $\# \ins_1 = 2 \# \ins_2$,
then $\# \ins_1$ would grow unboundedly in stable configurations,
hence $\ins_1 \in \Gamma$.
Note, however, that $\ins_2 \in \Delta$ since it would have count 0 in all such stable configurations.
This helps to see how $\Delta$ depends on the choice of infinite sequences of inputs;
if instead
$\# \ins_1 = \# \ins_2 / 2$ in all such inputs,
then $\ins_1 \in \Delta$ but $\ins_2 \in \Gamma$.
If $\# \ins_1 = \# \ins_2$,
then $\ins_1, \ins_2 \in \Delta$.

We choose such inputs $\vm$ as in
Lemma~\ref{lem:infinitely-many-m-doubled-different-outputs}
to be such that either
$\phi(\vm) \neq \phi(2\vm)$,
or
$\phi(\vm) \neq \phi(\vm+\vu_i)$
and
$\phi(\vm) \neq \phi(2(\vm + \vu_i))$,
where $\vu_i \in \{0,1\}^k$.
In the first case,
we don't need any inputs to be in $\Delta$;
we get a contradiction since Corollary~\ref{cor:push-Delta-simplified} with $\vd^\Delta = \vec{0}$
gives us a way to drive from $2\vm$ to a stable configuration very close to
(therefore having the same output as)
twice the stable configuration reached from $\vm$,
which gives a contradiction.
In the second case,
the contradiction arises between inputs $\vm$ and $2(\vm + \vu_i)$,
but unlike the first case,
we need to apply Corollary~\ref{cor:push-Delta-simplified}
with $\vd^\Delta \neq \vec{0}$.
The $\vd^\Delta$ we choose corresponds to the positive entries of $\vu_i$.
The assumption that $\phi(\vm) \neq \phi(\vm + \vu_i)$
is used to justify that those input states that are positive in $\vu_i$
are in fact contained in $\Delta$,
so that we are able to send their counts to 0 via Corollary~\ref{cor:push-Delta-simplified}
and conclude via Corollary~\ref{cor:unbdd-state-configs-stable} that the resulting configuration is stable.

\begin{proof}
    Suppose for the sake of contradiction that $\calD$ takes expected time $o(n)$.
    Since $\calD$ stably decides $\phi$,
    $\phi$ is semilinear~\cite{AngluinAE2006semilinear},
    so by Corollary~\ref{cor:infinitely-many-m-doubled-different-outputs}
    there is $\alpha>0$
    and an infinite subset $D \subseteq D^k_\alpha$
    so that one of the following two conditions holds.
    \begin{enumerate}
        \item
        For all $\vm \in D$,
        $\phi(\vm) \neq \phi(2\vm)$.

        \item
        There is $i \in \{1,\ldots,k\}$
        such that for all $\vm \in D$,
        $\phi(\vm) \neq \phi(\vm+\vu_i)$
        and
        $\phi(\vm) \neq \phi(2(\vm + \vu_i))$.
    \end{enumerate}
    % If $\phi(\vm) \neq \phi(2\vm)$,
    % define $\vu = \vec{0}$;
    % otherwise define $\vu = \vu_i$.
    % In either case note that $\phi(\vm) \neq \phi(2\vm + \vu)$.
    % In the latter case let
    % $\os_i$ denote the $i$'th input corresponding to the vector $\vu_i$ above.

    Let $I = \{ \vi_\vm\in\N^\Lambda \mid \vm\in D\}$
    denote the set of initial configurations corresponding to $D$.
    Let $S = \{ \vo \mid (\exists \vi \in I)\ \vi \reach \vo$ and $\vo$ is stable$\}$
    be the set of stable configurations reachable from some initial configuration in $I$.
    By assumption we have that for each $\vi \in I$, $\time{\vi}{S} = o(n)$.

    Apply Corollary~\ref{cor:push-Delta-simplified}.
    \corPushDeltaSimplifiedConclusion

    Let $b = 2$; below we ensure that $\max(\vd^\Delta) \leq b$.
    Let $\vm = \vi_{n_b} \rest \Sigma$ and $\vo = \vo_{n_b}$.

    Suppose we are in case~\eqref{cond:not-densely-almost-constant-1}
    of Lemma~\ref{lem:infinitely-many-m-doubled-different-outputs};
    then $\phi(\vm) \neq \phi(2\vm)$.
    Let $\vv^\Gamma = 2 \vo_{n_b}^\Gamma + \tvD_1.\vo_{n_b}^\Delta$.
    Then applying the above reachability argument with $\vd^\Delta = \vec{0}$
    gives that $2 \vi_{n_b} \reach \vv^\Gamma$,
    which is stable and has the same output as is correct for $\vm$, a contradiction since it is reachable from $2 \vi_{n_b}$ representing input $2 \vm$,
    which should have the opposite output.

    Suppose we are in case~\eqref{cond:not-densely-almost-constant-2}
    of Lemma~\ref{lem:infinitely-many-m-doubled-different-outputs};
    then there is $i \in \{1,\ldots,k\}$
    such that for all $\vm \in D$,
    $\phi(\vm) \neq \phi(\vm+\vu_i)$
    and
    $\phi(\vm) \neq \phi(2(\vm + \vu_i))$.
    Let $\ins_i$ be the input corresponding to unit vector $\vu_i$.
    We claim that $\ins_i \in \Delta$.
    To see why,
    recall that $\phi(\vm) \neq \phi(\vm+\vu_i)$.
    If $\ins_i \not\in\Delta$,
    then Corollary~\ref{cor:unbdd-state-configs-stable}
    tells us that $\vo + \{\ins_i\}$ is stable
    (hence the same output $\phi(\vm)$ as $\vo$).
    However,
    $\vi_{n_b} + \{\ins_i\} \reach \vo + \{\ins_i\}$,
    a contradiction since
    $\phi(\vm) \neq \phi(\vm + \vu_i)$.
    This shows that $\ins_i \in \Delta$.

    Let $\vd^\Delta = \{ 2 \ins_i \} \in \N^\Delta$.
    By the above,
    letting
    $\vv^\Gamma = 2 \vo_{n_b}^\Gamma + \tvD_1.\vo_{n_b}^\Delta + \tvD_2.\vd^\Delta$,
    we have that
    $2 \vi_{n_b} + \vd^\Delta \reach \vv^\Gamma$.
    By Corollary~\ref{cor:unbdd-state-configs-stable},
    $\vv^\Gamma$ is stable and reachable from the initial configuration
    $2 \vi_{n_b} + \vd^\Delta$
    corresponding to input $2\vm+\vu_i$,
    which should have the opposite output as $\vm$.
    This contradicts the correctness of $\calD$.
\end{proof}

\section{Definition of function computation with population protocols}
\label{sec:function-computation-defn}

In this section we formally define error-free (a.k.a.~\emph{stable}) function computation by population protocols.
This mode of computation was discussed briefly in the first population protocols paper\cite[Section 3.4]{angluin2006passivelymobile},
which focused more on Boolean predicate computation,
and more extensively in the more general model of chemical reaction networks\cite{CheDotSolDetFuncNaCo, DotHajLDCRNNaCo}.

\subsection{Issues with definition of approximation and computation}
\label{sec:discuss-issues-with-defn}
There are subtle issues in formalizing these definitions for population protocols,
as well as extending from exact to approximate computation, 
which we discuss before stating our definitions.

We focus on single-output functions $f:\N^k\to\N$.
It is simple to apply all of our results to multi-output functions $f:\N^k\to\N^l$,
simply by considering $f$ as the result of combining $l$ different single-output functions all computed in parallel by independent protocols.

%\todoi{DD: mention justification for making $q_0$ a linearly bounded function, since stably computable functions are themselves linearly bounded so there is no need to make it bigger just to represent output, and also it requires the preparer of the initial configuration to compute a non-semilinear function just to know whether the initial configuration is valid or not}

The model of chemical reaction networks allows the creation and destruction of molecules via reactions such as $X \to 2Y$ and $2X \to Y$.
However, all population protocol transitions occur between two agents, so to compute a function such as $f(m) = 2m$ via the transition $\ins,\qs \to \outs,\outs$ requires starting with $m$ agents in an input state $\ins$ and at least $m$ additional agents in a ``quiescent'' state $\qs$, to ensure that there are enough agents to represent the output of the function.
%\todo{Argue that we are not artificially restricting the function in this way.}

Next, the time complexity models are slightly different between chemical reaction networks and population protocols.
In addition to the distinction between continuous time in chemical reaction networks and discrete time in population protocols, the former model has an extra parameter called the \emph{volume}, which dictates the rate of bimolecular reactions.
To make the models have the same expected time, the ``volume'' of a population protocol is implicitly the total number of agents $n$.
However, as we noted above, the total number of agents $n$ in some cases necessarily differs from the number of agents representing \emph{input}.
This leads to a definitional quandary: 
is it more ``fair'' to measure the time as a function of the input size $m$ (number of agents representing input), 
or as a function of the total system size $n \geq m$ (number of total agents, including non-input agents)?
We will handle this by requiring the protocol to work no matter how large $n$ is compared to $m$, but will measure expected time as a function of $n$, 
and our main theorem applies to show an expected time \emph{lower} bound only when $n = O(m)$.
This will lead to the same \emph{asymptotic} time measurement within a multiplicative constant,
whether time is measured as a function of $m$ or $n$.

Finally, given the specific nature of the problem we study, that of \emph{approximating} functions with population protocols, we must take care in what are defined as valid initial configurations.
We observed that the protocol $\as,\ins \to \ans{b},\outs$ and $\ans{b},\ins \to \as,\qs$ 
can approximate the function $f(m) = \lfloor m/2 \rfloor$ within multiplicative factor $\gamma$
from the initial configuration $\vi_{m,\gamma}$ defined by $\vi_{m,\gamma}(\ins) = m$ and $\vi_{m,\gamma}(\as) = \floor{\gamma m}$.
The initial count of state $\as$ serves to control the approximation factor;
setting it lower makes the approximation better but the stabilization time longer.
What is the appropriate way to generalize this to a complete definition of function approximation?
Note that the following does \emph{not} work: 
declaring the set the valid initial configurations to be $\{\vi_{m,\gamma} \mid m\in\N\}$, where $\gamma$ is a fixed constant.
For then, we could simply set $\gamma=\frac{1}{2}$ and let $\outs=\as$, and have no transitions whatsoever: 
the initial configuration would already contain the correct amount of $\outs$, 
with the computation having been done not by the protocol itself, 
but merely by the specification of the initial configuration.
To avoid this sort of cheating by 
``sneaking computation into the initial configuration,''
we allow the designer of the protocol to set a constant lower bound
(not dependent on the input value),
but \emph{not} an upper bound, 
on how many agents have initial state $\as$.
% Also, our proof technique applies only to protocols whose lower bound 
% (as well as the related lower bound on the initial count of quiescent agents in state $\qs$)
% is $O(m)$ (thus $n=O(m)$ since $n = \vi(\as)+\vi(\qs) + \vi(\text{inputs})$).

% The expected time for a transition between states $\os_1$ and $\os_2$ scales with $n / (\# \os_1 \cdot \# \os_2)$.
% Without the requirement that $n=O(m)$, even a single transition between any two agents that are both not $\qs$ would require potentially unbounded time.

Informally, a population protocol exactly computes a function $f:\N^k \to \N$ if, 
starting in a configuration with counts of agents in ``input'' states $\ins_1,\ldots,\ins_k$ described by a vector $\vm \in \N^k$, 
and sufficiently large counts of other agents in a ``quiescent'' state $\qs$, 
the protocol is guaranteed to stabilize to exactly $f(\vm)$ agents in ``output'' state $\outs$.
Unlike predicate computation with a Boolean output (as studied in the foundational population protocol papers~\cite{angluin2006passivelymobile, angluin2006fast, angluin2007computational, AngluinAE2006semilinear}),
it is necessary to allow initial agents in the quiescent state $\qs$, beyond those representing the input, if $f(\vm) > \vm(1) + \ldots + \vm(k)$, to ensure there are enough agents to represent the output.
For example, the function $f(m) = 2m$ is computable via the transition $\ins,\qs \to \outs,\outs$ if $\#\qs \geq \#\ins$ initially.
However, the protocol must work in any sufficiently large population, i.e., for any sufficiently large initial count of $\qs$.

Informally, a protocol approximates a function if it is guaranteed to get $\# \outs$ ``close'' to the correct output value, 
where the closeness is controlled by the initial count of an ``approximation'' state $\as$.
(Actually, there is no requirement that the approximation error \emph{must} depend on $\# \as$,
but the definition is motivated by the protocols of Theorem~\ref{thm:positive}, 
which do use the approximation state in this way.)
In particular, the protocol may stabilize to different values of $\# \outs$ on different runs starting from the same initial configuration 
(so the protocol may not stably compute any function).

However, the protocol is required to stabilize its output count to \emph{some} value on each run.
An alternative formulation would relax this requirement, and merely require that $\# \outs$ eventually enters, 
and never again leaves, an interval surrounding the correct value,
while continuing to change the value of $\# \outs$ indefinitely.
However, our proof technique is based on the idea that certain 
(carefully chosen) 
input states must be absent, 
or very low count, in stable configurations.
Under this more relaxed definition, this would not be the case. 
For example, for any $\gamma>0$, there is a $t \in \R_{\geq 1}$ such that, for any $m$, starting with $\# \ins = m$, 
the transition $\ins,\ins \to \outs,\qs$ in expected time $t$ gets $\# \outs$ to the interval $[m/2 - \gamma m/2,m/2]$, 
so in \emph{constant} time the protocol ``stabilizes to that interval'',
even though the value of $\# \outs$ may still change within that interval, 
so it is not stable by the stricter definition.
However, at the time that $\# \outs$ enters the interval, $\# \ins = \gamma m$, 
whereas our proof technique requires $\# \ins = O(1)$ in any stable configuration.
In this example, 
once $\# \outs$ stabilizes (i.e., stops changing),
then $\# \ins$ is either 0 or 1.

Sections~\ref{subsec:exact-function-computation-defn} and~\ref{subsec:approximate-function-computation-defn} give formal definitions of these concepts.

\subsection{Definition of exact function computation}
\label{subsec:exact-function-computation-defn}

A \emph{function-computing leaderless population protocol} is a tuple $\calC = (\Lambda,\delta,\Sigma,\outs,\qs)$, where $(\Lambda,\delta)$ is a population protocol, $\Sigma = \{\ins_1,\ldots,\ins_k\} \subset \Lambda$ is the set of \emph{input states}, $\outs \in \Lambda$ is the \emph{output state}, and $\qs \in \Lambda \setminus \Sigma$ is the \emph{quiescent state}.
%Note that we require the quiescent state (but not the output state) to be unequal to any input states, since the protocol must work in the presence of arbitrarily large counts of the quiescent state, without the represented input changing.
%For $\vc\in\N^\Lambda$, define $\post_\calC(\vc) = \post_{(\Lambda,\delta)}(\vc)$.
We say that a configuration $\vo\in\N^\Lambda$ is \emph{stable} if, for all $\vo' \in \post(\vo)$, $\vo(\outs) = \vo'(\outs)$, i.e., the count of $\outs$ cannot change once $\vo$ is reached.

Let $f:\N^k \to \N$, $\vi \in \N^\Lambda$, and let $\vm = \vi \rest \Sigma$.
We say that $\calC$ \emph{stably computes $f$ from $\vi$} if, for all $\vc\in\post(\vi)$, there exists a stable $\vo\in\post(\vc)$ such that $\vo(\outs) = f(\vm)$, i.e., $\calC$ stabilizes to the correct output from the initial configuration $\vi$.
However, for any input $\vm \in \N^k$, there are many initial configurations $\vi \in \N^\Lambda$ representing it (i.e., such that $\vi \rest \Sigma = \vm$).
We now formalize what sort of initial configurations $\calC$ is required to handle.

We say a function $q_0:\N^k\to\N$ is \emph{linearly bounded} 
if there is a constant $c\in\N$ such that, 
for all $\vm\in\N^k$,
$q_0(\vm) \leq c \|\vm\|$.
We say that $\calC$ \emph{stably computes} $f$ if there is a linearly bounded function $q_0:\N^k\to\N$ such that,
for any $\vi\in\N^\Lambda$,
defining $\vm = \vi \rest \Sigma$, 
if $\vi(\qs) \geq q_0(\vm)$
and $\vi(\os) = 0$ for all $\os \in \Lambda \setminus (\Sigma \cup \{\qs\})$, 
then $\calC$ stably computes $f$ from $\vi$.
\newcommand{\equivProbOne}{It is well-known\cite{angluin2007computational} that this is equivalent to requiring, under the randomized model in which the next interaction is between a pair of agents picked uniformly at random, that the protocol stabilizes on the correct output with probability 1. More formally, given $f:\N^k \to \N$ and $\vm \in \N^k$, defining $S_{f,\vm}^\calC = \{ \vo\in\N^\Lambda \mid \vo \text{ is stable and } \vo(\outs) = f(\vm) \}$, $\calC$ stably computes $f$ if and only if, for all $\vm$, defining $\vi$ with $\vi \rest \Sigma = \vm$ as above with $\vi(\qs)$ sufficiently large, $\Pr\left[\vi \reach S_{f,\vm}^\calC \right] = 1$. It is also equivalent to requiring that every fair infinite execution leads to a correct stable configuration, where an execution is \emph{fair} if every configuration infinitely often reachable appears infinitely often in the execution.}
We say that an initial configuration $\vi$ so defined is \emph{valid}. 
% \opt{sub}{\footnote{\equivProbOne}} 
Since all semilinear functions are linearly bounded~\cite{CheDotSolDetFuncNaCo},
a linearly bounded $q_0$ suffices to ensure there are enough agents to represent the output of a semilinear function,
even if we choose $\vi(\qs) = q_0(\vi\rest\Sigma)$.
If $q_0$ were \emph{not} linearly bounded, and thus a super-linear count of state $\qs$ is required,
we would essentially need to do non-semilinear computation just to initialize the population protocol.

% \opt{normal}{\equivProbOne}

\equivProbOne

Let $f:\N^k\to\N$ and $t:\N\to\N$.
Given a function-computing leaderless population protocol $\calC$ that stably computes $f$,
we say $\calC$ \emph{stably computes $f$ in expected time $t$} if,
for all valid initial configurations $\vi$ of $\calC$, 
letting $\vm = \vi \rest \Sigma$,
$\time{\vi}{S_{f,\vm}^\calC} \leq t(n)$.

Note that unstability is closed upwards in $\N^\Lambda$.
In other words, for any $\vc\in\N^\Lambda$, 
if $\vc$ is not stable, then no $\vc' \geq \vc$ is stable either.
This is because $\vc$ is unstable if and only if there is a path $p$ such that $\vc \reach_p \vd$ and $\vc(\outs) \neq \vd(\outs)$.
Thus $p$ is applicable to any $\vc' \geq \vc$, 
also changing $\# \outs$ by the amount $\vd(\outs)-\vc(\outs)$, 
so $\vc'$ is also unstable. 
By contrapositive, the set of stable configurations is then closed \emph{downward}:
for any stable $\vo$, if $\vo' \leq \vo$, then $\vo'$ is also stable.

% Angluin, Aspnes, and Eisenstat~\cite{AngluinAE2006semilinear} observed that this can be used together with Dickson's Lemma to show that the set of unstable configurations is a finite union of \emph{cones}, 
% sets of the form $\{\vc'\in\N^\Lambda \mid \vc' \geq \vc\}$ for some $\vc$.
% In particular, they showed 
% (for a related notion of stability that applies easily to the definition of stability in this paper)
% that this implies that stable configurations,
% in addition to being closed downwards,
% are \emph{closed upwards under addition of states that are already large}.
% (How large exactly depends on the cones.)
% More precisely, we have the following observation.

% \todo{DD: I'm not sure if we used this Observation. I included it thinking we might need it to state explicit bounds without appealing to infinite sequences, but we went with the infinite sequences anyway. I do think it helps to give early on an intuition for why we focus on ``large count'' states.}

% \begin{obs}[adapted from~\cite{AngluinAE2006semilinear}]  \label{obs:stable-closed-up}
%   For each population protocol $\calP=(\Lambda,\delta)$, 
%   there is a constant $b\in\N$ such that, 
%   for every stable configuration $\vo\in\N^\Lambda$,
%   every $\os\in\Lambda$ such that $\vo(\os) \geq b$,
%   and every $m\in\N$,
%   $\vo + m \cdot \{\os\}$ is also stable.
% \end{obs}

\subsection{Definition of function approximation}
\label{subsec:approximate-function-computation-defn}

A \emph{function-approximating leaderless population protocol} is a tuple 
$\calA = (\Lambda,\delta,\Sigma,\outs,\qs,\as)$, 
where $(\Lambda,\delta,\Sigma,\outs,\qs)$ 
is a function-computing population protocol and 
$\as \in \Lambda \setminus \left( \Sigma \cup \{\outs,\qs\} \right)$ is the \emph{approximation state}.
Let $\epsilon,\tau \in \N$; intuitively $\tau$ represents the ``target'' (or ``true'') function output, 
and $\epsilon$ represents the allowed approximation error.
We say that a configuration $\vo \in \N^\Lambda$ is \emph{$\epsilon$-$\tau$-correct} if $|\vo(\outs) - \tau| \leq \epsilon$.%\footnote{In this definition, the error $\epsilon$ is \emph{additive}, unlike the example in the abstract, in which $\epsilon$ was a multiplicative error.}
%We say $\vo$ is \emph{$\epsilon$-$t$-stably correct} if it is stable and $\epsilon$-$t$-correct.

Let $f:\N^k \to \N$, $\epsilon\in\N$, $\vi \in \N^\Lambda$, and let $\vm = \vi \rest \Sigma$.
We say that $\calA$ \emph{stably $\epsilon$-approximates $f$ from $\vi$} if, 
for all $\vc\in\post(\vi)$, there exists a $\vo\in\post(\vc)$ that is stable and $\epsilon$-$f(\vm)$-correct, 
i.e., from the initial configuration $\vi$, $\calA$ gets the output to stabilize to a value at most $\epsilon$ from the correct output.
Let $S_{f,\vm,\epsilon}^\calA = \{ \vo\in\N^\Lambda \mid \vo$ is stable and $\epsilon$-$f(\vm)$-correct $\}$.
Note that $\calA$ stably $\epsilon$-approximates $f$ from $\vi$ if and only if $\Pr \left[ \vi \reach S_{f,\vm,\epsilon}^\calA \right] = 1$.

Let $\calE:\N\to\N$; the choice of $\calE$ as a function instead of a constant reflects the idea that the approximation error is allowed to depend on the initial count $\vi(\as)$ of the approximation state $\as$,
i.e., $\calE(\vi(\as))$ is the desired approximation error.
We say that $\calA$ \emph{stably $\calE$-approximates} $f$ 
if there are $a_0\in\N$ and linearly bounded
$q_0:\N^{k+1}\to\N$ such that,
for any $\vi\in\N^\Lambda$,
defining $\vm = \vi \rest \Sigma$, 
if
$\vi(\as) \geq a_0$, 
$\vi(\qs) \geq q_0(\vm,\vi(\as))$, 
and $\vi(\os) = 0$ for all $\os \in \Lambda \setminus \left( \Sigma \cup \{\qs,\as\} \right)$,
then $\calA$ stably $\calE(\vi(\as))$-approximates $f$ from $\vi$.\footnote{I.e., the initial count $\vi(\as)$ can influence the initial required count $\vi(\qs)$, since adding more initial $\as$ may imply that more quiescent agents are required as ``fuel''. However, $a_0$ is constant, not a function of $\vm$.}
An initial configuration $\vi$ so defined is \emph{valid}.
% We call the lower bound $a_0$ above the \emph{approximation threshold},
% we call the lower bound function $q_0$ above the \emph{quiescent threshold}.

\newcommand{\discussionApproximationState}{
    Note that we allow the protocol to require at least a certain initial amount of $\as$ in order to function correctly.
    For example, the protocol with transitions $\as,\ins \to \ans{b},\outs$ and $\ans{b},\ins \to \as,\qs$ stably $\calE$-approximates $f(m) = \floor{m/2}$, where $\calE:\N\to\N$ is the identity function, \emph{if} at least one $\as$ is present initially.
    However, setting $a_0$ does not imply the protocol is able to use $\as$ as a leader:
    since $a_0$ is just a lower bound.
    The protocol must work correctly for \emph{any} initial value of $\vi(\as) \geq a_0$,
    where $\calE(\vi(\as))$ defines how close the output must be to $f(\vm)$ to be ``correct''.
    
    Note also that in this example, the approximation error increases with $\vi(\as)$ (i.e., $\calE$ is monotonically increasing), while the expected time to stabilization decreases with $\vi(\as)$.
    It is conceivable for the approximation error to \emph{decrease} with $\vi(\as)$,
    or even not to depend on $\vi(\as)$, 
    although we do not know of any examples of such protocols.
    Our main theorem lower bounds the approximation error as a linear function of the count of the lowest-count state present in the initial configuration, 
    whether that is $\as$ or not, so our proof works regardless of the precise form of $\calE$.
}

\discussionApproximationState

\section{Sublinear-time, sublinear-error approximation of linear functions with negative or non-integer coefficients is impossible}
\label{sec:negative-results}

% below from icalp paper

% \todo{AB: below appeared to be added from ICALP paper. Ensure not discussed elsewhere}

We say a function $f: \N^k \to \N$ is \emph{$\N$-linear} 
if there are $c_1,\ldots,c_k \in \N$ 
such that for all $\vm\in\N^k$,
$f(\vm) = \sum_{i=1}^k c_i \vm(i)$.

As we consider leaderless population protocols, we need to make sure that $\as$ does not act as a small count ``leader''. 
Consistent with the rest of this paper, we reason about initial configurations with $\vi(\as) \geq \alpha n$ for some $\alpha>0$ to ensure $\alpha$-density.

Let $f:\N^k\to\N$.
In defining running time for function-approximating population protocols, 
we express the expected time as a function of \emph{both} the total number of agents $n=\|\vi\|$ 
\emph{and} the initial count $\vi(\as)$ of approximation states.
Let $\calE:\N\to\N$ and $t:\N^2\to\N$.
Given a function-approximating population protocol $\calA$ that $\calE$-approximates $f$,
we say $\calA$ \emph{\emph{$\calE$-approximates} $f$ in expected time $t$} if,
for all valid initial configurations $\vi$ of $\calA$, 
letting $\vm = \vi \rest \Sigma$,
$\time{\vi}{S_{f,\vm,\calE(\vi(\as))}^\calA} \leq t(n,\vi(\as))$.

The following theorem states that given any linear function $f$ and any population protocol $\calP$,
if $f$ has a non-integer or negative coefficient, 
then $\calP$ requires at least linear time to approximate $f$ with sublinear error.
It states this by contrapositive: if the protocol takes sublinear time, 
then the error $\calE:\N\to\N$ must grow at least linearly with the initial count of approximation state $\as$. 
In particular, the initial configurations $\vi$ (letting $n=\|\vi\|$) on which our argument maximizes the error have $\vi(\as) = \Omega(n)$.
Thus, the fact that $\calE(a) \geq \gamma a$ implies that on these $\vi$, 
the error is $\Omega(n)$.

\begin{theorem} \label{thm:main}
  Let $f:\N^k\to\N$ be a linear function that is not $\N$-linear.
  Let $\calE:\N\to\N$.
  Let $\calA$ be a function-approximating leaderless population protocol that stably $\calE$-approximates $f$ in expected time $t$, 
  where for some $\alpha>0$, $t(n, \alpha n) = o(n)$.
  Then there is a constant $\gamma>0$ such that,
  for infinitely many $a \in\N$,
  $\calE(a) \geq \gamma a$.
\end{theorem}

A protocol stably computing $f$ also stably $\calE$-approximates $f$ for $\calE(a) = 0$, so we have:

\begin{cor} \label{cor:main}
  Let $f:\N^k\to\N$ be a linear function $f(\vm) = \sum_{i=1}^k \floor{c_i \vm(i)}$, 
  where $c_i \not\in\N$ for some $i \in \{1,\ldots,k\}$.
  Let $\calC$ be a function-computing leaderless population protocol that stably computes $f$.
  Then $\calC$ takes expected time $\Omega(n)$.
\end{cor}

This gives a complete classification of the asymptotic efficiency of computing linear functions $f(\vm) = \sum_{i=1}^k \floor{c_i \vm(i)}$ with population protocols.
If $c_i\in\N$ for all $i\in\{1,\ldots,k\}$, 
then $f$ is stably computable in logarithmic time by Observation~\ref{obs:stably-compute-positive-integer-coefficient}.
Otherwise, $f$ requires linear time to stably compute by Corollary~\ref{cor:main}, 
or even to approximate within 
sublinear error 
by Theorem~\ref{thm:main}.
% $f$ \emph{can} be approximated within linear error in logarithmic time by Theorem~\ref{thm:positive}.
The remainder of Section~\ref{sec:negative-results} is devoted to proving Theorem~\ref{thm:main}.

\todoi{DD: make sure floors are handled uniformly... also can we justify why we took the floor of each term in a linear sum, rather than taking the sum and then a floor? I just decided that arbitrarily, but if it makes a difference on some corner cases we should figure it out.}

Theorem~\ref{thm:main} follows directly from the two theorems in this section.

%\todoi{DD: generalize these proofs to the full class of functions the theorems mention, or come up with a better way of explaining why we can assume ``without loss of generality'' that we are either multiplying by a single rational coefficient, or that we are just doing subtraction $m_1-m_2$}

First, we show that subtraction takes linear time to approximate with sublinear error.

\begin{theorem} \label{thm:subtraction-slow}
    Let $f:\N^k\to\N$ be a linear function $f(\vm) = \sum_{i=1}^k \floor{c_i \vm(i)}$, 
    where $c_i < 0$ for some $i \in \{1,\ldots,k\}$.
    Let $\calE:\N\to\N$ and $t:\N^2\to\N$.
    Let $\calA$ be a function-approximating leaderless population protocol that stably $\calE$-approximates $f$ in expected time $t$.
    \todo{DD: not too comfortable with handling $t$ this way; the density of different configs below is different, so they may not all be the same $\alpha$-density}
    Suppose there is $\alpha > 0$ such that $t(n,\alpha n) = o(n)$.
    Then there is a constant $\gamma>0$ such that,
    for infinitely many $a \in\N$,
    $\calE(a) \geq \gamma a$.
\end{theorem}

\begin{proof}
    % \todo{DD: I think we should be able to do a ``reduction'' to show that if $f(m_1,m_2)=m_1-m_2$ could be done quickly, then so could subtraction with larger coefficients, so I'll make the assumption that we are just concerned about $m_1-m_2$.}
    
    Assume without loss of generality that $f(\vm) = c_1 \vm(1)- c_2\vm(2)$ for $c_1,c_2 > 0$;
    a function with more inputs can have those inputs set to 0, 
    and the remaining two re-ordered to be the first two,
    to result in this $f$.
    Assume for notational ease that $c_1=c_2=1$.
    The extension to other coefficients in $\Q_{> 0}$ is routine,
    requiring us only to modify the set $I_\alpha^=$ below to contain configurations $\vi$ with $c_1 \vi(x_1) = c_2 \vi(x_2)$ instead of merely $\vi(x_1) = \vi(x_2)$.
    
    % the extension to $c_1 \in \Q_{\geq 0} \setminus \N$ is handled by Theorem~\ref{thm:division-slow}
    % (by restricting to inputs where $\vm(2)=0$),
    % and the extension to $c_2  \in \Q_{\geq 0} \setminus \N$ can be handled by a routine modification of the techniques of the proof of Theorem~\ref{thm:division-slow}.

    Let $\calE:\N\to\N$ and $t:\N^2\to\N$.
    Let $\calC$ be a function-approximating population protocol that stably $\calE$-approximates $f$ in expected time $t$.
    Let $\alpha > 0$ be such that $t(n,\alpha n) = o(n)$ and,
    for all $m\in\N$,
    there is a valid $\alpha$-dense initial configuration $\vi\in\N^\Lambda$ such that $\vi(\ins_1)=\vi(\ins_2)=m$.
    Such an $\alpha$ exists since the quiescent threshold $q_0:\N^k\to\N$ is linearly bounded.
    % Since $\calC$ allows dense inputs, 
    % there exists $\alpha > 0$ such that $t(n,\alpha n) = o(n)$
    % and for all inputs $\vm\in\N^2$, 
    % there is a valid $\alpha$-dense initial configuration $\vi$ with $\vi\rest\Sigma = \vm$.
    
    Let $I_\alpha^=$ be the set of all such $\alpha$-dense valid $\vi$, where $\vi(x_1) = \vi(x_2)$
    (i.e., the set of $\alpha$-dense initial configurations representing an input $\vm$ to $f$ such that $f(\vm)=0$).
    Let $S = \{ \vo \mid (\exists \vi \in I_\alpha^=)\ \vi \reach \vo$ and $\vo$ is stable$\}$
    be the set of stable configurations reachable from some initial configuration in $I_\alpha^=$.
    By assumption we have that for each $\vi \in I_\alpha^=$, $\time{\vi}{S} = o(n)$.
    
    Apply Corollary~\ref{cor:push-Delta-simplified}.
    \corPushDeltaSimplifiedConclusion

    Let $b = 1$; below we ensure that $\max(\vd^\Delta) \leq b$.
    
    There are two cases:
    \begin{description}
    \item[\underline{$\outs \in \Delta$}:]
        We claim that $\ins_1 \in \Delta$.
        To see why,
        observe that increasing the initial amount of $\ins_1$ by $1$ 
        should increase the output by $1$.
        If we had $\ins \in \Gamma$,
        then by Corollary~\ref{cor:unbdd-state-configs-stable},
        for any stable configuration $\vo$,
        the configuration $\vo + \{\ins_1\}$ would also be stable,
        but would have the same output.
        Let $\vi$ be any initial configuration representing an input with $x_1=x_2$,
        such that $\vi \reach \vo$.
        Then we should have $\vo(\outs) = 0$.
        Then $\vi + \{\ins_1\} \reach \vo + \{\ins_1\}$,
        a contradiction since the correct output to compute from $\vi + \{\ins_1\}$ is $1$, but the output is also $0$ in stable configuration $\vo + \{\ins_1\}$.
        This shows the claim that $\ins_1 \in \Delta$.
        
        Let $\vd^\Delta=\{\ins_1\}$.
        Let $\vw = 2 \vo_{n_b}^\Gamma + \tvD_1.\vo_{n_b}^\Delta$
        and let $\vc_1 = \tvD_2.\vd^\Delta$.
        The above implies that 
        $2 \vi_{n_b} +\{\ins_1\} \reach \vw+\vc_1$.
        For all $l\in\N$,
        let $\vv_l = l (\vw + \vc_1)$ and $\vi'_l = l(2\vi_{n'_0}+\{\ins_1\})$.
        Thus $\vi'_l \reach \vv_l$.
        
        Let $\vm=(2lm+l,2lm)=\vi'_l\rest\Sigma$,
        with correct output $f(2lm+l,2lm)=l$, but $\vv_l(\outs) = 0$.
        
    \item[\underline{$\outs \in \Gamma$}:]
        Then for all sufficiently large $n \geq n_b$, 
        $2 \vo_{n}(\outs) + \left( \min\limits_{n'\in N} \left( \tvD_1.\vo_{n'}^\Delta \right)(\outs) \right) > 0$.
        Fix such an $n$.
        Let $\vw = 2 \vo_{n}^\Gamma + \tvD_1.\vo_{n}^\Delta$; note that $\vw(\outs) > 0$.
        Letting $\vd^\Delta=\vec{0}$,
        the above implies that 
        $2 \vi_{n} \reach \vw$.
        For all $l\in\N$,
        let $\vv_l = l \vw$ and $\vi'_l = 2l\vi_{n}$.
        Thus $\vi'_l \reach \vv_l$.
        
        Let $\vm=(2lm,2lm)=\vi'_l\rest\Sigma$,
        with correct output $f(2lm,2lm)=0$, but $\vv_l(\outs) \geq l$.
    \end{description}
    In each case $\vv_l$ is stable by Observation~\ref{obs:unbdd-state-configs-stable}
    and reachable from an initial configuration $\vi'_l$ representing input $\vm$,
    with $|\vv_l(\outs) - f(\vm)| \geq l$,
    so $\calE(2l\vi_{n_b}(\as)) \geq l$.
    Since $2 \vi_{n_b}(\as)$ is constant with respect to $l$, 
    choosing $\gamma = \frac{1}{2 \vi_{n_b}(\as)}$ proves the theorem.
\end{proof}

% \opt{normal}{\begin{proof}\proofThmSubtractionSlow\end{proof}}

Now we show that division by a constant takes linear time to approximate with sublinear error.

\begin{theorem} \label{thm:division-slow}
    Let $f:\N^k\to\N$ be a linear function $f(\vm) = \sum_{i=1}^k \floor{c_i \vm(i)}$, 
    where $c_i \not\in\Z$ for some $i \in \{1,\ldots,k\}$.
    Let $\calE:\N\to\N$ and $t:\N^2\to\N$.
    Let $\calA$ be a function-approximating leaderless population protocol that stably $\calE$-approximates $f$ in expected time $t$.
    \todo{DD: not too comfortable with handling $t$ this way; the density of different configs below is different, so they may not all be the same $\alpha$-density}
    Suppose there is $\alpha > 0$ such that $t(n,\alpha n) = o(n)$.
    Then there is a constant $\gamma>0$ such that,
    for infinitely many $a \in\N$,
    $\calE(a) \geq \gamma a$.
\end{theorem}

\begin{proof}
    If $c_i<0$ for some $i$, then Theorem~\ref{thm:subtraction-slow} applies and we are done, so assume all $c_i \geq 0$.
    Assume without loss of generality that $f(m) = \floor{c m}$, where $c \in \Q_{> 0} \setminus \N$;
    a function with more inputs can have those inputs set to 0 to result in this $f$.
    Write $c$ in lowest terms as $\frac{p}{r}$ for $p,r \in \Z_{\geq 1}$, and note that $r \geq 2$ since $c \not \in \Z$.
    Let the input state $\ins_1$ be denoted simply $\ins$.
    
    Let $\calE:\N\to\N$.
    Let $\delta>0$ and let $\calA$ be a function-approximating population protocol that stably $\calE$-approximates $f$ in expected time $o(n)$.
    Let $\alpha > 0$ be such that $t(n,\alpha n) = o(n)$ and,
    for all $m\in\N$,
    there is a valid $\alpha$-dense initial configuration $\vi\in\N^\Lambda$ with $\vi(\ins) = m$.
    Such an $\alpha$ exists since the quiescent threshold $q_0:\N^k\to\N$ is linearly bounded.
    
    Let $I_\alpha$ be the set of all such $\alpha$-dense valid $\vi$.
    Let $S = \{ \vo \mid (\exists \vi \in I_\alpha)\ \vi \reach \vo$ and $\vo$ is stable and $\calE(\vi(\as))$-$f(\vi(\ins))$-correct $\}$ be the set of stable ``$\epsilon$-$t$''-correct configurations reachable from some configuration $\vi \in I_\alpha$,
    and where the choice of $\epsilon$ and $t$ to define $\epsilon$-$t$-correct depends on $\vi$.
    By hypothesis we have that for each $\vi \in I_\alpha$, $\time{\vi}{S} = o(n)$.
    
    Apply Corollary~\ref{cor:push-Delta-simplified}.
    \corPushDeltaSimplifiedConclusion

    Note that $\outs \in \Gamma$ since $f(m)$ grows unboundedly with $m$ and $m=\vi_n(\ins)$ grows unboundedly with $n$.
    We claim that $\ins \in \Delta$.
    To see why,
    observe that increasing the initial amount of $\ins$ by $r$ 
    should increase the output by $p$.
    If we had $\ins \in \Gamma$,
    then by Corollary~\ref{cor:unbdd-state-configs-stable},
    for any stable configuration $\vo$,
    the configuration $\vo + \{r \ins\}$ would also be stable,
    but would have the same output.
    Let $\vi$ be any initial configuration
    such that $\vi \reach \vo$.
    Then $\vi + \{r \ins\} \reach \vo + \{r \ins\}$,
    a contradiction since the correct output to compute from $\vi + \{r \ins\}$ is $p$ larger than the correct output for $\vi$.
    This shows the claim that $\ins \in \Delta$.
    
    Let $b = 1$; below we ensure that $\max(\vd^\Delta) \leq b$.
    Let 
    $\vw = r \left( 2 \vo_{n_b}^\Gamma + \tvD_1.\vo_{n_b}^\Delta \right)$ 
    and 
    $\vc_1 = \tvD_2.\vd^\Delta$.
    The above argument shows that $2r \vi_{n_b} \reach \vw$
    (letting $\vd^\Delta = \vec{0}$)
    and $2r \vi_{n_b}  + r\{\ins\} \reach \vw + r\vc_1$
    (letting $\vd^\Delta = \{\ins\}$).
    Since $\vw \in\N^\Gamma$ and $\vc_1\in\N^\Gamma$, both $\vw$ and $\vw+r\vc_1$ are stable by Observation~\ref{obs:unbdd-state-configs-stable}.
    Recall $f(2rm) = \floor{2rm p/r} = 2m p$.
    We have two cases:
    \begin{description}
    \item[\underline{$\vw(\outs) \neq 2m p$}:]
        Thus $|\vw(\outs) - 2m p|\geq 1$.
        For all $l\in\N$,
        let $\vv_l = l \vw$
        and $m_l = 2lrm$.
        Then 
        $2lr \vi_{n_b} \reach \vv_l$, and 
        $$|\vv_l(\outs) - f(m_l)| 
        = |l \vw(\outs) - 2lm p|
        = l |\vw(\outs) - 2m p| \geq l.$$
        
    \item[\underline{$\vw(\outs) = 2m p$}:]
        For all $l\in\N$,
        let $\vv_l = l (\vw + r\vc_1)$
        and $m_l = l (2rm+r)$.
        Thus, 
        $l(2r \vi_{n_b} + r\{\ins\}) 
        \reach \vv_l$.
        Also, $f(m_l) = \floor{l p(2rm+r)/r} = 2lm p + lu$.
        Note that $\vc_1(\outs) \in \Z$.
        There are two subcases:
        \begin{description}
        \item[\underline{$\vc_1(\outs) \geq 1$}:]
            Then $\vv_l(\outs) 
            = l (\vw(\outs) + r\vc_1(\outs))
            \geq l ( 2m p + r )
            = 2lm p + lr$,
            so
            \begin{align*}
                \vv_l(\outs) - f(m_l)
                &\geq  
                2lm p + lr - \left( 2lm p + l \right)
                %\\&
                =
                lr - l
                \geq
                l.
            \end{align*}
            
        \item[\underline{$\vc_1(\outs) \leq 0$}:]
            Then $\vv_l(\outs) 
            = l (\vw(\outs) + r\vc_1(\outs))
            \leq 2lm p$,
            so
            \begin{align*}
                f(m_l) - \vv_l(\outs)
                &\geq  
                2lm p + l - 2lm p
                =
                l.
            \end{align*}
        \end{description}
    \end{description}
    In each case $\vv_l$ is stable by Observation~\ref{obs:unbdd-state-configs-stable} and reachable from an initial configuration representing input $m_l$,
    with count $2lr \vi_{n_b}(\as)$ of $\as$,
    but $|\vv_l(\outs)-f(m_l)| \geq l$,
    so $\calE(2lr \vi_{n_b}(\as)) \geq l$.
    Since $2r \vi_{n_b}(\as)$ is constant with respect to $l$, 
    choosing $\gamma = \frac{1}{2r \vi_{n_b}(\as)}$ proves the theorem.
\end{proof}

\section{Logarithmic-time, linear-error approximation of linear functions with nonnegative rational coefficients is possible}
\label{sec:positive-results}

Recall the definition of $\N$-linear function from Section~\ref{sec:negative-results}.
It is easy to see that any $\N$-linear function $f$ can be stably computed in logarithmic time.
Recall that $\ins,\qs \to \outs,\outs$ stably computes $f(m) = 2m$ in expected time $O(\log n)$.
The extension to larger coefficients, e.g., $f(m)=4m$, uses a series of transitions:
\begin{align*}
  \ins,\qs &\to \outs,\ins'
  \\
  \ins',\qs &\to \outs,\ins''
  \\
  \ins'',\qs &\to \outs,\outs
\end{align*}
The extension to multiple inputs, e.g., $f(m_1,m_2,m_3) = 4m_1+m_2+2m_3$, uses similar transitions for each input:
\begin{align*}
  \ins_1,\qs &\to \outs,\ins_1'
  \\
  \ins_1',\qs &\to \outs,\ins_1''
  \\
  \ins_1'',\qs &\to \outs,\outs
  \\
  \ins_2,\qs &\to \outs,\qs
  \\
  \ins_3,\qs &\to \outs,\outs
\end{align*}
We summarize this in the following observation.

\begin{obs} \label{obs:stably-compute-positive-integer-coefficient}
  Let $f:\N^k\to\N$ be an $\N$-linear function.
  There is a function-computing leaderless population protocol that stably computes $f$ in expected time $O(\log n)$.
\end{obs}

A function is \emph{$\Q_{\geq 0}$-linear} 
if there are $c_1,\ldots,c_k \in \Q_{\geq 0}$
such that for all $\vm\in\N^k$,
$f(\vm) = \sum_{i=1}^k \floor{c_i \vm(i)}$.
We now describe how to stably approximate $\Q_{\geq 0}$-linear functions 
with a linear approximation error, in logarithmic time.
(It is open to do this for negative coefficients, e.g., $f(m_1,m_2) = m_1-m_2$).
Recall the following simple example of a population protocol that approximately divides by 2 
(that is, with probability 1 it outputs a value guaranteed to be a certain distance to the correct output), 
with a linear approximation error,
and is fast ($O(\log n)$ time) with initial counts $\# \ins = m$, $\# \as = \gamma m$, and $\# \qs = \# \outs = 0$:
\begin{align*}
  \as,\ins &\to \ans{b},\outs
  \\
  \ans{b},\ins &\to \as,\qs
\end{align*}
which stabilizes $\# \outs$ to somewhere in the interval $\{m/2,m/2+1,\ldots, m/2 + \gamma m\}$.

To see that the protocol is correct,
note that the transition sequence can make $\# \outs$ closer to one endpoint of the interval or the other depending on which transitions are chosen to consume the \emph{last} $\gamma m$ of $\ins$, 
but no matter what, the first transition executes at least as many times as the second, 
but not more than $\gamma m$ times more.

If $\#\as=1$ initially, the above protocol stably computes $\floor{m/2}$ (taking linear time just for the last transition; and in total takes $\Theta(n \log n)$ time, by a coupon collector argument).

To see that the protocol takes $O(\log n)$ time if $\# \as = \gamma m$ initially,
note $n = m + \gamma m \leq 2m.$
Observe that $\#\as + \#\ans{b} = \gamma m$ in any reachable configuration.
Thus the probability any given interaction is one of the above two transitions is $\approx \frac{\gamma m \#\ins}{n^2}$,
so the expected number of interactions until such a transition occurs is $\frac{n^2}{\gamma m \#\ins}$.
After $m$ such transitions occur, all the input $\ins$ is gone and the protocol stabilizes, 
which by linearity of expectation takes expected number of interactions
$$
  \sum_{\#\ins = 1}^m \frac{n^2}{\gamma m \#\ins} 
  = 
  \frac{n^2}{\gamma m} \sum_{\#\ins = 1}^m \frac{1}{\#\ins} 
  \approx 
  \frac{n^2}{\gamma m} \ln m 
  \leq 
  \frac{n^2}{\gamma n/2} \ln n 
  =
  \frac{2 n}{\gamma} \ln n,
$$
i.e., expected parallel time $\frac{2}{\gamma} \ln n$.
Thus this shows a tradeoff between accuracy and speed in a single protocol,
adjustable by the initial count of $\as$.
In this case, the approximation error increases, and the expected time to stabilization decreases, with increasing initial $\# \as$.

% As example of computing a linear function with \emph{negative} coefficients, 
% consider $f(m_1,m_2) = m_1-m_2$.
% Viewed as a function $f:\N^2\to\N$, it is not well defined since $m_1-m_2$ can be negative.
% We could handle this by talking instead about the piecewise linear function $f(m_1,m_2) = \max(0,m_1-m_2)$.
% Then $f$ is stably computed by the protocol with transitions $\ins_1,\qs \to \outs,\qs$ and $\ins_2,\outs \to \qs,\qs$.
% Alternately, we could say that $f$ is linear when restricted to the domain $\{(m_1,m_2) \mid m_1 \geq m_2 \}$, 
% and that the above protocol stably computes the linear function $f$, 
% as long as we restrict to initial configurations $\vi\in\N^\Lambda$ such that $\vi(\ins_1) \geq \vi(\ins_2)$.
% This latter convention will make for less awkward theorem statements, so we choose this interpretation.

% The expected time, in the worst case over all inputs, is $\Omega(n)$.
% The worst case occurs when $m_1=m_2$, in which case the final transition $\ins_2,\outs \to \qs,\qs$ occurs when $\#\ins_2=1$ and $\#\outs=1$, hence takes expected time $\Omega(n)$.
% However, modify the protocol to start with $\vi(\as) = \gamma m_1$, and add the transition $\as,\qs \to \ins_1,\qs$.
% Now, once the final transition completes (in expected time $\approx \ln \vi(\as)$),
% we have $\# \ins_1 \geq \gamma m_1$ in all reachable configurations.
% Thus, the transition $\ins_2,\outs \to \qs,\qs$ requires at most $\approx \frac{1}{\gamma} \ln n$ expected time to complete.
% %The protocol then stably $\vi(\as)$-approximates $f$ in expected time $$

More generally, we can prove the following.
In particular, if $a = \Omega(n)$,
then $t(n,a) = O(\log n)$.
Also, if $a = o(n)$, then the approximation error is $o(n)$, and if $a = \omega(\log n)$, then the expected time is $o(n)$ also.
This does not contradict Theorem~\ref{thm:main} since setting $a=o(n)$ implies the initial configurations are not all $\alpha$-dense for a fixed $\alpha>0$.

\newcommand{\thmPositive}{
  Let $f:\N^k\to\N$ be a $\Q_{\geq 0}$-linear function.
  Let $\calE: \N \to \N$ be the identity function.
  Define $t:\N^2\to\N$ by $t(n,a) = \frac{n}{a} \log n.$
  Then there is a function-approximating leaderless population protocol $\calA$ that $\calE$-approximates $f$ in expected time $O(t)$.
}

\begin{theorem}\label{thm:positive}
  \thmPositive
\end{theorem}

\newcommand{\proofThmPositive}{
  Write each rational coefficient $c_i = \frac{p_i}{r_i}$, where $p_i,r_i \in \Z_{\geq 1}$.
  (Since $f$ is linear, the case $p_i=0$ is easy to handle by simply ignoring $\ins_i$ as an input, so we assume all coefficients are strictly positive.)
  %\todo{DD: note that the amount of $\qs$ needed depends on the amount of $\as$ supplied, hence the reason for the hard-to-parse dependence of $q_0(\vm,a)$ on $a$ in the defn of function approximation}
  The initial configuration $\vi$ has $\vi(\ins_i) = \vm(i)$ for each $i \in \{1,\ldots,k\}$, $\vi(\as) > 1$ and $\vi(\qs) \geq 2k \vi(\as) + \sum_{i=1}^k p_i \vi(\ins_i)$.
  
  First, for each $i \in \{1,\ldots,k\}$, we have transitions that multiply $\# \ins_i = \vm(i)$ by $p_i$.
  If $p_i=1$, we have the transition
  $
    \ins_i,\qs \to \outs_i,\qs.
  $
  If $p_i=2$, we have the transition
  $
    \ins_i,\qs \to \outs_i,\outs_i.
  $
  If $p_i \geq 3$, we have a sequence of transitions that multiply $\# \ins_i$ by $p_i$:
  \begin{align*}
      \ins_i,\qs &\to \outs_i,\ins_i^{(p_i-1)}
      \\
      \ins_i^{(p_i-1)},\qs &\to \outs_i,\ins_i^{(p_i-2)}
      \\
      \ins_i^{(p_i-2)},\qs &\to \outs_i,\ins_i^{(p_i-3)}
      \\
      \ldots
      \\
      \ins_i^{(2)},\qs &\to \outs_i,\outs_i
  \end{align*}
  This ensures that for each $i \in \{1,\ldots,k\}$, eventually $\# \outs_i = p_i \vi(\ins_i)$.
  Note that each agent in state $\ins_i$ eventually causes $p_i$ agents in state $\qs$ to be consumed as inputs to the above transitions, 
  hence at least $\sum_{i=1}^k p_i \vm(i)$ agents in state $\qs$ are required initially for the above to complete.
  
  Next, we split up the approximation states to have a specific one for each input $\ins_i$.
  In other words, the initial configuration has $\vi(\as)$ agents in state $\as$,
  and we want to reach a configuration with 
  $\vi(\as)$ agents in state $\as_1$, 
  $\vi(\as)$ other agents in state $\as_2$, 
  $\ldots$, 
  $\vi(\as)$ other agents in state $\as_k$.
  If $k=1$ then we simply use $\as$ directly in place of the state $\as_{1}^{(1)}$ below.
  Otherwise, let $l\in\N$ be such that $2^{l-1} < k \leq 2^l$ (i.e., $2^l$ is the next power of $2$ at least $k$).
  
  We have the transitions that make copies of $\as$ for each input:
  \begin{align*}
      \as,\qs &\to \as_0,\as_1
      \\
      \as_0,\qs &\to \as_{00},\as_{01}
      \\
      \as_1,\qs &\to \as_{10},\as_{11}
      \\
      \as_{00},\qs &\to \as_{000},\as_{001}
      \\
      \as_{01},\qs &\to \as_{010},\as_{011}
      \\
      \as_{10},\qs &\to \as_{100},\as_{101}
      \\
      \as_{11},\qs &\to \as_{110},\as_{111}
      \\
      \as_{000},\qs &\to \as_{0000},\as_{0001}
      \\
      &\ldots
  \end{align*}
  and so on,
  up to transitions whose output (right-hand side) states are binary strings of length $l$.
  Also, for each $i \in \{1,\ldots,k\}$, we have the transition $\as_{b(i)},\qs \to \as_{i}^{(1)},\qs$, 
  where $b(i)$ is the $i$'th string in the lexicographical ordering of $\{0,1\}^l$.
  If $k$ is not a power of $2$, then we will simply let $\as_{b(i)}$ go unused for $i \in \{k+1,\ldots,2^l\}$.
  
  Note that each agent in state $\as$ eventually causes $\sum_{j=1}^l 2^j = 2^{l+1}-1 < 2k$ agents in state $\qs$ to be consumed as inputs to the above transitions,
  hence at least $\vi(\as) (2^{l+1}-1)$ agents in state $\qs$ are required initially for the above to complete.
  Combined with the lower bound on $\vi(\qs)$ associated to the first set of transitions, at least $\vi(\as) (2^{l+1}-1) + \sum_{i=1}^k p_i \vi(\ins_i) \leq 2k \vi(\as) + \sum_{i=1}^k p_i \vi(\ins_i)$ agents in state $\qs$ are required initially.
  Note that this is $O(\| \vi \rest (\Sigma \cup \{\as\}) \|).$
  
  Next, we approximately divide $\# \outs_i$ by $r_i$, using the state $\as_i^{(1)}$ produced by the second sequence of transitions.
  If $r_i=1$, this is easily handled (exactly) by the transition $\outs_i,\qs \to \outs,\qs$.
  Otherwise, for each $i$ such that $r_i \geq 2$, we have the transitions that divide $\# \outs_i$ by $r_i$:
  \begin{align*}
      \as_i^{(1)}, \outs_i &\to \as_i^{(2)}, \outs
      \\
      \as_i^{(2)}, \outs_i &\to \as_i^{(3)}, \qs
      \\
      \as_i^{(3)}, \outs_i &\to \as_i^{(4)}, \qs
      \\
      \ldots
      \\
      \as_i^{(r_i)}, \outs_i &\to \as_i^{(1)}, \qs
  \end{align*}
  which consumes $r_i$ copies of $\outs_i$ (of which there are $p_i \vm(i)$) for each $\outs$ produced.
  Since for all $i$, the first transition above outputs the same state $\outs$, the final count of $\outs$ will be the sum $\sum_{i=1}^k \floor{\frac{p_i}{r_i} \vm(i)}$, i.e., the value $f(\vm)$.
  
  It remains to prove the stated bound on expected time.
  Above we stated a lower bound on $\vi(\qs)$ required for the protocol to be correct.
  To obtain the stated expected time, we double this bound and add $\|\vi \rest (\Sigma \cup \{\as\})\|$, 
  which ensures that $\# \qs$ is always at least $n/2$ in any reachable configuration.
  Note that $n \geq \sum_{i=1}^k p_i \vm(i)$.
  Let $\hat{p} = \max_{i} p_i$.
  
  We analyze a simpler process that stochastically dominates the actual Markov process.
  We assume that the first set of transitions (multiplying $\# \ins_i$ by $p_i$) 
  completes before the start of the second (making copies of $\as$ for each input), 
  and that these then complete before the start of the third group (dividing $\# \outs_i$ by $r_i$).
  Also, transitions that can go in parallel within these groups 
  (such as the first set of transitions for different values of $i$)
  can have their expected times analyzed independently and summed to obtain a loose bound on the time for them all to terminate.
  
  %For this analysis we rely on derivations developed in other papers on population protocols and chemical reaction networks~\cite{angluin2006passivelymobile, CheDotSolDetFuncNaCo, DotHajLDCRNNaCo}.
  Let $i \in \{1,\ldots,k\}$.
  For the first group, we assume that for each $\ins_i^{(p_i-j)},\qs \to \outs_i,\ins_i^{(p_i-j-1)}$ completes (consuming all available $\ins_i^{(p_i-1)}$ states) before the next $\ins_i^{(p_i-j-1)},\qs \to \outs_i,\ins_i^{(p_i-j-2)}$ starts, since this process stochastically dominates the actual process that allows the transitions to go in parallel.
  For notational brevity we just analyze the first transition $\ins_i,\qs \to \outs_i,\ins_i^{(p_i-1)}$.
  By our choice of $\vi(\qs)$, each interaction has probability at least $\frac{1}{2}$ to have a $\qs$ state, and conditioned on this, probability at least $\frac{\# \ins_i}{n}$ for the other state to be $\ins_i$, so probability at least $\frac{\# \ins_i}{2 n}$ to be of the form $\ins_i,\qs \to \outs_i,\ins_i^{(p_i-1)}$.
  Thus the expected number of interactions until this transition happens is at most $\frac{2 n}{\# \ins_i}$.
  By linearity of expectation, the expected number of interactions until all $\ins_i$ are consumed is at most $\sum_{\# \ins_i = 1}^{\vm(i)} \frac{2 n}{\# \ins_i} \approx 2 n \ln \vm(i))$.
  Thus the expected parallel time to complete these transitions is $2 \ln \vm(i)$.
  Thus for each $i$, the expected time for all $p_i$ of these transitions to complete is at most $2 p_i \ln \vm(i)$,
  so the expected time for all to complete is at most $2 \sum_{i=1}^k p_i \ln \vm(i) \leq 2 k \hat{p} \ln n$.
  
  The second group of transitions can be analyzed similarly, 
  by assuming they each complete (consume all non-$\qs$ input states) before the next starts.
  Each then takes $\ln \vi(\as)$ expected time to complete.
  There are $2^{l+1} - 1 < 2k$ total transitions so $2 k \ln \vi(\as) \leq 2 k \ln n$ expected time is required.
  
  The third group of transitions is a bit different.
  Since we do not begin analysis until the second group of transitions completes, 
  we start with $\# \as_i^{(1)} = \vi(\as)$ and $\# \outs_i = p_i \vm(i)$.
  Also note that $\vi(\as) = \sum_{j=1}^{r_i} \# \as_i^{(j)}$ in any subsequently reachable configuration.
  Thus, in any such configuration the probability that the next transition involves an $\as_i^{(j)}$ state is $\frac{\vi(\as)}{n}$, 
  and conditioned on this, the probability that the \emph{other} input state is $\outs_i$ is $\frac{\#\outs_i}{n}$, 
  so probability $\frac{\vi(\as) \# \outs_i}{n^2}$ that the next transition is \emph{one} of the $r_i$ transitions.
  Thus the expected number of transitions until such a transition happens is $\frac{n^2}{\vi(\as) \# \outs_i}$.
  By linearity of expectation, the expected number of interactions until all $\outs_i$ are consumed is at most 
  $\sum_{\# \outs_i = 1}^{p_i \vm(i)} \frac{n^2}{\vi(\as) \# \outs_i} \approx \frac{n^2}{\vi(\as)} \ln(p_i \vm(i))$,
  so expected parallel time $\frac{n}{\vi(\as)} \ln(p_i \vm(i))$.
  Thus the expected time for all $k$ groups of transitions to complete is at most 
  $\frac{n}{\vi(\as)} \sum_{i=1}^k \ln(p_i \vm(i)) = \frac{n}{\vi(\as)} k \ln(\hat{p} n)$.
  
  Thus, summing the above three bounds, the total expected time to stabilization is at most 
  $2 k \hat{p} \ln n + 2 k \ln n + \frac{n}{\vi(\as)} k \ln(\hat{p} n)$.
  Since $k$ and $\hat{p}$ are constant with respect to $n$ and $\vi(\as)$, 
  the bound is dominated by the last term, which is $O\left( \frac{n}{\vi(\as)} \ln n \right)$.
}

\begin{proof}\proofThmPositive\end{proof}
\section{Exact computation of nonlinear functions}
\label{sec:nonlinear-functions}

\todo{DD: in this section and the predicate section, the definitions of eventually $\N$-linear function and eventually constant predicate only talk about inputs with all coordinates positive. We should generalize this to allow some of them to be 0; since that would preserve $\alpha$-density of the initial configurations}

%\todo{AB: fix reference}
%In Section~\ref{sec:approx-negative-results}, 
In Section~\ref{sec:negative-results}, 
we obtained a precise characterization of the linear functions stably computable in sublinear time by population protocols
and furthermore show that those not exactly computable in sublinear time are not even approximable with sublinear error in sublinear time.
However, the class of functions stably computable (in any amount of time) by population protocols is known to contain non-linear functions such as $f(m_1,m_2)=\max(m_1,m_2)$, or $f(m) = m$ if $m$ is even and $f(m) = 2m$ if $m$ is odd.
In fact a function is stably computable by a population protocol if and only if its \emph{graph} $\{(\vm,f(\vm)) \mid \vm\in\N^{k} \}$ is a semilinear set~\cite{AngluinAE2006semilinear, CheDotSolDetFuncNaCo}.
A set $A \subseteq \N^k$ is \emph{semilinear} if and only if~\cite{ginsburg1966} 
it is expressible as a finite number of unions, intersections, and complements of sets of one of the following two forms: 
\emph{threshold sets} of the form $\{ \vx \mid \sum_{i=1}^k a_i \cdot \vx(i) < b \}$ 
for some constants $a_1,\ldots,a_k,b \in \Z$ 
or 
\emph{mod sets}  of the form $\{ \vx \mid \sum_{i=1}^k a_i \cdot \vx(i) \equiv b \mod c \}$ 
for some constants $a_1,\ldots,a_k,b,c\in\N$.

Say that a set $P \subseteq \N^k$ is a \emph{periodic coset} if there exist $\vb,\vp_1,\ldots,\vp_l \in \N^k$
such that $P = \{ \vb + n_1 \vp_1 + \ldots + n_l \vp_l \mid n_1,\ldots,n_l \in \N \}$.
(These are typically called ``linear'' sets, but we wish to avoid confusion with linear functions.)
Equivalently, a set is semilinear if and only if it is a finite union of periodic cosets.

Although our technique fails to completely characterize the efficient computability of all semilinear functions,
we show that a wide class of semilinear functions cannot be stably computed in sublinear time:
functions that are not eventually $\N$-linear.
The only exceptions,
for which we cannot prove linear time is required, 
yet neither is there known a counterexample protocol stably computing the function in sublinear time,
are functions whose ``non-integral-linearities are near the boundary of $\N^k$''.
For example, the function $f(m) = 0$ if $m \leq 3$ and $f(m) = m$ otherwise 
is non-linear (although it is semilinear, so stably computable), 
but restricted to the domain of inputs $> 3$,
it is linear with positive integer coefficients.
Thus it is an example of a function whose ``population protocol time complexity'' is unknown. %\footnote{Being semilinear, it is computable (in linear time) by the following population protocol. Define transitions $\ins,\qs \to \outs,\ins'$ and $\ins',TODO$}

Corollary~\ref{cor:main} and Observation~\ref{obs:stably-compute-positive-integer-coefficient} 
imply that a linear function is stably computable in sublinear time by a population protocol 
if and only if it is $\N$-linear.
Theorem~\ref{thm:negative-result-eventually-positive-integral-linear} 
generalizes the forward direction (restricted to nonlinear functions) to eventually $\N$-linear functions.

\subsection{Eventually affine functions}

% We need a technical lemma, proven in~\cite[Lemma 4.3]{CheDotSolDetFuncNaCo}.
Say that a function $f:\N^k\to\N$ 
is \emph{eventually $\N$-affine} if there exist
$n_0,c_1,\ldots,c_k \in \N$
and
$b \in \Z$
% and $q_1,\ldots,q_k \in \Q$
such that, for all $\vm\in \N_{\geq n_0}^k$,
$f(\vm) = b + \sum_{i=1}^k c_i \vm(i)$.

The function $f:\N^k\to\Q$ is affine 
if and only if all points on the graph of $f$ lie on a $k$-dimensional hyperplane.\footnote{
    We let the range be $\Q$ instead of $\N$ here to get a bidirectional implication,
    but the direction we are interested in requires only range $\N$:
    if $f:\N^k\to\N$ is affine,
    then all points lie on a $k$-dimensional hyperplane.
}
This holds if and only if,
for all $\vm,\vv \in \N^k$,
$f(\vm+\vv) - f(\vm) = f(\vm+2\vv) - f(\vm+\vv)$.
In other words,
the change in output resulting by moving by a vector $\vv$,
starting from $\vm$,
is the same if we move by $\vv$ a second time.
The next lemma, due to Sungjin Im~\cite{sungjinpersonal}, 
shows that a function $f:\N^k \to \N$ is eventually $\N$-affine if the above holds for any sufficiently large input $\vm$ and any \emph{0/1-valued} vector $\vv \in \{0,1\}^k$.
In other words,
if $f$ ``looks $\N$-affine'' when moving by small amounts (each component at most 1),
then it is $\N$-affine.
This appears obvious,
but some care is needed:
for example, 
it is false if we assume that $\vv$ are only unit vectors. 
For example, the non-affine function 
$f(m_1,m_2) = m_1 \cdot m_2$ 
obeys 
$f(\vm+\vu) - f(\vm) = f(\vm+2\vu) - f(\vm+\vu)$ 
for each $\vm \in \N^2$ and each $\vu \in \{(0,1),(1,0)\}$, 
but fails if $\vu = (1,1)$.
%It generalizes this slightly to consider functions that are \emph{eventually}-$\N$-affine and that \emph{eventually} have the stated property.

\begin{lemma} \label{lem:constant-difference-implies-affine}
    Let $f:\N^k \to \N$.
    Suppose there exists $n_0 \in \N$ such that for all 
    $\vm\in\N^k_{\geq n_0}$ and $\vv\in\{0,1\}^k$,
    $f(\vm+\vv)-f(\vm) = f(\vm+2\vv)- f(\vm+\vv)$. 
    Then $f$ is eventually $\N$-affine:
    there are $n_0,b,c_1,\ldots,c_k \in \N$ such that, 
    for all $\vm\in\N^k_{\geq n_0}$,
    $f(\vm) = b + \sum_{i=1}^k c_i \vm(i)$.
\end{lemma}

\begin{proof}
Assume the hypothesis and let $\vv \in \{0,1\}^k$ and $\vm \in \N^k_{\geq n_0}$.
A straightforward induction shows that, for all $n \geq 1$,
$f(\vm+n \vv)-f(\vm+(n-1)\vv) = f(\vm+\vv)- f(\vm)$, and thus
\begin{align} \label{eq:affineind}
f(\vm + n \vv) = f(\vm) + n \cdot ( f(\vm + \vv) - f(\vm)).
\end{align}

Let $i \in \{1,\ldots,k\}$,
fix values of $m_{i+1}',\ldots,m_{k}' \geq n_0$, 
and define $f_i:\N^{i} \to \N$ by 
$f_i(m_1,\ldots,m_i) = f(m_1,\ldots,m_{i-1},m_i,m_{i+1}',\ldots,m_k')$.
We show by induction on $i$ that $f_i$ is $\N$-affine on inputs $m_1,\ldots,m_i \geq n_0$,
i.e., $f(m_1,\ldots,m_i) = b + c_1 m_1 + \ldots + c_i m_i$
for some $b, c_1,\ldots,c_i \in \N$.

\paragraph{Base case:}
    For any fixed $m_2', \ldots ,m_{k}' \geq n_0$, 
    let $f_1(m_1) = f(m_1,m_2', \ldots ,m_k')$.
    We must show $f_1$ is an $\N$-affine function of $m_1$ (assuming $m_1 \geq n_0$).
    Let $\vv=(1,0, \ldots ,0)$ and let $\vm=(n_0,m_2' \ldots ,m_{k}')$. 
    Then using eq.~\eqref{eq:affineind}, we obtain that for any $m_1 \geq n_0$, 
    \begin{eqnarray}
        f_1(m_1) &=&
        f(n_0,m_2', \ldots ,m_{k}') + 
        \nonumber
        \\&&
        (m_1 - n_0) \cdot \left[ f(n_0+1,m_2', \ldots ,m_{k}') -  f(n_0,m_2', \ldots ,m_{k}') \right],
        \nonumber
    \end{eqnarray}
    which is an affine function $b + c m_1$ of $m_1$, 
    with offset $b = f(n_0,m_2', \ldots ,m_{k}') -  $\\
    $n_0 \cdot \left[ f(n_0+1,m_2', \ldots ,m_{k}') - 
    f(n_0,m_2', \ldots ,m_{k}') \right]$ 
    and coefficient $c = f(n_0+1,m_2', \ldots ,m_{k}') -  f(n_0,m_2', \ldots ,m_{k}')$.

    % Then by the hypothesis of the lemma,
    %   \begin{eqnarray*}
    %   f(n_0+1, m_2', \ldots ,m_{k}') -  f(n_0, m_2', \ldots ,m_{k}')
    %   &=&
    %   f(n_0+2, m_2', \ldots ,m_{k}') -  f(n_0+1, m_2', \ldots ,m_{k}')
    %   \\&=&
    %   f(n_0+3, m_2', \ldots ,m_{k}') -  f(n_0+2, m_2', \ldots ,m_{k}') 
    %   \\&=&
    %   \ldots
    %   \end{eqnarray*}
    % Since the function output is nonnegative, the above difference must be nonnegative.
    % Thus for any $m_1 \geq n_0$, 
    % $$
    %     f_1(m_1)
    %     =
    %     f(n_0,m_2', \ldots ,m_{k-1}') + m_1 \cdot \left[ f(n_0+1,m_2', \ldots ,m_{k}') -  f(n_0,m_2', \ldots ,m_{k}') \right],
    % $$
    % which is an affine function $b + c m_1$ of $m_1$, 
    % with nonnegative integer offset $b = f(n_0,m_2', \ldots ,m_{k-1}')$ and coefficient $c = f(n_0+1,m_2', \ldots ,m_{k}') -  f(n_0,m_2', \ldots ,m_{k}')$.

\paragraph{Inductive case:}
    Fix values $m_{i+1}', \ldots ,m_k' \geq n_0$.
    By the inductive hypothesis, 
    $f_{i}(m_{1}, \ldots ,m_i) = f(m_1, \ldots ,m_i,m_{i+1}', \ldots ,m_k')$ 
    is an $\N$-affine function of $m_{1}, \ldots ,m_i$ (if all $m_j \geq n_0$). 
    We want to show that
    $f_{i+1}(m_{1}, \ldots ,m_{i+1}) = f(m_1, \ldots ,m_{i+1},m_{i+2}' \ldots ,m_k')$ 
    is an $\N$-affine function of $m_1,\ldots,m_{i+1}$ (if all $m_j \geq n_0$).

    Let $m_1,\ldots,m_{i+1} \geq n_0$. 
    By eq.~\eqref{eq:affineind}, letting 
    $\vm=(m_1,\ldots,m_{i},n_0,m_{i+2}', \ldots ,m_k')$ 
    and
    $\vv=(\underbrace{0,\ldots,0}_{i},1,0,\ldots,0)$,
    we obtain that:
    \begin{eqnarray}
        && f_{i+1}(m_{1}, \ldots ,m_{i+1})                    \nonumber
        \\&=& f(m_1, \ldots ,m_i,m_{i+1},m_{i+2}', \ldots ,m_k') \nonumber
        \\&=& 
        f(m_1, \ldots ,m_i,n_0,m_{i+2}', \ldots ,m_k') +               \label{eq:1}
        \\&& 
        (m_{i+1} -  n_0) \cdot \left[ f(m_1, \ldots m_i,n_0+1,m_{i+2}' \ldots ,m_k') -  f(m_1, \ldots m_i,n_0,m_{i+2}', \ldots ,m_k') \right]  \nonumber
    \end{eqnarray}
    By the inductive hypothesis, defining $f_{i,0},f_{i,1}:\N^i \to \N$ by
    $f_{i,0}(m_1,\ldots,m_i) = f(m_1, \ldots m_i,n_0,m_{i+2}', \ldots ,m_k')$ 
    and 
    $f_{i,1}(m_1,\ldots,m_i) = f(m_1, \ldots m_i,n_0+1,m_{i+2}', \ldots ,m_k')$,
    $f_{i,0}$ and $f_{i,1}$ are both $\N$-affine.
    Thus for some $a_{1}, \ldots a_i, b \in \N$ and $c_{1}, \ldots c_i, d \in \N$, 
    \begin{equation} \label{eq:2}
        f_{i,0}(m_1, \ldots, m_i) = b + a_{1} m_{1} + \ldots + a_i m_i
    \end{equation}
    \begin{equation} \label{eq:3}
        f_{i,1}(m_1, \ldots, m_i) = d + c_{1} m_{1} + \ldots + c_i m_i
    \end{equation}
    
    Substituting \eqref{eq:2} and \eqref{eq:3} into \eqref{eq:1}, 
    defining $f_{i+1}:\N^{i+1}\to\N$ for all $m_1,\ldots,m_{i+1}$ by
    \begin{eqnarray*}
        f_{i+1}(m_1,\ldots,m_{i+1}) 
        &=& 
        f(m_1, \ldots ,m_i,m_{i+1},m_{i+2}', \ldots ,m_k')
        \\&=&
        (b + a_{1} m_{1} + \ldots + a_i m_i) +
        \\&&
        (m_{i+1} -  n_0) \cdot \left[ (d + c_{1} m_{1} + \ldots + c_i m_i) -  (b + a_{1} m_{1} + \ldots + a_i m_i) \right].
    \end{eqnarray*}
    Let $a'_{i+1} = d-b$ and $b' = b - a'_{i+1} n_0$.
    For $j \in \{1,\ldots,i\}$, let
    $e_j = c_j-a_j$ and
    $a'_j = a_j-e_jn_0$.
    Then
    \begin{eqnarray}
        &&f_{i+1}(m_1, \ldots ,m_{i+1})   \nonumber 
        \\&=&  \nonumber
        b + a_{1} m_{1} + \ldots + a_i m_i +
        (m_{i+1} -  n_0) \cdot \left[ a_{i+1} + e_{1} m_{1} + \ldots + e_i m_i \right]
        \\&=& 
        b + a_{1} m_{1} + \ldots + a_i m_i +
        a_{i+1} m_{i+1}
        +
        e_{1} m_{1} m_{i+1} + e_{2} m_{2} m_{i+1} + \ldots + e_i m_i  m_{i+1}
        \nonumber
        \\&&
        -  a_{i+1} n_0 -  e_1 m_1 n_0 -  e_2 m_2 n_0 - \ldots -  e_i m_i n_0
        \nonumber
        \\&=& 
        b - a_{i+1} n_0
        + \sum_{j=1}^{i+1} a_{j} m_{j} 
        + \sum_{j=1}^{i} e_{j} m_{j} m_{i+1} 
        - \sum_{j=1}^{i} e_j m_j n_0
        \nonumber
        \\&=& 
        b'
        + \sum_{j=1}^{i+1} a'_{j} m_{j} 
        + \sum_{j=1}^{i} e_{j} m_{j} m_{i+1} 
        \label{eq:quad}
    \end{eqnarray}
    
    To prove $f_{i+1}$ is $\N$-affine, we must show that $e_{1}=e_2=\ldots =e_i = 0$ and that $a'_{i+1} \geq 0$ (the inductive hypothesis impies that $b,a_1, \dots, a_i \geq 0$, so if each $e_j = 0$, then $a'_j = a_j-e_jn_0 = a_j \geq 0$).
    First, we claim that $e_1,\ldots,e_i \geq 0$.
    To see why, 
    suppose for the sake of contradiction that some $e_l < 0$, 
    and consider fixing $m_j$ for each $j \not\in\{l,i+1\}$ and taking the limit of $m_l = m_{i+1} = n$ as $n\to\infty$.
    Equation~\eqref{eq:quad} will be dominated by the quadratic term in $n$, i.e., $e_l m_l m_{i+1} = e_l n^2$, and thus $f_{i+1}$ would be negative for large enough $n$, a contradiction.

    We will eventually show that $e_1,\ldots,e_i = 0$.
    Once this is shown, a similar argument shows that $a'_{i+1} \geq 0$, or else with increasing $m_{i+1}$, 
    $f_{i+1}$ would become negative.
    (The assumption $a'_{i+1} \geq 0$ is not used subsequently in the proof of the inductive case; 
    it is only needed at the conclusion.)
    
    To show that $e_{1}=e_2=\ldots =e_i = 0$,
    define $\vv = (\underbrace{1,\ldots,1}_{i+1},\underbrace{0,\ldots,0}_{k-i-1}) \in \N^k$.
    Let $\vm = (\underbrace{n_0,\ldots,n_0}_{i+1},m_{i+2}',\ldots,m_k')$.
    Let $\vm_{i+1} = (\underbrace{n_0,\ldots,n_0}_{i+1}) \in \N^{i+1}$ be the first $i+1$ coordinates of $\vm$.
    Similarly, let $\vv_{i+1} = (\underbrace{1,\ldots,1}_{i+1}) \in \N^{i+1}$ be the first $i+1$ coordinates of $\vv$.
    Then $f(\vm) = f_{i+1}(\vm_{i+1})$, $f(\vm+\vv) = f_{i+1}(\vm_{i+1}+\vv_{i+1})$, and $f(\vm+2\vv) = f_{i+1}(\vm_{i+1}+2\vv_{i+1})$, 
    Applying the hypothesis of the lemma and these identities, 
    $$
        f_{i+1}(\vm_{i+1}+\vv_{i+1}) - f_{i+1}(\vm_{i+1})
        =
        f_{i+1}(\vm_{i+1}+2\vv_{i+1}) - f_{i+1}(\vm_{i+1}+\vv_{i+1})
    $$
    Substituting eq.~\eqref{eq:quad}, 
    \begin{eqnarray*}
        f_{i+1}(\vm_{i+1}+\vv_{i+1}) - f_{i+1}(\vm_{i+1})
        &=&
        \left[
        b'
        + \sum_{j=1}^{i+1} a'_{j} (n_0+1) + \sum_{j=1}^{i} e_{j} (n_0+1)^2
        \right]
        \\&&
        -
        \left[
        b'
        + \sum_{j=1}^{i+1} a'_{j} n_0 + \sum_{j=1}^{i} e_{j} n_0^2 
        \right]
        \\&=&
        \sum_{j=1}^{i+1} a'_j + \sum_{j=1}^{i} e_j (2n_0 + 1)
    \end{eqnarray*}
    Similarly, the difference $f(\vm+2\vv) - f(\vm+\vv)$ can be expressed as
    \begin{eqnarray*}
        f_{i+1}(\vm_{i+1}+2\vv_{i+1}) - f_{i+1}(\vm_{i+1}+\vv_{i+1})
        &=&
        \left[
        b' 
        + \sum_{j=1}^{i+1} a'_{j} (n_0+2) + \sum_{j=1}^{i} e_{j} (n_0+2)^2 
        \right]
        \\&&
        -
        \left[
        b'
        + \sum_{j=1}^{i+1} a'_{j} (n_0+1) + \sum_{j=1}^{i} e_{j} (n_0+1)^2 
        \right]
        \\&=&
        \sum_{j=1}^{i+1} a'_j + \sum_{j=1}^{i} e_j (2n_0 + 3)
    \end{eqnarray*}
    Since the first difference equals the second, 
    $
        \sum_{j=1}^{i+1} e_j (2n_0 + 1) = \sum_{j=1}^{i+1} e_j (2n_0 + 3).
    $
    Since each $e_j \geq 0$, this implies each $e_j=0$, so $a'_j = a_j$.
    Thus 
    $
        f_{i+1}(m_1, \ldots ,m_{i+1})
        =
        b'+\sum_{j=1}^{i+1} a_{j} m_{j},
    $
    an $\N$-affine function, proving the inductive case.
\end{proof}

The contrapositive of Lemma~\ref{lem:constant-difference-implies-affine}
states that if $f$ is not eventually $\N$-affine, then 
for all $m_0 \in \N$, 
there is 
$\vv\in\{0,1\}^k$ and $\vm\in\N^k_{\geq m_0}$
such that 
$f(\vm+\vv)-f(\vm) \neq f(\vm+2\vv)- f(\vm+\vv)$.
Since $\{0,1\}^k$ is finite, 
by the pigeonhole principle, 
some infinite subset of such $\vm$'s can be found
that agree on the same $\vv$.
Thus we have the following corollary.

\begin{corollary} \label{cor:constant-difference-implies-affine}
    Let $f:\N^k\to\N$.
    If $f$ is not eventually $\N$-affine,
    then there is $\vv\in\{0,1\}^k$ 
    such that
    for all $m_0 \in \N$, 
    there is $\vm\in\N^k_{\geq m_0}$
    such that 
    $f(\vm+\vv)-f(\vm) \neq f(\vm+2\vv)- f(\vm+\vv)$.
\end{corollary}

Finally, we need not only that we can find an infinite set of ``arbitrarily large counter-examples to affine''
given by Corollary~\ref{cor:constant-difference-implies-affine},
but \emph{furthermore} that this set is $\alpha$-dense for some $\alpha>0$.
This is not true for arbitrary functions.
For example, consider the function $f(m_1,m_2) = m_1+m_2$ unless $m_1 = 2^{m_2}$, in which case $f(m_1,m_2)=0$ instead.
However, if $f$ is a semilinear function, then we can find an infinite set of ``counter-examples to affine''
that \emph{is} $\alpha$-dense for some $\alpha>0$.

\todo{DD: this is the only point where we really rely on the stably computable functions being limited to semilinear; maybe point this out and factor out a separate theorem that assumes $\alpha$-density of counterexamples, but dispenses with the need to rely on the theorem telling us that $f$ must be semilinear if it is stably computable}

\begin{lemma} \label{lem:constant-difference-implies-affine-and-dense}
    Let $f:\N^k\to\N$ be semilinear but not eventually $\N$-affine.
    Then there are 
    $\vv\in\{0,1\}^k$, 
    $\alpha>0$, 
    and infinitely many $\alpha$-dense $\vm\in\N^k$
    such that 
    $f(\vm+\vv)-f(\vm) \neq f(\vm+2\vv)- f(\vm+\vv)$.
\end{lemma}

\begin{proof}
    For all $\vv \in \{0,1\}^k$,
    let $C_\vv = \{ \vm\in\N^k \mid f(\vm+\vv)-f(\vm) \neq f(\vm+2\vv)- f(\vm+\vv) \}$.
    Corollary~\ref{cor:constant-difference-implies-affine} tells us that
    there is $\vv\in\{0,1\}^k$ 
    such that,
    for all $m \in \N$, 
    we have
    $\N^k_{\geq m} \cap C_\vv \neq \emptyset$.

    This implies that for all $m \in \N$,
    $|C_\vv \cap \N^k_{\geq m}| = \infty$.
    Otherwise, since $\N^k_{\geq m} \subset \N^k_{\geq m+1}$ for all $m$,
    if 
    $|C_\vv \cap \N^k_{\geq m}| < \infty$,
    then for sufficiently large $m'$,
    $C_\vv \cap \N^k_{\geq m'} = \emptyset$,
    contradicting the fact that $C$ intersects \emph{all} $\N^k_{\geq m'}$.

    We denote the graph of $f$ by $G(f) = \{ (\vm,f(\vm)) \mid \vm \in \N^k \} \subset \N^{k+1}$.
    Since $f$ is semilinear, $G(f)$ is a semilinear set.
    Thus it is a finite union 
    $G(f) = \bigcup_{j=0}^{p-1} P_j$
    of $p$ periodic cosets $P_0,\ldots,P_{p-1}$. 
    Intuitively, the lemma will be proven as follows.
    $C_\vv$, the ``counterexamples to $\N$-affineness'', occur in these periodic cosets.
    Adding period vectors to a point keeps it in the same periodic coset.
    However, if we add the same multiple of period vectors each time, and if their sum is positive on all coordinates, then this sets a minimum $\alpha$-denseness that we cannot fall below.
    This is how we will show that $C_\vv$ has infinitely many $\alpha$-dense points.
        
    For all $\vm \in \N^k$, %and $i \in \{0,1,2\}$,
    let 
    % $\vm_i = \vm + i \vv$,
    % and let
    $\vm^f = (\vm,f(\vm)) \in \N^{k+1}$ denote the point in $G(f)$ corresponding to input $\vm$.
    % $\vm^f_1 = (\vm+\vv,f(\vm+\vv))$,
    % and
    % $\vm^f_2 = (\vm+2\vv,f(\vm+2\vv))$.
    By the pigeonhole principle, 
    there are $x,y,z \in \{0,\ldots,p-1\}$
    and an infinite subset $C_P \subseteq C_\vv$ 
    such that,
    for all $m_0$, 
    there is $\vm \in C_P \cap \N^k_{\geq m_0}$
    such that 
    $\vm^f \in P_x$,
    $(\vm+\vv)^f \in P_y$,
    $(\vm+2\vv)^f \in P_z$.
    For notational convenience we assume that $x=0,y=1,z=2$.\footnote{
        This can be assumed without loss of generality even if some of $x,y,z$ are equal, 
        since we can have ``duplicate'' periodic cosets that are equal despite having different subscripts.
    }
    In other words, we can find an infinite subset $C_P$ of $C_\vv$ in which,
    in the finite union of periodic cosets defining $G(f)$,
    $\vm^f$ is always part of the same periodic coset $P_0$, 
    $(\vm+\vv)^f$ is always part of the same periodic coset $P_1$,
    and
    $(\vm+2\vv)^f$ is always part of the same periodic coset $P_2$.

    \newcommand{\vmlarge}{\vec{m}_\mathsf{L}}
    \newcommand{\vmsmall}{\vec{m}_\mathsf{S}}
    
    \todo{DD: This argument is very symbolic and technical. A figure might help.}
    Let $\vmsmall^f,\vmlarge^f \in C_P$ such that
    $\min(\vmlarge - \vmsmall) \geq 1$;
    i.e., $\vmlarge$ is strictly larger than $\vmsmall$ on all coordinates.
    By the fact that $C_P$ contains arbitrarily large points 
    (i.e., points in $\N^k_{\geq m}$ for all $m$), 
    such $\vmsmall$ and $\vmlarge$ must exist.
    
    Let $\vb_0,\vp_0^{(1)},\ldots,\vp_0^{(l_0)} \in\N^{k+1}$ be such that 
    $$P_0 = \left\{ \left. \vb_0 + n^{(1)} \vp_0^{(1)} + \ldots + n^{(l_0)} \vp_0^{(l_0)} \right| n^{(1)},\ldots,n^{(l_0)}\in\N \right\},$$
    and similarly for 
    $\vb_1,\vp_1^{(1)},\ldots,\vp_1^{(l_1)}, \vb_2,\vp_2^{(1)},\ldots,\vp_2^{(l_2)} \in\N^{k+1}$ 
    and $P_1,P_2$,
    respectively.
    
    For $i\in\{0,1,2\}$, 
    let $\vd_i^f = (\vmlarge+i\vv)^f - (\vmsmall+i\vv)^f$,
    and let $\vd_i$
    denote $\vd_i^f$ restricted to the first $k$ coordinates,
    so that $\vd_i = (\vmlarge+i\vv) - (\vmsmall+i\vv) = \vmlarge-\vmsmall$.
    In other words, $\vd_0^f,\vd_1^f,\vd_2^f$ differ only on their last coordinate,
    representing $f(\vmlarge+i\vv) - f(\vmsmall+i\vv)$.
    Then $\vd_0^f = \sum_{j=1}^{l_0} n^{(j)}_{0} \vp_0^{(j)}$ 
    for some $n^{(1)}_{0},\ldots,n^{(l_0)}_{0} \in \N$
    since $\vmsmall^f,\vmlarge^f \in P_0$.
    Similarly, there must exist 
    $n^{(1)}_{1},\ldots,n^{(l_1)}_{1} \in \N$ and 
    $n^{(1)}_{2},\ldots,n^{(l_2)}_{2} \in \N$
    such that 
    $\vd_1^f = \sum_{j=1}^{l_1} n^{(j)}_{1} \vp_1^{(j)}$
    and
    $\vd_2^f = \sum_{j=1}^{l_2} n^{(j)}_{2} \vp_2^{(j)}$.

    For all $n\in\N$, 
    define 
    $\vm_n^f = \vmsmall^f + n \vd_0^f$
    (note that $\vmlarge^f = \vm_1^f$ and $\vmsmall^f = \vm_0^f$),
    letting $\vm_n$ be $\vm_n^f$ restricted to the first $k$ coordinates.
    By the definition of $P_0$, 
    for all $n\in\N$,
    $\vm_n^f \in P_0$,
    since $\vmsmall^f \in P_0$ and $\vd_0^f$ is a nonnegative multiple of period vectors in $P_0$.
    Our goal next is to show that all sufficiently large $\vm_n^f$ are in $C_P \subseteq C_\vv$.
    After showing that, we argue that $\vm_n$ are all $\alpha$-dense for some $\alpha > 0$, proving the lemma.
    
    For $i\in\{0,1,2\}$, let $\Delta y_i = \vd_i^f(k+1) = f(\vmlarge+i\vv) - f(\vmsmall +i\vv)$.
    Note that for all $n\in\N$,
    % $\Delta y_0 = \vq_{n+1}^f(k+1) - \vq_{n}^f(k+1)$,
    % $\Delta y_1 = \vr_{n+1}^f(k+1) - \vr_{n}^f(k+1)$,
    % and
    % $\Delta y_2 = \vs_{n+1}^f(k+1) - \vs_{n}^f(k+1)$,
    % or put another way,
    % for each $i\in\{0,1,2\}$,
    $f(\vmsmall+i\vv+n\vd_i) = f(\vmsmall+i\vv) + n\Delta y_i$.
    
    We claim that for all $n\in\N$,
    $\vm_n \in C_\vv$.
    Let $c_0 = f(\vmsmall+\vv) - f(\vmsmall)$
    and $c_1 = f(\vmsmall+2\vv) - f(\vmsmall+\vv)$.
    Note that $c_0 \neq c_1$ since $\vmsmall \in C_\vv$.
    For all $n\in\N$,
    \begin{eqnarray*}
        f(\vm_n+\vv) - f(\vm_n)
        &=&
        f(\vmsmall+\vv + n \vd_1) - f(\vmsmall + n \vd_0)
        \\&=&
        f(\vmsmall+\vv) + n \Delta y_1 - (f(\vmsmall) + n \Delta y_0)
        \\&=&
        % f(\vmsmall+\vv) - f(\vmsmall) + n (\Delta y_1 - n\Delta y_0)
        % \\&=&
        c_0 + n (\Delta y_1 - \Delta y_0),
    \end{eqnarray*}
    and
    \begin{eqnarray*}
        f(\vm_n+2\vv) - f(\vm_n+\vv)
        &=&
        f(\vmsmall+2\vv + n \vd_2) - f(\vmsmall+\vv + n \vd_1)
        \\&=&
        f(\vmsmall+2\vv) + n \Delta y_2 - (f(\vmsmall+\vv) + n \Delta y_1)
        \\&=&
        % f(\vmsmall+2\vv) - f(\vmsmall+\vv) + n (\Delta y_2 - \Delta y_1)
        % \\&=&
        c_1 + n (\Delta y_2 - \Delta y_1).
    \end{eqnarray*}
    Thus 
    \begin{eqnarray*}
        [f(\vm_n+\vv) - f(\vm_n)] - [f(\vm_n+2\vv) - f(\vm_n+\vv)]
        &=&
        [c_0 + n (\Delta y_1 - \Delta y_0)] - [c_1 + n (\Delta y_2 - \Delta y_1)]
        \\&=&
        (c_0-c_1) + n (2 \Delta y_1 - \Delta y_2 - \Delta y_0).
    \end{eqnarray*}
    There are two cases to complete the claim that $\vm_n \in C_\vv$:
    \begin{description}
        \item[\underline{$2 \Delta y_1 - \Delta y_2 - \Delta y_0 = 0$}:]
        Then for all $n$, the above difference is $c_0 - c_1 \neq 0$, 
        so $f(\vm_n+\vv) - f(\vm_n) \neq f(\vm_n+2\vv) - f(\vm_n+\vv)$,
        implying $\vm_n \in C_\vv$.
        
        \item[\underline{$2 \Delta y_1 - \Delta y_2 - \Delta y_0 \neq 0$}:]
        Then for all sufficiently large $n$,
        $(c_0-c_1) + n (2 \Delta y_1 - \Delta y_2 - \Delta y_0) \neq 0$,
        so $f(\vm_n+\vv) - f(\vm_n) \neq f(\vm_n+2\vv) - f(\vm_n+\vv)$,
        implying $\vm_n \in C_\vv$.
    \end{description}
    
    Letting $\alpha' = \min(\vd_0) / \| \vd_0 \|$, we have that for all $n\in\N$, $n \vd_0$ is $\alpha'$-dense.
    Note that $\alpha' > 0$ since we chose $\vd_0 = \vmlarge - \vmsmall$ to be positive on all coordinates.
    Let $\alpha = \alpha'/2$.
    Then for all sufficiently large $n$,
    $\vm_n = \vmsmall + n \vd_0$ is $\alpha$-dense.
    Since $\vm_n \in C_P \subseteq C_\vv$, 
    this proves the lemma.
\end{proof}

\subsection{Time lower bounds for stably computing nonlinear functions}

Recall that definition of $\N$-linear functions from Section~\ref{sec:negative-results}.
We say a function $f: \N^k \to \N$ is \emph{eventually $\N$-linear} 
if it is eventually $\N$-affine with offset $b=0$,
i.e.,
if there are $c_1,\ldots,c_k \in \N$ and $m_0 \in \N$
such that for all $\vm\in\N^k_{\geq m_0}$,
$f(\vm) = \sum_{i=1}^k c_i \vm(i)$.
% We say a function $f: \N^k \to \N$ is \emph{eventually $\N$-linear} 
% if it is eventually $\N$-affine with offset $b=0$, i.e., $f(\vec{0}) = 0$.
% We say the function is \emph{$\N$-linear} if it is eventually $\N$-linear with $m_0 = 0$.
% Similarly, a function is \emph{$\Q_{\geq 0}$-linear} 
% if there are $c_1,\ldots,c_k \in \Q_{\geq 0}$
% such that for all $\vm\in\N^k$,
% $f(\vm) = \sum_{i=1}^k \floor{c_i \vm(i)}$.

The following lemma is a special case of the main result of this Section,
Theorem~\ref{thm:negative-result-eventually-positive-integral-linear},
showing that the non-linear functions in a sense ``closest'' to being eventually $\N$-linear,
the eventually $\N$-affine functions,
are not stably computable in sublinear time.
% Note that a function $f:\N^k\to\N$
% is affine but not linear if there are constants 
% $b,c_1,\ldots,c_k \in \N$, not all $0$,
% and
% $q_1,\ldots,q_k \in \Q$
% such that, for all $\vm\in\N^k$,
% $f(\vm) = b + \sum_{i=1}^k q_i (\vm(i)-c_i)$.
This includes functions as simple as $f(m)=m-1$
(computable slowly via transition $\ins,\ins \to \ins,\outs$).
%In fact we can easily generalize to functions that are \emph{eventually}-$\N$-affine but not eventually $\N$-linear.

\begin{theorem} \label{thm:affine-not-sublinear-computable}
    Let $f:\N^k\to\N$ be eventually $\N$-affine but not eventually $\N$-linear.
    Then no function-computing leaderless population protocol stably computes $f$ in sublinear time.
\end{theorem}

\begin{proof}
    Since $f$ is eventually $\N$-affine,
    there $m_0\in\N$ and $b,c_1,\ldots,c_k\in\N$ such that,
    for all $\vm\in\N^k_{\geq m_0}$,
    $f(\vm) = b + \sum_{i=1}^k c_i \vm(i)$.
    Since $f$ is not eventually $\N$-linear, $b \neq 0$.
    Thus,
    for all $\vm\in\N^k_{\geq m_0}$,
    $f(2\vm) \neq 2 f(\vm)$.
    
    Let $I$ be the set of all $\alpha$-dense valid initial configurations $\vi$ representing input in $\N^k_{\geq m_0}$
    such that $f(2 \vm) > 0$.
    Let $S = \{ \vo \mid (\exists \vi \in I_{\alpha})\ \vi \reach \vo$ and $\vo$ is stable$\}$.
    By hypothesis we have that for each $\vi \in I$, $\time{\vi}{S} = o(n)$.
    
    Apply Corollary~\ref{cor:push-Delta-simplified}.
    \corPushDeltaSimplifiedConclusion

    Let $b = 0$; below we ensure that $\max(\vd^\Delta) \leq b$.
    Let $\vm = \vi_{n_b}\rest\Sigma$.
    % Note that $\outs \in \Gamma$ since $f(\vm)$ grows unboundedly with $\|\vm\|$.
    Let $\vs_0 = 2 \vo_{n_b}^\Gamma + \tvD_1.\vo_{n_b}^\Delta$.
    Letting $\vd^\Delta = \vec{0}$,
    the above argument shows that $2 \vi_{n_b} \reach \vs_0$.
    Since $\vs_0 \in\N^\Gamma$, it is stable by Observation~\ref{obs:unbdd-state-configs-stable}.
    Since the protocol is correct, $\vs_0(\outs) = f(2\vm) > 0$.
    Then $4 \vi_{n_b} \reach 2\vs_0$, also stable, with $2\vs_0(\outs) = 2 f(2\vm) \neq f(4 \vm)$,
    a contradiction.
\end{proof}

Finally, we can prove the main result of this section, 
that any protocol requires linear expected time to stably compute a non-eventually-$\N$-linear function.

Intuitively, the proof has a similar shape to that of  Theorem~\ref{thm:predicates}.
However, applying the surgery lemmas to fool the population protocol is more difficult,
since with the predicate output convention, 
it is immediate that if two configurations have a well-defined output and share at least one state in common, they must have the same output (equal to the vote of that state).
With the integer-count output convention, things are trickier.
The surgery of Corollary~\ref{cor:push-Delta-simplified} 
to consume additional input states affects the count of the output state---how do we know that the effect of the surgery on the output is not consistent with the desired output of the function?
In order to arrive at a contradiction we develop two techniques. 
The first involves showing that the slope of the change in the count of the output state as a function of the input states is inconsistent.
The second involves exposing the semilinear structure of the graph of the function being computed, and forcing it to enter the ``wrong piece'' (i.e., periodic coset).

\begin{theorem} \label{thm:negative-result-eventually-positive-integral-linear}
    Let $f: \N^k \to \N$, 
    and let $\mathcal{C}$ be a function-computing leaderless population protocol that stably computes $f$.
    If $f$ is not eventually $\N$-linear then 
    $\calC$ takes expected time $\Omega(n)$. 
\end{theorem}

\begin{proof}
    If $f$ is eventually $\N$-affine but not eventually $\N$-linear, 
    then Theorem~\ref{thm:affine-not-sublinear-computable}
    implies that $f$ requires expected time $\Omega(n)$ to stably compute.
    So assume that $f$ is not eventually $\N$-affine.
    
    Since $f$ is semilinear, 
    Lemma~\ref{lem:constant-difference-implies-affine-and-dense}
    tells us that
    there is $\vv\in\{0,1\}^k$ and $\alpha>0$
    such that
    % for all $m_0 \in \N$, there is 
    there are infinitely many
    $\alpha$-dense $\vm\in\N^k$
    such that 
    $f(\vm+\vv)-f(\vm) \neq f(\vm+2\vv)- f(\vm+\vv)$.
    Let $C$ be the set of all such $\vm$.
    
    % For each point $\vm\in\N^k$ and $d\in\N$,
    % define the set $C^d_{\vm} = \{\vm + \sum_{i=1}^k c_i \vu_i \mid c_1,\ldots,c_k \in \{0,1,\ldots,d\} \}$,
    % the ``cone of length $d$ above $\vm$'', 
    % which constitutes the $(d+1)^k$ points obtainable by adding up to $d$ to each of the $k$ coordinates of $\vm$.
    
    Note that for every point $\vm'\in\N^k$,
    there exists $\vm\in\N^k$ and $\vu \in \{0,1\}^k$ such that $\vm' = 2\vm + \vu$;
    i.e., every point $\vm'$ is equal some point $2\vm$ with all even coordinates, plus some binary vector $\vu$ (to reach the odd coordinates in $\vm'$).
    
    Let $I$ be a set of $\alpha$-dense valid initial configurations $\vi$
    representing an input $\vm$ such that there is some $\vm' \in C$ and $\vu \in \{0,1\}^k$
    such that $\vm' = 2\vm + \vu$.
    Let $S = \{ \vo \mid (\exists \vi \in I)\ \vi \reach \vo$ and $\vo$ is stable$\}$
    be the set of stable configurations reachable from some initial configuration in $I$.
    Assume for the sake of contradiction that for each $\vi \in I$, $\time{\vi}{S} = o(n)$.
    
    Apply Corollary~\ref{cor:push-Delta-simplified}.
    \corPushDeltaSimplifiedConclusion

    Some of the inputs $\ins_1,\ldots,\ins_k$ are in $\Delta$,
    and others are in $\Gamma$.
    Let 
    $\vu^\Delta = \vu \rest \Delta$,
    $\vu^\Gamma = \vu \rest \Gamma$,
    $\vv^\Delta = \vv \rest \Delta$,
    and
    $\vv^\Gamma = \vv \rest \Gamma$.
    % Let $\vv^\Delta = \vv \rest \Delta$
    % denote the counts in $\vv$ of inputs in $\Delta$,
    % and let $\vv^\Gamma = \vv \rest \Gamma$ denote the remaining counts in $\vv$.
    %
    Let $b = 3$; below we ensure that $\max(\vd^\Delta) \leq b$.
    Let $\vm\in\N^k$ be the input represented by $\vi_{n_b}$.
    Let $\vm' \in C$ and $\vu \in \{0,1\}^k$ such that $\vm' = 2\vm + \vu$.
    Let $\vs_0 = 2 \vo_n^\Gamma + \tvD_1.\vo_n^\Delta + \tvD_2.\vu^\Delta$.
    Letting $\vd^\Delta = \vu^\Delta$ in Lemma~\ref{lem:push-Delta-simplified}, % and $\vq = \tvD_2.\vd^\Delta$.
    we have $2\vi_{n_b}+\vu^\Delta \reach \vs_0$.
    By correctness, since initial configuration $2\vi_{n_b}+\vu$ represents input $\vm'$,
    we have $\vs_0(\outs) = f(\vm')$.
    
    Let $\vs_1 = \vs_0 +   \tvD_2.\vd^\Delta$.
    Then letting $\vd^\Delta = \vu^\Delta + \vv^\Delta$ in Lemma~\ref{lem:push-Delta-simplified},
    we have that
    $2\vi_{n_b}+\vu^\Delta+\vd^\Delta \reach \vs_1$.
    By additivity this implies that
    $2\vi_{n_b}+\vu^\Delta+\vu^\Gamma+\vv^\Gamma \reach \vs_1 + \vu^\Gamma + \vv^\Gamma$.
    By correctness, $\vs_1(\outs) = f(\vm'+\vv)$.
    By Corollary~\ref{cor:unbdd-state-configs-stable},
    $\vs_1 + \vu^\Gamma + \vv^\Gamma$ is stable.
    
    Let $\vs_2 = \vs_0 + 2 \tvD_2.\vd^\Delta$.
    Then letting $\vd^\Delta = \vu^\Delta + 2 \vv^\Delta$ in Lemma~\ref{lem:push-Delta},
    we have that
    $2\vi_{n_b}+\vu^\Delta+2\vv^\Delta \reach \vs_2$.
    By additivity this implies that
    $2\vi_{n_b}+\vu^\Delta+2\vv^\Delta+\vu^\Gamma+2\vv^\Gamma \reach \vs_2 + \vu^\Gamma + 2 \vv^\Gamma$.
    By Corollary~\ref{cor:unbdd-state-configs-stable},
    $\vs_1 + \vu^\Gamma + 2 \vv^\Gamma$ is stable.
    
    By linearity,
    $\vs_2(\outs) = \vs_0(\outs) + 2 \vq(\outs)$.
    However, since 
    $f(\vm'+\vv)-f(\vm') \neq f(\vm'+2\vv)- f(\vm'+\vv)$,
    this implies $\vs_2(\outs) \neq f(\vm'+2\vv)$.
    However, since $\vs_2$ is stable and reachable from $2\vi_{\vm} + \vu + 2\vv$, representing input $\vm'+2\vv$,
    this implies that the protocol stabilizes on the wrong output.
\end{proof}

\section{Conclusion}
\label{sec:conclusion}

% \todoi{Summarize results}
Some interesting questions remain open.

\paragraph{Time complexity of non-eventually-$\N$-linear functions and non-eventually-constant predicates.}
The most obvious open question is to determine the optimal stabilization time complexity of computing semilinear functions and predicates
not satisfying the hypotheses of 
Theorems~\ref{thm:predicates}
and~\ref{thm:negative-result-eventually-positive-integral-linear};
namely the \emph{eventually $\N$-linear} functions,
(e.g., $f(m)=0$ if $m<3$ and $f(m)=m$ otherwise)
and \emph{eventually constant} predicates
(e.g., $\phi(m)=1$ iff $m\geq 2$).
(See Sections~\ref{sec:nonlinear-functions} and~\ref{sec:predicates} for formal definitions.)
The only known examples of eventually $\N$-linear functions computable in sublinear time 
are the $\N$-linear functions
(e.g., $f(m)=2m$),
computable in logarithmic time by Observation~\ref{obs:stably-compute-positive-integer-coefficient}.
The only known examples of eventually constant predicates that are computable in sublinear time 
are the \emph{detection predicates} studied in~\cite{SpeedFaultsDIST}
(and observed to be decidable in logarithmic time):
predicates whose value depends only on the presence or absence of certain inputs, 
but not on their exact positive values
(e.g., $\phi(m)=1$ iff $m \geq 1$).

\paragraph{Allowing more than $O(1)$ states.}
Alistarh, Aspnes, Eisenstat, Gelashvili, and Rivest~\cite{alistarh2017timespace} 
showed time lower bounds for leader election and majority with superconstant states
(i.e., the number of states is allowed to grow with the population size). 
In this paper we have shown time lower bounds for more general function and predicate computation with a \emph{constant} set of states.
It is natural to ask whether more general function and predicate computation time lower bounds 
can be proven for superconstant states using similar techniques.
Notably, Chatzigiannakis, Michail, Nikolaou, Pavlogiannis, and Spirakis~\cite{CMNPS11} 
have shown that if the number of states $\lambda_n = o(\log n)$,
then the protocol's computational power remains limited to computing semilinear predicates, 
the same limitation that applies when $\lambda_n = O(1)$.\footnote{The bound on states is described in~\cite{CMNPS11} in terms of space available to a Turing machine that computes the transition function, permitting $o(\log \log n)$ space.
Since a Turing machine with space $s$ has $2^{O(s)}$ configurations, this is equivalent to requiring that the number of states be at most $o(\log n)$ (and furthermore that the transition function be uniformly computed by a single Turing machine.)}
However, with $\lambda_n = \Omega(\log n)$, 
the computational ability of population protocols moves beyond semilinear predicates~\cite{CMNPS11}.
Thus, there is a wider class of functions and predicates under this relaxed constraint,
which may require new techniques to analyze their population protocol time complexity.

\paragraph{Convergence time without a leader.}
Although we measure computation time with respect to stabilization---the ultimate goal of stable computation---some work uses a different goalpost for completion.
Consider a protocol stably computing a function,
and consider one particular transition sequence that describes its history.
We can say the transition sequence \emph{converged}
at the point when the output count is the same in every subsequently reached configuration. 
In contrast, recall that the point of stabilization is when the output count is the same in every subsequently \emph{reachable} configuration 
(whether actually reached in the transition sequence or not).
In other words, after stabilization, even an adversarial scheduler cannot change the output.
Measuring time to stabilization in the randomized model, as we do here, 
measures the expected time until the probability of changing the output becomes $0$.
% All subsequent references in this paper to expected time for population protocol computation refer to time until stabilization.

Our proof shows only that stabilization must take expected $\Omega(n)$ time for ``most'' predicates and functions if there is no initial leader.
However, convergence could occur much earlier in a transition sequence than stabilization.
Indeed,
Kosowski and Uzna\'{n}ski~\cite{kosowski2018brief} show that all semilinear predicates can be computed without an initial leader,
converging in $O(\polylog~n)$ time if a small probability of error is allowed,
and converging in $O(n^\epsilon)$ time with probability 1, 
where $\epsilon$ can be made arbitrarily close to 0 by changing the protocol.

% We leave as an open question whether there is a protocol that converges in expected $o(n)$ time 
% and stably computes a function not of the form $f(\vm) = \sum_{i=1}^k c_i \vm(i)$ where all $c_i \in \N$.
% We reiterate that there are protocols that work with a leader to stably compute semilinear predicates with convergence time $O(\log^5 n)$~\cite{angluin2006fast},
% whereas absent an initial leader, the best known convergence time is $O(n)$~\cite{angluin2006fast, DotHajLDCRNNaCo}.
% Thus if stable leader election can converge in expected sublinear time (itself an open question),
% by coupling the two protocols it might be possible to achieve stable computation of arbitrary semilinear predicates and functions with sublinear convergence time, even from a leaderless initial configuration.

\paragraph{Stabilization time with a leader.}
It remains open to determine the optimal stabilization time for stably computing semilinear predicates and functions \emph{with} an initial leader.
The stably computing protocols converging in $O(\log^5 n)$ time~\cite{angluin2006fast, CheDotSolDetFuncNaCo} 
provably require expected time $\Omega(n)$ to stabilize, and it is unknown whether faster stabilization is possible.

\paragraph{High-probability computation of non-semilinear functions/predicates without a leader.}
Going beyond stable computation, the open question of Angluin, Aspnes, and Eisenstat~\cite{angluin2006fast} 
asks whether their efficient high-probability simulation of a space-bounded Turing machine 
by a population protocol could remove the assumption of an initial leader.
That simulation has some small probability $\epsilon > 0$ of failure, 
so if one could elect a leader with a small probability $\epsilon' > 0$ of error 
and subsequently use it to drive the simulation, 
by the union bound the total probability of error  would be at most $\epsilon + \epsilon'$ (i.e., still close to 0).
Recent work by Kosowski and Uzna\'{n}ski~\cite{kosowski2018brief} may be relevant,
which shows a protocol for leader election converging in $O(\polylog~n)$ time with a small probability of error.
% However, it remains an open question whether the necessary protocol exists.

\paragraph{Non-dense initial configurations without a leader.}
Our general negative result applies to $\alpha$-dense initial configurations.
However, is sublinear time stable computation possible from other kinds of initial configurations that satisfy our intuition of not having preexisting leaders?
In particular, our technique for proving time lower bounds fails when the input states themselves have counts sufficiently skewed to prevent $\alpha$-denseness, even without any other state initially present.
For example, given any function $f:\N\to\N$ computable quickly with a leader,
imagine a function $f_2:\N^2\to\N$ such that
$f_2(m,1) = f(m)$ for all $m\in\N$.
It is difficult to rule out the possibility of a protocol using the unique agent in state $x_2$ as a leader 
to simulate the fast protocol computing $f$.
However, it is not clear is how to ensure that $f_2$ is computed correctly on other inputs where $m_2 \neq 1$,
since it is unknown how to compute whether $m_2=1$ in sublinear time.

\paragraph{Functions approximable in unbounded time.}
% Our main negative result shows that linear functions with some coefficient not in $\N$ cannot even be approximated in sublinear time with sublinear error.
% In contrast, we have a positive result showing 
We showed in Theorem~\ref{thm:positive} that linear functions all coefficients in $\Q_{\geq 0}$ \emph{can} be approximated in logarithmic time with linear error,
but we also showed in Theorem~\ref{thm:main} that linear functions with some coefficient not in $\N$ cannot even be approximated in sublinear time with sublinear error.
However, this is the first study of approximate function computation with population protocols, 
so it is not even clear what functions can be approximated at all with unbounded time. 
Clearly it is not limited to semilinear functions, or even Turing-computable functions.
For example, the identity function is a close approximation of the uncomputable function
$f(m) = m$ if the $m$'th Turing machine halts and $f(m)=m+1$ otherwise. 

\todo{DD: I added the remainder recently (Oct 2018). Is it reasonable?}
A sensible way to formulate the question might be this:
require (as we do) for a protocol on input $\vm \in \N^k$ to stabilize,
but nondeterministically to one of a set of possible outputs $Y_\vm \subseteq \N$.
Define the nondeterministic function it computes to be $f:\N^k \to \mathcal{P}(\N)$ defined by $f(\vm) = Y_\vm$.
What range of such functions can be computed?
This is applicable to approximation when the output set is a small interval.

It might also be interesting to consider multi-valued nondeterministic functions,
where the protocol has $\ell$ different output states $\outs_1,\ldots,\outs_\ell$ and computes a function 
$f:\N^k \to \mathcal{P}(\N^\ell)$.
In fact, this question is interesting even when there is no input,
in the generalized model of chemical reaction networks that permit some reactions (transitions) 
such as $\ins \to \outs,\outs$ or $\ins_1,\ins_2 \to \ins_1$
to alter the population size.
The system starts with a single $\ins$,
and it nondeterministically stabilizes to some counts of $\outs_1,\ldots,\outs_\ell$.
It might appear as though the set of counts that can be reached is semilinear, 
but in fact non-semilinear sets are possible~\cite{hopcroft1979on}.

\paragraph{Alternate Boolean output conventions.}
%\todo{DD: we may want to leave this out}
The standard model by which a population protocol reports a Boolean output is by consensus:
each state ``votes'' 0 or 1, and output is defined only if all states present in a configuration vote unanimously.
Brijder, Doty, and Soloveichik~\cite{recrn} showed that under some reasonable alternative output conventions,
the class of Boolean predicates stably computable remains the semilinear predicates.
However, it may not be the case that the \emph{efficiency} of predicates remains unchanged.
For example, one alternative is the \emph{democratic} output convention:
the output is $b\in\{0,1\}$ if the $b$ voters outnumber the $1-b$ voters.
In this case, the majority predicate becomes trivially computable in constant time,
merely by having each input state $\ins_1$ and $\ins_2$ vote for itself,
and no transitions are even required: 
the initial configuration already reports the correct answer.

\paragraph{Acknowledgements.}
The authors are grateful to Sungjin Im for the proof of Lemma~\ref{lem:constant-difference-implies-affine}.
We also thank anonymous reviewers for their helpful comments.

\bibliographystyle{plain}
\bibliography{div}

\end{document}